\documentclass[article, prx, 12pt,tightenlines,nofootinbib,longbibliography,floatfix,nobalancelastpage,superscriptaddress]{revtex4-2}

\usepackage[utf8]{inputenc}
\usepackage{float}
\usepackage{graphicx}
\usepackage{physics}
\usepackage{esint}
\usepackage{amssymb}
\usepackage{mathtools}
\usepackage{amsmath}
\usepackage{amsthm}
\usepackage{mathrsfs}
\usepackage[colorlinks=true, 
    linkcolor=violet, 
    citecolor=violet,
    urlcolor=violet, plainpages=false, pdfpagelabels]{hyperref}
\usepackage[dvipsnames]{xcolor}
\usepackage{bm}
\usepackage[most]{tcolorbox}
\usepackage{enumitem}
\usepackage{fancyhdr}
\usepackage{tcolorbox}

\counterwithin{section}{part}

\newtcolorbox[auto counter]{example}[1]{%
  colback=violet!2!,colframe=violet!95!black,fonttitle=\bfseries,
  title={Example~\thetcbcounter: #1},enhanced,breakable}
\newtcolorbox[auto counter,number format=\Roman]{nutshell}[1]{%
  colback=black!3!,colframe=black!,fonttitle=\bfseries,
  title={Part~{\thetcbcounter} in a Nutshell},enhanced,breakable}

\newtheorem{theorem}{Theorem}

\numberwithin{equation}{part}
\makeatletter
\renewcommand{\l@subsection}[2]{}   
\renewcommand{\l@subsubsection}[2]{}   

\makeatother

\begin{document}


\title{\LARGE{Benasque Lectures on Gaussian Bosonic Systems and Analogue Gravity}}

\author{Anthony J.~Brady}
\email{ajbrad4123@gmail.com}\thanks{These notes were initiated during my postdoctoral appointment at the University of Arizona in 2023 and were substantially expanded and refined in 2025 during my appointment at the Joint Center for Quantum Information and Computer Science (QuICS), NIST \& University of Maryland.}


\affiliation{
Joint Center for Quantum Information and Computer Science (QuICS), 
NIST \& University of Maryland, College Park, MD 20742, USA
}

\date{\today}

\begin{abstract}
    These notes are adapted from six lectures that I delivered at \textit{Analogue Gravity in Benasque 2023}. They present the unified Gaussian (phase-space) framework to describe linear bosonic quantum systems---the standard tool in quantum optics and continuous-variable quantum information---emphasizing its simplicity and platform independence, with applications to semi-classical black holes and analogue gravity. Parts (\ref{lecture:1}-\ref{lecture:3}) develop the formalism: from harmonic dynamics and Gaussian transformations to state characterization via moments, Wigner functions, and entanglement measures. Part (\ref{lecture:4}) applies these tools to semi-classical black holes---discussing Hawking radiation and quantum superradiance in rotating black holes---and laboratory analogues in light-matter systems via toy models.

\end{abstract}

\maketitle

\newpage
\renewcommand{\baselinestretch}{.9}
\tableofcontents
\newpage
\renewcommand{\baselinestretch}{1.0}


\part*{Introduction}
\addcontentsline{toc}{part}{Introduction}

\begin{quote}
    The career of a young theoretical physicist consists of treating the harmonic oscillator in ever-increasing levels of abstraction. ---Sidney Coleman
\end{quote}
A thorough understanding of quantum harmonic oscillators---their dynamics, environmental interactions, and measurement---is indispensable to any modern physicist. Linear bosonic quantum systems, modeled as coupled networks of quantum oscillators, describe phenomena ranging from quantum information processing to black-hole evaporation. Remarkably, a universal framework exists for their analysis: the \emph{phase-space formalism}, also known as the \emph{Gaussian} or \emph{continuous-variable (CV)} formalism; we use these terms interchangeably. These notes develop this formalism systematically and demonstrate its application to the Hawking effect in semi-classical black holes and laboratory analogue systems.

\vspace{4em}

\begin{figure}[h]
\centering
\includegraphics[width=0.99\linewidth]{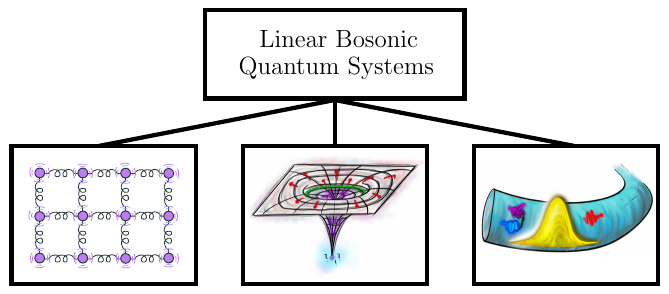}
\caption{A network of coupled quantum harmonic oscillators, Hawking radiation from a black-hole event horizon, and photonic quantum fluctuations propagating near an analogue white-black hole in a dielectric medium---all fall under the umbrella of linear bosonic systems and can be described through the phase-space formalism.}
\label{fig:flowchart}
\end{figure}

\clearpage


\section*{Motivation and Scope}

\begin{table}
\label{table:hilbert2phaseSpace}
\renewcommand{\arraystretch}{1.5}
{
\begin{tabular}{c c c } 
\hline\hline 
Hilbert Space & \hspace{5em} & Phase Space \\
\hline   

$\mathscr{H}^{\otimes N}$ & & $\mathbb{R}^{2N}$
\\ 
$\rho\in\mathscr{H}^{\otimes N}$ & & $(\bm\mu,\bm\sigma)\in\left(\mathbb{R}^{2N}, \mathbb{R}^{2N\times2N}\right)$
\\ 
$U:\mathscr{H}^{\otimes N}\rightarrow\mathscr{H}^{\otimes N}$ & & $\bm S\in{\rm Sp}(2N, \mathbb{R})$ \\ 
\hline\hline
\end{tabular}
}
\caption{From Hilbert space to phase space.}
\end{table}

The phase-space formalism offers a transformative approach to transition from a complex (infinite dimensional) Hilbert space description of $N$ quantum harmonic oscillators $\mathscr{H}^{\otimes N}$ to the more intuitive phase space---the $2N$-dimensional real space $\mathbb{R}^{2N}$~\cite{serafini2017book, weedbrook2012rmp}. Rather than being described by a density matrix $\rho$ in an infinite dimensional Hilbert space, a Gaussian quantum state of the system of oscillators is described by its first-moment vector $\bm \mu$ and its covariance matrix $\bm \sigma$, much like classical statistical mechanics. Similarly, unitary evolution of a quantum state by the (formally infinite dimensional) unitary operator $U$ is replaced by finite-dimensional matrix multiplication via $\bm S\in{\rm Sp}(2N,\mathbb{R})$, where ${\rm Sp}(2N,\mathbb{R})$ denotes the (matrix representation of the) $2N$-dimensional real symplectic group. Due to this simplification, several generic assertions can be formulated and various quantities can be readily computed---an explicit expression for the entropy of the system can be calculated, the entanglement structure can be characterized, and a simple description of system-environment interactions can be provided.

Beyond conceptual simplicity, the formalism captures a broad class of physical phenomena. Key effects in quantum field theory in curved spacetime (QFCS)---including black hole evaporation and ergoregion-induced quantum superradiance of rotating astrophysical bodies---reduce to Gaussian evolution. By extension, analogue gravity, which emulate aspects of QFCS with light-matter systems engineered in the lab, can be likewise described. Importantly, practical matters, such as background noise in an experiment or inefficiency of a detection system, are incorporated as Gaussian channels and handled with relative ease. The overarching objective of these notes is to integrate all these facets into a unified narrative and equip the reader with the essential tools to analyze linear bosonic systems generically, and more specifically, in the contexts of QFCS and analogue gravity.

\section*{Outline and Further Reading}

These notes originate from six lectures delivered at the \href{https://www.benasque.org/2023ag/}{\textit{Analogue Gravity in Benasque 2023}} workshop. The audience comprised researchers in quantum optics, condensed-matter physics, and gravitational theory, motivating a presentation that is both general and computationally concrete. I organize the material into four parts:
\begin{itemize}
    \item[(\ref{lecture:1}-\ref{lecture:3})] \emph{Gaussian formalism}: Dynamics, state characterization, and entanglement in linear bosonic networks, including environmental coupling.
    \item[(\ref{lecture:4})] \emph{The Hawking effect}: Mode decomposition of quantum fields in curved spacetime, black-hole evaporation, and toy models for laboratory analogues.
\end{itemize}
The computational tools developed in Parts (\ref{lecture:1}-\ref{lecture:3}) aim to enhance the efficiency and clarity of analysis in Part (\ref{lecture:4}) and, moreover, can be used in ongoing research in this field. For illustrative and intuitive purposes, we introduce diagrammatic methods---in the form of \emph{symplectic diagrams} (or circuits)---which allows us to visualize the physics at play without sacrificing rigor. Worked examples (purple boxes) illustrate explicit calculations throughout; brief ``lecture summaries'' (gray boxes) distill core results at the end of each Part. A biased and compressed bibliography is provided---mainly pedagogical but often historical in nature.

Background on quantum mechanics, as well as minimal working knowledge of general relativity and quantum field theory at the level of free scalar fields, is recommended but not completely necessary. Recommended further reading and background material are:
\begin{itemize}
    \item \textbf{Quantum Continuous Variables}: Most of the material presented in Parts (\ref{lecture:1}-\ref{lecture:3}) is standard and can be found in Serafini's book~\cite{serafini2017book}, or the review~\cite{weedbrook2012rmp}; the latter having a more quantum-information flavor to it. My presentation relies heavily on Serafini's~\cite{serafini2017book}. Useful background knowledge also includes quantum optics~\cite{Chiao2014:QuOptics, Kok2010:QuOpProcess} and quantum noise theory~\cite{Clerk2010:QuNoise}.
    \item \textbf{Quantum Fields in Curved Space}: Conventional references for QFCS are~\cite{birrelldavies1982,Wald:1995yp,Jacobson2005qfcs}; Jacobson's notes~\cite{Jacobson2005qfcs} represent a particularly gentle introduction to essential concepts. 
    \item \textbf{Black Hole Evaporation}: For semi-classical black holes, tons of insight and formulae can be mined from The Bible of black hole physics~\cite{Frolov1998bible}; see also Fabbri and Navarro-Salas's book~\cite{fabbri05} on black-hole evaporation.
    \item \textbf{Analogue Gravity}: This is a niche area of ongoing research. So I point only to a few historically insightful papers~\cite{unruh81analogue, Visser2003essential}, semi-recent overviews and literature surveys~\cite{jacquet2020nextGen, Jacquet2020:PolaritonAnalogues, Almeida2023:AnalogueGravHistory}, and the living review~\cite{Barcel2005:LivRvwAgrav}. 
\end{itemize}

\clearpage


\part{Feel the Vibrations}\label{lecture:1}

\begin{figure}[h]
    \centering
    \includegraphics[width=.99\linewidth]{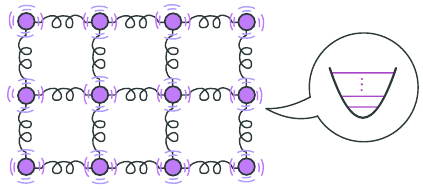}
    \caption{Network of interacting quantum harmonic oscillators. Each mode has equally spaced energy levels, with the ground state (vacuum) containing zero quanta. Bilinear coupling, external driving, excitation loss, heating, and linear measurements, such as homodyne detection, are all described within the phase-space framework.}
    \label{fig:oscillators}
\end{figure}

In this first part, I introduce the quantum harmonic oscillator in textbook fashion and transition smoothly into the Gaussian phase-space formalism. We begin with a single oscillator (or mode\footnote{We use the terms oscillator and mode interchangeably, since the oscillations might be that of an optical or microwave field, a breathing mode of a mechanical membrane, or the motion of a trapped ion.}), extend to $N$ coupled oscillators (Fig.~\ref{fig:oscillators}), and develop the symplectic transformations that governs their collective dynamics. The discussion culminates in the structure of thermal (Gibbs) Gaussian states---a ubiquitous class in quantum field theory, continuous-variable quantum information, and analogue gravity systems.

\vspace{4em}

\clearpage

\section{Quantum Oscillators}

This first section provides a crash course on the single quantum harmonic oscillator (single mode) in a textbook description. We then move to $N$ harmonic oscillators (multiple modes). We introduce useful, compact machinery that prove useful throughout. 

Consider a quantum harmonic oscillator with (dimensionless) position and momentum operators $\hat{q}$ and $\hat{p}$. When referring to an electromagnetic field, the $q$'s and $p$'s correspond to the quadrature operators---i.e., the orthogonal Fourier components of the electric field. The canonical operators obey the canonical commutation relations (CCR)
\begin{equation}
    \comm{\hat{q}}{\hat{p}}=i\hat{I},
\end{equation}
where $\hat{I}$ denotes the identity on the single-mode bosonic Hilbert space.\footnote{I will drop the identity henceforth.} Instead of the quadrature operators, we often use the annihilation and creation operators, $\hat{a}$ and $\hat{a}^\dagger$. These relate to the canonical operators $\hat{q}$ and $\hat{p}$ via
\begin{equation}
    \hat{a}=\frac{1}{\sqrt{2}}\left(\hat{q}+i\hat{p}\right),
\end{equation}
and obey $\comm{\hat{a}}{\hat{a}^\dagger}=\hat{I}$. The term annihilation (creation) refers to the fact that $\hat{a}$ ($\hat{a}^\dagger$) destroys (creates) one quanta of energy. 

Free evolution of an oscillator, with oscillation frequency $\omega$, is governed by the Hamiltonian,
\begin{align}
    \hat{H}_{\rm qho}&= \frac{\hbar\omega}{2}\left(\hat{q}^2+\hat{p}^2\right)\\
    &=\hbar\omega(\hat{n}+1/2),
\end{align}
where $\hbar\omega/2$ is the zero-point energy. We drop this additive term throughout. The operator $\hat{n}\triangleq\hat{a}^\dagger\hat{a}$ is the number operator and counts the number of quanta. Specifically, we can have an oscillator with a definite number of quanta $n\in\mathbb{N}$. Quantum states with a definite number of quanta are \emph{number (or Fock) states} and defined as
\begin{equation}
    \ket{n}\triangleq\frac{\left(\hat{a}^{\dagger}\right)^n}{\sqrt{n!}}\ket{0}.
\end{equation}
Here $\ket{0}$ is the ground state (or vacuum state) of zero quanta, ${\hat{n}\ket{0}=0}$. Fock states are eigenstates of the number operator (and by extension, the free Hamiltonian) with eigenvalue $n\in\mathbb{N}$. We destroy and create quanta of a Fock state via 
\begin{align}
    \hat{a}\ket{n}&=\sqrt{n}\ket{n-1},\\
    \hat{a}^\dagger\ket{n}&=\sqrt{n+1}\ket{n+1}.
\end{align}
The set of Fock states form an orthonormal basis (the Fock basis) in the bosonic Hilbert space of a single mode, $\mathscr{H}$ (the Fock space):
\begin{equation}
    \innerproduct{m}{n}=\delta_{mn} \qq{and}
   \sum_{n=0}^\infty\dyad{n}=\hat{I}
\end{equation}
where $\delta_{mn}$ is the Kronecker delta.

In the phase-space formalism, it is more convenient to refer to the position and momentum, $\hat{q}$ and $\hat{p}$, which take on real values, rather than the discrete number of quanta. The canonical operators $\hat{q}$ and $\hat{p}$ are Hermitian, canonically-conjugate operators and have non-normalizable eigenstates, $\ket{q}$ and $\ket{p}$ such that $\hat{q}\ket{q}=q\ket{q}$, $\hat{p}\ket{p}=p\ket{p}$ and $q,p\in\mathbb{R}$. These eigenstates resolve the identity in the continuum, 
\begin{equation}
\frac{1}{2\pi}\int_{q\in\mathbb{R}}\dd{q}\dyad{q}= \frac{1}{2\pi}\int_{p\in\mathbb{R}}\dd{p}\dyad{p}=\hat{I}.
\end{equation}
Due to $q$ and $p$ being conjugate, Fourier transform relates the position and momentum eigenstates relate:
\begin{equation}
    \ket{q}=\frac{1}{\sqrt{2\pi}}\int_{p\in\mathbb{R}}\dd{p}{\rm e}^{-ipq}\ket{p}.
\end{equation}
Note we use the convention $\delta(x-x^\prime)\triangleq\int\dd{k}{\rm e}^{i(x-x^\prime)k}/2\pi$.

We can ``shift'' (or displace) the position of the oscillator $q \to q+ x$ by, e.g., driving the oscillator with a classically coherent current $J$ through the interaction Hamiltonian $H = g J \hat{q}$, where $g$ represents a coupling parameter, such that $x=gJ$. Generically, the displacement by $x$ is accomplished through the unitary ${\rm e}^{-ix\hat{p}}$. Likewise we can ``boost'' the momentum $p \to p+k$ via ${\rm e}^{ik\hat{q}}$. We can shift and boost the oscillator through the generalized displacement operator,
\begin{equation}\label{eq:displace_1mode}
    D_{x,k}={\rm e}^{i(k\hat{q}-x\hat{p})}.
\end{equation}
Let $\alpha=(x+ik)/\sqrt{2}$ represent the displacement on the complex plane. Then:
\begin{equation}
    D_\alpha={\rm e}^{\alpha\hat{a}^\dagger-\alpha^*\hat{a}}.
\end{equation}
Displacing the vacuum by an amount $\alpha$ leads to a displaced vacuum (or \emph{coherent state})
\begin{equation}\label{eq:coherent_state}
    \ket{\alpha}\triangleq D_\alpha\ket{0}=e^{-\frac{\abs{\alpha}^2}{2}}\sum_{n=0}^\infty\frac{\alpha^n}{\sqrt{n!}}\ket{n}.
\end{equation}
Coherent states are regarded as a ``classical'' states because their time evolution follows the classical equations of motion, and they minimize Heisenberg's uncertainty principle for position and momentum, meaning they can be predicted with the same accuracy as a classical state. To a good approximation, a stable laser is described by a coherent state. The coherent state is an eigenstate of the annihilation operator $\hat{a}$ with eigenvalue $\alpha$.\footnote{Thus, the annihilation operator does not annihilate photons in a laser!} Finally, coherent states form an overcomplete basis in $\mathscr{H}$:
\begin{equation}\label{eq:alpha_completeness}
\int_{\alpha\in\mathbb{C}}\frac{\dd{\alpha}}{\pi}\dyad{\alpha}=\hat{I}.
\end{equation}

We now extend the discussion to a set of $N$ oscillators and introduce convenient notation to highlight the underlying geometrical (symplectic) structure of the $N$-mode phase space. Consider a $N$ mode bosonic quantum system with a Hilbert space $\mathscr{H}^{\otimes N}$, where $\mathscr{H}$ denotes the single-mode Hilbert space. The system can be associated with the phase space of $\mathbb{R}^{2N}$ induced by the CCR of the $q$'s and $p$'s of the modes. We pack all the information about the modes into a vector of canonical operators,
\begin{equation}
    \hat{\bm r}^\top\triangleq\left(\hat{q}_1,\hat{p}_1,\dots,\hat{q}_N,\hat{p}_N\right),
\end{equation}
and express the CCR relations compactly by the relation,
\begin{equation}\label{eq:ccr}
    [\hat{\bm r}_k,\hat{\bm r}_j]={\rm i} \bm{\Omega}_{kj},
\end{equation}
where $\bm\Omega$ represents the $N$-mode \emph{symplectic form} ($2N\times2N$, anti-symmetric form),
\begin{equation}\label{eq:symplectic_form}
    \bm\Omega = \bigoplus_{i=1}^N\bm\Omega_1 \qq{with} 
    \bm\Omega_1=
    \begin{pmatrix}
        0 & 1 \\
        -1 & 0
    \end{pmatrix},
\end{equation}
such that $\bm\Omega\bm\Omega^\top=\bm I_{2N}$. We will learn more about the symplectic structure as we go along. As an aside, we quote a useful relation between the canonical operators in this form and the total number operator $\hat{N}=\sum_{i=1}^N\hat{n}_i$,
\begin{equation}\label{eq:multimode_numop}
    \hat{N}=\frac{1}{2}\hat{\bm r}^\top\hat{\bm r}-\frac{N}{2}.
\end{equation}

We can shift the canonical coordinates by a constant amount $\hat{\bm r}\rightarrow\hat{\bm r}-\bm\zeta$ via displacement operators, similar to before. The unitary transformation for a shift is given by the multimode \emph{displacement (or Weyl) operator},
\begin{equation}\label{eq:weyl_op}
    {D}_{\bm\zeta}\triangleq\exp\left({\rm i}{\bm\zeta}^\top\bm\Omega\hat{\bm r}\right),
\end{equation}
such that $D_{\bm\zeta}^\dagger\hat{\bm r}D_{\bm\zeta}=\hat{\bm r}-\bm\zeta$. Displacement operators are crucial and useful tools for describing quantum theory in phase space; so we highlight their main properties. Observe that $\mathcal{D}_{\bm\zeta}=\bigotimes_{i=1}^N\mathcal{D}_{\bm\zeta_i}$, where $\bm\zeta_i\in\mathbb{R}^2$ are single-mode displacements; in other words, displacements are always local. 

Importantly, the set of displacement operators comprise an operator basis in the space of bounded operators $\mathcal{B}(\mathscr{H}^{\otimes N})$:\footnote{A bounded operator $\hat{A}\in\mathcal{B}(\mathscr{H}^{\otimes N})$ is such that, for any normalizable state, $\psi$, $\norm*{\hat{A}\psi}\leq\norm{\psi}a$, with $a<\infty$, and the norm is defined as $\norm{\phi}=\abs{\ip{\phi}}$. An example of an unbounded operator is $\hat{n}$. Contrariwise, $\exp(-\lambda\hat{n})$ is bounded for all $\lambda>0$.}
\begin{equation}\label{eq:weyl_opbasis}
    \Tr({D}_{\bm\zeta}{D}_{-\bm\nu})=(2\pi)^N\delta^{2N}(\bm\zeta-\bm\nu),
\end{equation}
where $\delta^{2N}(\cdot)$ is the $2N$-dimensional Dirac-delta distribution. One can derive this by first evaluating the trace of the single-mode displacement~\eqref{eq:displace_1mode} in the position or momentum basis and showing that it is proportional to the Dirac-delta distribution. The result can then be easily extended to $N$ modes by the locality condition $\mathcal{D}_{\bm\zeta}=\bigotimes_{i=1}^N\mathcal{D}_{\bm\zeta_i}$. The composition rule~\eqref{eq:weyl_comp} is useful in the derivation. Thus, given some general operator function $f(\hat{\bm r})$ of canonical operators,
\begin{equation}\label{eq:weyl_transform}
    f(\hat{\bm r})=\frac{1}{(2\pi)^N}\int_{\bm\zeta\in\mathbb{R}^{2N}}\dd{\bm\zeta}\Tr(f(\hat{\bm r})D_{\bm\zeta}^\dagger)D_{\bm\zeta}.
\end{equation}

Displacement operators satisfy the composition rule, 
\begin{equation}
    {D}_{\bm\zeta+\bm\nu}={\rm e}^{{\rm i}\omega(\bm\zeta,\bm\nu)/2}{D}_{\bm\zeta}{D}_{\bm\nu},\label{eq:weyl_comp}
\end{equation}
where $\omega(\bm\zeta,\bm\nu)\triangleq\bm\zeta^\top\bm\Omega\bm\nu$ is the \emph{symplectic product}. Since $\omega(\bm\zeta,\bm\nu)=\sum_{i=1}^N\bm\zeta_{q_i}\bm\nu_{p_i}-\bm\nu_{q_i}\bm\zeta_{p_i}$, we can geometrically interpret the symplectic product as the area between $\bm\zeta$ and $\bm\nu$ in phase space. The symplectic product is anti-symmetric in its arguments, $\omega(\bm\nu,\bm\zeta)=-\omega(\bm\zeta,\bm\nu)$, and is invariant under symplectic transformations $\bm S \in {\rm Sp}(2N, \mathbb{R})$ (see next section for details), $\omega(\bm S\bm\zeta,\bm S\bm\nu)=\omega(\bm\zeta,\bm\nu)$. By swapping $\bm\zeta$ and $\bm\nu$ in Eq.~\eqref{eq:weyl_comp}, it follows that ${D}_{\bm\zeta}{D}_{\bm\nu}={\rm e}^{{\rm i}\omega(\bm\nu,\bm\zeta)}{D}_{\bm\nu}{D}_{\bm\zeta}$. 

We can the single-mode coherent state to a multimode coherent state described by the phase-space displacement $\bm{\mu}\in\mathbb{R}^{2N}$:
\begin{equation}
    \ket{\bm{\mu}}= D_{\bm \mu} \ket{0}.
\end{equation}
Likewise, we extend the single-mode completeness relation~\eqref{eq:alpha_completeness} to multiple modes:
\begin{equation}
    \frac{1}{(2\pi)^N}\int\dd{\bm\mu} D_{\bm \mu}\dyad{0}D_{\bm \mu}^\dagger=\hat{I}.
\end{equation}


\section{Linear Evolution}
We focus on physical systems, such as a finite set of bosonic fields $\varphi_J$ labeled by $J$, that couple through bilinear interactions like $\varphi_I\varphi_J$. These interactions arise in the Lagrangian description of the system and lead to linear equations of motion for the fields (often second-order wave equations), viz., the Euler-Lagrange equations of motion. Generally,
\begin{equation}
    \mathcal{K}(\varphi_J) + \sum_I V_{JI} \varphi_{I} = 0
\end{equation}
where $V_{JI}$ denotes a (multiplicative) interaction term between $\varphi_I$ and $\varphi_J$ and $\mathcal{K}(\cdot)$ is a linear differential operator such that $\mathcal{K}(\varphi_1 +\varphi_2)=\mathcal{K}(\varphi_1)+ \mathcal{K}(\varphi_2)$; for example, $\mathcal{K}= (\beta \partial_t^2 + \alpha \partial_x^2)$. Due to the linearity of the equations of motion, the ``out'' operators at time $t$ relate to the ``in'' operators at time $t_0<t$ via linear (matrix) transformations. We consider specific examples later. 

For now, we simplify matters drastically and phrase our discussion in terms of the canonical operators of $N$ oscillators, as opposed to continuous fields.

\subsection*{Unitary evolution}
Linear equations of motion for a network of $N$ oscillators lead to linear transformations of the oscillators' canonical variables: $\hat{\bm r}^{\rm out}=\bm S\hat{\bm r}^{\rm in}$ where $\bm S$ is a real $2N\times2N$ matrix. To preserve the CCR, the transformation $\bm S$ must satisfy the \emph{symplectic condition} 
\begin{equation}
    \bm S\bm\Omega\bm S^\top=\bm\Omega,
\end{equation}
with $\det\bm S=1$. Any transformation satisfying this condition is called a \emph{symplectic transformation} such that $\bm S\in{\rm Sp}(2N,\mathbb{R})$, where ${\rm Sp}(2N,\mathbb{R})$ is a matrix representation of the real symplectic group, a non-compact Lie group. A $2N\times2N$ matrix is generally characterized by $4N^2$ parameters, however the symplectic condition on $\bm S$ introduces $N(2N-1)=2N^2-N$ constraints due to anti-symmetry of the symplectic form. The number of free parameters of $\bm S$ is then $4N^2-(2N^2-N)=2N^2+N$. Hence $|{\rm Sp}(2N,\mathbb{R})|=2N^2+N$ since $\bm S$ is arbitrary.

We provide a more physical picture of symplectic transformations via Heisenberg evolution. Consider a generic bilinear interaction 
\begin{equation}
    \hat{g}=\tfrac{1}{2}\hat{\bm r}^\top\bm g\hat{\bm r}
\end{equation}
where $\bm g$ is a real symmetric matrix. From which we construct the unitary $U_\lambda=\exp(-i\hat{g}\lambda)$. In other words, $\hat{g}$ generates translations in $\lambda$. Given $\hat{\bm r}(\lambda) = U_\lambda^\dagger \hat{\bm r}(0) U_\lambda$, it follows that:
\begin{align}
    \dv{\hat{\bm r}_\ell}{\lambda}&=i\comm{\hat{g}}{\hat{\bm r}_\ell}\nonumber\\
    &=i\frac{\bm g_{jk}}{2}\comm{\hat{\bm r}_j\hat{\bm r}_k}{\hat{\bm r}_\ell}\nonumber\\
    &=-\frac{\bm g_{jk}}{2}\left(\hat{\bm r}_j\bm\Omega_{k\ell}+\bm\Omega_{j\ell}\hat{\bm r}_k\right)\nonumber\\
    &=(\bm\Omega\bm g\hat{\bm r})_\ell.\nonumber
\end{align}
To go from the second to the third line, we use $\comm*{\hat{A}\hat{B}}{\hat{C}}=\hat{A}\comm*{\hat{B}}{\hat{C}}+\comm*{\hat{A}}{\hat{C}}\hat{B}$ and the CCR~\eqref{eq:ccr}. In the final equality, we use symmetry of $\bm g$ together with anti-symmetry of $\bm\Omega$. This first-order differential equation has a simple solution:
\begin{equation}
    \hat{\bm r}(\lambda)= U_\lambda^\dagger\hat{\bm r}(0)U_\lambda=\exp(\lambda\bm{\Omega}\bm g)\hat{\bm r}(0).
\end{equation}
We see that $\bm S\triangleq\exp(\lambda\bm{\Omega}\bm g)$ is the symplectic transformation that we were looking for.\footnote{One can check that, to first order in $\lambda$, the symplectic condition is indeed satisfied.} To make the connection to the symplectic transformation explicit, we often denote the unitary associated with $\bm S$ as $U_{\bm S}$, rather than using the parametric notation $U_\lambda$. 

In practice, the operator $\hat{g}$ may or may not be a physical microscopic Hamiltonian but, rather, represents a useful characterization of the physics. This is the case in various scattering scenarios where $\hat{g}$ simply generates the transformation from the ``in'' operators in the asymptotic past to the ``out'' operators in the asymptotic future.

Before making further generic comments about symplectic transformations, let us look at several useful examples.
\begin{example}{Handy symplectic transformations}
    \begin{itemize}
        \item \textit{Displacement transformation:} Consider the displacement operator $D_{\bm \mu}$~\eqref{eq:weyl_op}. Using $\hat{\bm r}\rightarrow\bm S\hat{\bm r}$, it is simple to show that
    \begin{equation}\label{eq:weyl_Stransform}
        U_{\bm S}D_{\bm \mu}U_{\bm S}^\dagger=D_{\bm S\bm\mu}.
    \end{equation}
        \item \textit{Single-mode transformations:} By the Bloch-Messiah decomposition [see Eq.~\eqref{eq:bloch_messiah} below], any single-mode symplectic transformation $\bm S$ can be decomposed into single-mode rotations (or phase shifts), $\bm R(\phi)$, and single-mode squeezers, $\textbf{Sq}(e^r)$, such that $\bm S=\bm R(\phi_2)\textbf{Sq}(e^r)\bm R(\phi_1)$. Phase rotations are generated by free evolution, viz., $\hat{g}_{\text{rot}}=\hat{n}$, with symplectic transformation
        \begin{equation}
            \bm R(\phi)=
            \begin{pmatrix}
            \cos\phi & \sin\phi\\
            -\sin\phi & \cos\phi
            \end{pmatrix}.
        \end{equation}
        This is just a $2\times2$ rotation matrix in the single-mode phase space. The input-output relation for the annihilation operator is simply $\hat{a}(\phi)=e^{i\phi}\hat{a}$. 
        
        The squeezing transformation is generated by $\hat{g}_{\text{sqz}}=(\hat{a}^2-\hat{a}^{\dagger\,2})/2=(\hat{q}\hat{p}+\hat{p}\hat{q})/2$, where $r$ is the squeezing strength. This leads to the symplectic transformation,
        \begin{equation}\label{eq:sqz_transform}
            \textbf{Sq}(e^r)=e^{r\bm Z},
        \end{equation}
        where $\bm Z={\rm diag}(1,-1)$ is the Pauli-Z matrix. Single-mode squeezing is sometimes referred to as phase-sensitive amplification because it treats the $q$ and $p$ quadratures differently---amplifying one while suppressing the other. In annihilation and creation variables, the input-output relation of a single-mode squeezer is
        \begin{equation}
            \hat{a}(r)=\hat{a}\cosh{r}+\hat{a}^\dagger\sinh{r}.
        \end{equation}
        Thus starting from the vacuum state (zero quanta), single-mode squeezing generates $\sinh^2{r}$ number of quanta.
        
        \item \textit{Beamsplitter:} A beamsplitter-like transformation is a two-mode ``passive'' (energy-preserving) transformation. Given modes $\hat{a}_1$ and $\hat{a}_2$, the generator of the beamsplitter interaction is $\hat{g}_{\text{bs}}=(\hat{a}_1\hat{a}_2^\dagger+{\rm h.c.})/2=(\hat{q}_1\hat{q}_2+\hat{p}_1\hat{p}_2)/2$. This leads to the symplectic transformation 
        \begin{equation}
            \bm B(\theta)=
            \begin{pmatrix}
                \cos\theta\bm I_2 & \sin\theta\bm I_2\\
                -\sin\theta\bm I_2 & \cos\theta\bm I_2
            \end{pmatrix},
        \end{equation}
        where $\theta$ parameterises the interaction strength and time. Observe that $\cos^2\theta$ is the transmission probability of the beamsplitter. The input-output relations for annihilation and creation variables is
        \begin{align}
        \hat{a}_1(\theta)&=\hat{a}_1\cos\theta+\hat{a}_2\sin\theta,\\
        \hat{a}_2(\theta)&=\hat{a}_2\cos\theta-\hat{a}_1\sin\theta.
        \end{align}
        Physically, we can interpret a general beamsplitter-like interaction as classical wave scattering that preserves the total intensity of the waves.

        \item \textit{Passive transformations:} General $N$-mode passive transformations refer to unitary operations that preserve the total photon number $\hat{N}$. Inserting the symplectic transformation $\hat{\bm r}\rightarrow\bm O\hat{\bm r}$ in the formula~\eqref{eq:multimode_numop}, we see that $\bm O^\top\bm O=\bm I$ must be satisfied in order to preserve the total photon number---i.e., $\bm O$ must be a (symplectic) orthogonal transformation. These transformations can be generated via $\hat{g}_{\bm O}=\sum_{i,j}h_{ij}\hat{a}_j^\dagger\hat{a}_k+{\rm h.c}$, where $h_{ij}\in\mathbb{C}$. A general $N$-mode passive transformation can further be decomposed into single-mode phase shifts and two-mode beamsplitters (\emph{Reck-Zeilinger decomposition}~\cite{reck94}).
        
        \item \textit{Two-mode squeezing:} Two-mode squeezing creates entangled particle-pairs and, thus, represents an ``active'' transformation. Two-mode squeezing is generated by the $\hat{g}_{\text{tms}}=(\hat{a}_1\hat{a}_2-{\rm h.c.})/2=(\hat{q}_1\hat{p}_2+\hat{q}_2\hat{p}_1)/2$. This leads to the symplectic transformation
        \begin{equation}\label{eq:tms_transform}
            \bm S(r)=
            \begin{pmatrix}
                \cosh{r}\bm I_2 & \sinh{r}\bm Z\\
                \sinh{r}\bm Z & \cosh{r}\bm I_2
            \end{pmatrix},
        \end{equation}
        where $r$ is the squeezing strength.
        The input-output relations for the annihilation operators is
        \begin{align}
            \hat{a}_1(r)&=\hat{a}_1\cosh{r}+\hat{a}_2^\dagger\sinh{r},\label{eq:a1_tms}\\
             \hat{a}_2(r)&=\hat{a}_2\cosh{r}+\hat{a}_1^\dagger\sinh{r}.\label{eq:a2_tms}
        \end{align}
        The number of particles generated by two-mode squeezing in the first (or second) mode is $\sinh^2{r}$. Assuming mode 1 is in a coherent state with intensity $I_{\rm in}$ while mode 2 is in vacuum, two-mode squeezing amplifies the input intensity $I_{\rm in}$ via $I_{\rm out}/I_{\rm in}=\cosh^2{r}\triangleq G$, where $G$ is the amplification gain. This process is called phase-insensitive amplification because it affects $q$ and $p$ quadratures identically.
        \item \textit{SUM-gate:} The SUM-gate is a CV analog of the standard CNOT gate that is used abundantly in computation. The SUM-gate is generated by $\hat{g}_{\text{sum}}=\hat{q}_1\hat{p}_2$, which leads to the symplectic transformation
        \begin{equation}\label{eq:sum_gate}
            \textbf{SUM}=
            \begin{pmatrix}
                \bm I_2 & -\bm\Pi_p\\
                \bm\Pi_q & \bm I_2
            \end{pmatrix},
        \end{equation}
        where $\bm\Pi_p={\rm diag}(0,1)$ and $\bm\Pi_q={\rm diag}(1,0)$ are projections along the $p$ and $q$ directions, respectively. The SUM-gate is useful for ancilla-assisted, non-destructive Gaussian measurements. The input-output relations for the canonical operators is,
        \begin{align}
        \begin{aligned}
     \hat{q}_1^\prime&=\hat{q}_1,& \qq{} \hat{p}_1^\prime&= \hat{p}_1-\hat{p}_2,&\\
     \hat{q}_{2}^\prime&=\hat{q}_{2}+\hat{q}_1,&\qq{}\hat{p}_{2}^\prime&=\hat{p}_{2}.&
        \end{aligned}
        \end{align}
    \end{itemize}
\end{example}

\begin{figure}
    \centering
    \includegraphics[width=\linewidth]{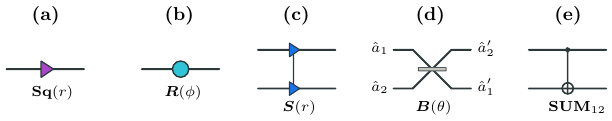}
    \caption{Symplectic diagrams: (a) Single-mode squeezer $\textbf{Sq}(r)$, (b) phase rotation $\bm R(\phi)$, (c) two-mode squeezer $\bm S(r)$, (d) beampslitter $\bm B(\theta)$, (d) SUM-gate $\textbf{SUM}_{12}$. Single- and two-mode squeezing lead to (phase-sensitive and phase-insensitive) amplification; our triangle notation is hence borrowed from classical circuit notation for linear amplifiers. Convention for input-output relations of the beamsplitter is shown, where we follow the modes through their transmission path. The SUM-gate is the CV equivalent of the CNOT gate. Input to output flows from left to right.}
    \label{fig:circuit_elements}
\end{figure}

In Fig.~\ref{fig:circuit_elements}, we present the symplectic transformations as basic elements in a circuit, where ``time'' flows from left to right and ``space'' is the vertical direction. We can stitch these basic elements together to form a connected \textit{symplectic diagram}. We generally refer to these types of diagrams as a continuous-variable (CV) quantum circuit, however we use the terminology symplectic diagram here due to the circuit elements being purely symplectic transformations deriving from bilinear interactions. As we will see, it is often handy to convey physical processes by symplectic diagrams to gain quick insight about the physics or flow of particles and entanglement in a system of oscillators. Importantly, these diagrams are``universal'', in the sense that they represent the underlying symplectic structure without referring directly to the physical platform. 

A sometimes useful (but always insightful) representation of $\bm S$ is the following: It turns out that sny symplectic matrix $\bm S$ can be diagonalized by passive transformations $\bm O$ and $\bm O^\prime$ (\emph{Bloch-Messiah decomposition}~\cite{braunstein05}),
\begin{equation}\label{eq:bloch_messiah}
    \bm S = \bm O^\prime\left(\bigoplus_{i=1}^N{\rm e}^{r_i\bm Z}\right)\bm O.
\end{equation}
This bears similarities to the singular value decomposition of $\bm S$. Observe that $\bm O^\prime,\bm O\in{\rm Sp}(2N,\mathbb{R})\cap{\rm SO}(2N)\approx U(N)$, where $U(N)$ is the unitary group of dimension $\abs{U(N)}=N^2$. Furthermore, the squeezing transformations have $N$ free parameters---a squeezing strength $r_i$ for each mode. Therefore, $\abs{{\rm Sp}(2N,\mathbb{R})}=2\abs{U(N)}+N=2N^2+N$, as we mentioned previously. Interestingly, we can further decompose the multimode passive elements, $\bm O$ and $\bm O^\prime$, into an array of beamsplitters and phase-shifts (\textit{Reck-Zeilinger decomposition}~\cite{reck94}). In this sense, we can always think of Gaussian evolution as two-modes scattering at a time in a sequential fashion, with particle creation (via single-mode squeezers) in the middle. This decomposition also allows us to buil' any symplectic diagram with basic elements of Fig.~\ref{fig:circuit_elements} if we so choose. A simple example is the decomposition of a two-mode squeezer with gain $G=\cosh^2{r}$, in which case $\bm S(r)=\bm B_{1/2}(\textbf{Sq}(e^r)\oplus\textbf{Sq}(e^{-r}))\bm B_{1/2}^\top$. 

\begin{example}{Squeezing is necessary for particle creation}
    Using the Bloch-Messiah decomposition, we can analytically show that squeezing is a necessary ingredient for particle creation. Consider $N$ modes with ``in'' canonical operators, $\hat{\bm r}^{\rm in}$. The input photon number can be written as $\hat{N}^{\rm in}=\frac{1}{2}\hat{\bm r}^{\rm in\,\top}\hat{\bm r}^{\rm in}-N/2$ per Eq.~\eqref{eq:multimode_numop}. A symplectic transformation takes the ``in'' operators to ``out'' operators, $\hat{\bm r}^{\rm out}=\bm S\hat{\bm r}^{\rm in}$. By the Bloch-Messiah decomposition, we have $\bm S=\bm O_2\bm D\bm O_1$, where $\bm O_i$ denote symplectic orthogonal transformations and $\bm D=\bigoplus_{i=1}^N\exp(r_i\bm Z)$ denotes a direct sum of single-mode squeezers. We compute the ``out'' number operator:
    \begin{align}
        \hat{N}^{\rm out}&=\frac{1}{2}\hat{\bm r}^{\rm out\,\top}\hat{\bm r}^{\rm out}-N/2 \\
        &=\frac{1}{2}\hat{\bm r}^{\rm in\,\top}\bm S^\top\bm S\hat{\bm r}^{\rm in}-N/2 \\
        &= \frac{1}{2}\hat{\bm r}^{\rm in\,\top}\bm O_2^\top\bm D^2\bm O_2\hat{\bm r}^{\rm in}-N/2,
    \end{align}
    where we used $\bm O_1^\top\bm O_1=\bm I$ to go from the second to the third line. In the absence of squeezing (i.e., $r_i=0\,\forall\,i$ $\implies\bm D=\bm I$), we see that $\hat{N}^{\rm out}=\hat{N}^{\rm in}$.
\end{example}

\subsection*{Open Gaussian quantum systems}

A generalization of unitary Gaussian evolution is to that of a Gaussian quantum channel, $\mathcal{E}_G$. Gaussian channel evolution is generally non-unitary, however by \textit{going to the Church of the Larger Hilbert Space}, we can describe evolution of the system in a unitary fashion by introducing an environment $\tau_E$, which can be taken as a pure state, that couples to the system through a unitary $U$. The unitary $U$ is said be a (non-unique) \textit{unitary dilation} of $\mathcal{E}_G$. For Gaussian quantum channels, the unitary is Gaussian $U_{\bm S}$ corresponding to a symplectic transformation $\bm S$, and the environment state is a Gaussian state $\tau_E$; we define Gaussian states formally in Part~\ref{lecture:2}. For now, suffice it to say that  evolution via $\mathcal{E}_G$ may be formally written as,\footnote{Dilations exist for arbitrary quantum channels, but we maintain the Gaussian character by emphasizing the symplectic transformation $\bm S$ and Gaussianity of $\tau_E$.}
\begin{equation}\label{eq:dilation_gauss}
    \mathcal{E}_G({\Psi})=\Tr_E\left[U_{\bm S}\left(\rho\otimes{\tau}_E\right)U_{\bm S}^\dagger\right].
\end{equation}
We discuss general aspects of Gaussian channels in Part~\ref{lecture:2}.

We would be remiss not to mention dynamical evolution of open quantum systems at this stage. An important class of Gaussian channels are those corresponding to open-system quantum evolution---particularly, Gaussian quantum-Markov semi-groups generated by (time-homogenous) Lindbladians. The master equation for this is
\begin{equation}
    \partial_t \rho = -i \comm{H}{\rho} + \sum_{j,k} \bm{\Gamma}_{jk} \left(\hat{\bm{r}}_j \rho \hat{\bm{r}}_k - \frac{1}{2}\acomm{\hat{\bm{r}}_k\hat{\bm{r}}_j}{\rho}\right),
\end{equation}
where 
\begin{equation}
    \hat{H} = \tfrac{1}{2}\hat{\bm{r}}^\top \bm{\mathcal{H}} \hat{\bm{r}}
\end{equation}
is the bilinear system Hamiltonian and $\bm{\Gamma}$ encodes the two-point correlators of the Gaussian environment. 

One route to this Lindblad master equation is to go to the Church of the Larger Hilbert space and introduce Markovian bath quadratures $\hat{\bm{\mathfrak r}}(t)$ that satisfy
\begin{equation}
    \comm*{\hat{\bm{\mathfrak r}}_k(t)}{\hat{\bm{\mathfrak r}}_j(t')}
    = i\bm{\Omega}_{kj}\,\delta(t-t'),
\end{equation}
with an environment state specified entirely by its two-point
correlators,
\begin{equation}
    \expval{\hat{\bm{\mathfrak r}}_k(t)\hat{\bm{\mathfrak r}}_j(t')}_E
    \propto \bm{\Gamma}_{kj}\,\delta(t-t').
\end{equation}
That is, the bath is memoryless (Markovian) and Gaussian.

We not delve further into the Lindbladian description. Instead, we illustrate the same physics directly at the operator level in the Heisenberg picture, for the concrete case of single-mode photon damping/loss. This leads to a Heisenberg-Langevin equation for a damped mode, which is particularly useful for building simple models for laboratory setups (e.g., stationary modes in an open quantum system). 

\begin{example}{Heisenberg-Langevin Equation: Damped Resonator}
\label{example:HL-eq}
For simplicity we work in the annihilation/creation picture. Consider a single stationary bosonic mode $\hat A(t)$ with $\comm*{\hat A(t)}{\hat A^\dagger(t)}=1$, coupled to a Markovian, traveling-bath mode $\hat e_t$ satisfying the white-noise commutation relation
\begin{equation}
    [\hat e_t,\hat e_{t'}^\dagger]=\delta(t-t').
\end{equation}
The system is governed by a (quadratic) Hamiltonian $\hat H_{\rm S}$ and couples to the bath through the interaction
\begin{equation}
    \hat H_{\rm int}(t)
    = i\sqrt{\gamma}\left(\hat e_t^\dagger\,\hat A(t)
    - \hat e_t\,\hat A^\dagger(t)\right),
\end{equation}
where $\gamma$ is the (energy) damping rate. The bath operator obeys
\begin{equation}
    \partial_t \hat e_{s}(t)
    = -i[\hat e_s(t),\hat H_{\rm int}(t)]
    = \sqrt{\gamma}\,\hat A(t)\,\delta(t-s),
\end{equation}
which integrates to the jump condition
\begin{equation}
    \hat e_{s}(t)
    = \hat e_{s}(0) + \sqrt{\gamma}\,\hat A(s)\,\Theta(t-s).
\end{equation}
Evaluating this at the interaction time from above, $t^+\triangleq s+0^+$,
and defining the incoming and outgoing bath fields as
\begin{equation}
    \hat e_{\rm in}(s)\triangleq \hat e_s(0),
    \qquad
    \hat e_{\rm out}(s)\triangleq \hat e_s(t^+),
\end{equation}
produces the famous input-output relation~\cite{Clerk2010:QuNoise}
\begin{equation}
    \hat e_{\rm out}(s) - \hat e_{\rm in}(s)
    = \sqrt{\gamma}\,\hat A(s).
\end{equation}
The Heisenberg equation for the system mode is
\begin{equation}
    \partial_t \hat A(t)
    = -i[\hat A(t),\hat H_{\rm S}]
    - i[\hat A(t),\hat H_{\rm int}(t)]
    = -i[\hat A(t),\hat H_{\rm S}]
    - \sqrt{\gamma}\,\hat e_t(t),
\end{equation}
where in the last step we used
$-i[\hat A, i\sqrt{\gamma}(\hat e_t^\dagger \hat A - \hat e_t\hat A^\dagger)]
= -\sqrt{\gamma}\,\hat e_t$.
To obtain a closed equation for $\hat A(t)$ in terms of the input field,
substitute the equal-time limit of the bath operator,
\begin{equation}
    \hat e_t(t)
    = \hat e_{\rm in}(t) + \sqrt{\gamma}\,\hat A(t)\,\Theta(0)
    = \hat e_{\rm in}(t) + \frac{\sqrt{\gamma}}{2}\,\hat A(t),
\end{equation}
where we use $\Theta(0)=\tfrac12$. This yields the
famous \emph{Heisenberg-Langevin} equation
\begin{equation}
\label{eq:HL-eqn}
    \partial_t \hat A(t)
    = -i[\hat A(t),\hat H_{\rm S}]
    - \frac{\gamma}{2}\,\hat A(t)
    - \sqrt{\gamma}\,\hat e_{\rm in}(t).
\end{equation}
The last two terms describe Markovian damping and the accompanying quantum-noise injection required to preserve the commutation relation of $A(t)$ for all $t$.
\end{example}


\section{Normal Modes and Gibbs Form}

Before introducing Gaussian states proper in Part~\ref{lecture:2}, we first prove a heuristic description in terms of normal mode decompositions and Gibbs (or thermal) forms of quantum states. 

Consider a physical bilinear Hamiltonians $\hat{H}_G=\hat{\bm{r}}^\top\bm{\mathcal{H}}\hat{\bm{r}}/2$ with $\bm{\mathcal{H}}>0$ denoting the \emph{Hamiltonian matrix}. The restriction to positive-definite Hamiltonian matrices implies that the Hamiltonian $\hat{H}_G$ has a lower bound on its energy and, thus, a ground (vacuum) state. We further assume that the Hamiltonian matrix is time-independent, or at least approximately so, such that a ground state is unambiguously identifiable.\footnote{For QFCS, this is a strong assumption, as the notion of a vacuum state is generally ambiguous unless time-translation symmetries, quasi-static approximations, or a well-defined scattering problem is defined.} Observe that we ignore a linear term $\propto\bm{\mu}^\top\bm\Omega\hat{\bm r}$ which physically arises by coupling the oscillators to a classical source with amplitude $\bm\mu$. However this linear term can always be subsumed into displacements $D_{\bm\mu}$, which we can tack on separately at the end. Said differently, we work in the displaced frame with the phase-space origin defined at $\bm\mu=\bm 0$.

Since $\bm{\mathcal{H}}>0$, there exists a symplectic transformation $\bm S$ that decouples the Hamiltonian into its normal modes (\emph{Williamson's theorem}), 
\begin{equation}\label{eq:normal_modes}
    \bm{\mathcal{H}}=\bm S^{-\top}\left(\bigoplus_{j=1}^N\omega_j\bm{I}_2\right)\bm S^{-1},
\end{equation}
where $\omega_j>0$ are the normal-mode frequencies. We will accept this decomposition unquestioningly without proof. This normal-mode decomposition actually holds for any positive definite matrix $\bm X>0$, in which case it is often referred to as \emph{symplectic diagonalization} of $\bm X$ and the mode frequencies are the \emph{symplectic eigenvalues} of $\bm X$. Practically, the symplectic eigenvalues can be found by taking the positive eigenvalues of the matrix $i\bm\Omega\bm X$. 

By the normal mode decomposition, we can write the Hamiltonian as follows:
\begin{align}
    \hat{H}_G&= \frac{1}{2}\hat{\bm{r}}^\top\bm{\mathcal{H}}\hat{\bm{r}}  \nonumber\\
    &=\frac{1}{2} \hat{\bm{r}}^\top \bm S^{-\top}\left(\bigoplus_{j=1}^N\omega_j\bm{I}_2\right)\bm S^{-1}\hat{\bm{r}}\nonumber\\
    &=\frac{1}{2}U_{\bm S}\hat{\bm{r}}^\top\left(\bigoplus_{j=1}^N\omega_j\bm{I}_2\right)\hat{\bm{r}}U_{\bm S}^\dagger \nonumber\\
    &=U_{\bm S}\left(\sum_{j=1}^N\frac{\omega_j}{2}(\hat{q}_j^2+\hat{p}_j^2)\right)U_{\bm S}^\dagger\nonumber\\
    &=U_{\bm S}\left(\sum_{j=1}^N\hat{H}_{{\rm qho},j}\right)U_{\bm S}^\dagger.
\end{align}
Therefore, we can think of the coupled network of oscillators as a set of initially independent oscillators that mix via $\bm S$.

Bilinear interactions between oscillators lead to \emph{Gaussian states}, whose name will become apparent in Part~\ref{lecture:2}. For now, we claim that a general Gaussian state $\rho_G$ of $N$ modes can be written in \emph{Gibbs (or thermal) form} as
\begin{equation}
    \rho_G=\frac{D_{\bm\mu}{\rm e}^{-\beta\hat{H}_G} D_{\bm\mu}^\dagger}{\Tr({\rm e}^{-\beta\hat{H}_G})},
\end{equation}
where $\beta$ is the inverse temperature and $\hat{H}_G=\hat{\bm{r}}^\top\bm{\mathcal{H}}\hat{\bm{r}}/2$. By the normal mode decomposition and applying the conjugation relation $Uf(\hat{\bm r})U^\dagger=f(U\hat{\bm r}U^\dagger)$, we can rewrite $\rho_G$ as
\begin{equation}\label{eq:rhoG}
     \rho_G=D_{\bm\mu}U_{\bm S}\left(\bigotimes_{j=1}^N\frac{{\rm e}^{-\beta\hat{H}_{{\rm qho}, j}}}{\Tr({\rm e}^{-\beta\hat{H}_{{\rm qho}, j}})}\right) U_{\bm S}^\dagger D_{\bm\mu}^\dagger.
\end{equation}
For any Gaussian state $\rho_G$, we can thus imagine preparing $\rho_G$ by coupling an independent set of $N$ thermally equilibrated oscillators via $\bm S$ and then displacing each oscillator by $\bm\mu_j$.

Pure Gaussian states $\ket{\psi_G}$ are recovered in the zero temperature limit, $\dyad{\psi_G}=\lim_{\beta\rightarrow\infty}\rho_G(\beta)$. Thus any pure Gaussian state can be written as $\ket{\psi_G}=D_{\bm\mu}U_{\bm S}\ket{0}$, where $\ket{0}$ is the vacuum state. From the Bloch-Messiah decomposition~\eqref{eq:bloch_messiah} and the fact that the vacuum is rotationally invariant, any single-mode pure Gaussian state is thus equivalent to a displaced squeezed vacuum state $\ket{\psi_G}=D_{\bm \mu}U_{\textbf{Sq}(r,\theta)}\ket{0}$, where $r$ is the squeezing strength and $\theta$ is the squeezing angle.

\begin{example}{Gaussian states in the Fock basis}\label{example:states_fock}
\begin{itemize} 
    \item \textit{Coherent state:} The coherent (or displaced vacuum) state $D_{\bm\mu}\ket{0}=\bigotimes_{j=1}^N D_{\bm\mu_j}\ket{0}$ is a pure Gaussian state with non-trivial mean. The variances of each mode are identical and equal to that of the vacuum state---i.e., coherent states saturate Heisenberg's uncertainty principle. A Fock basis representation of the coherent state is show in Eq.~\eqref{eq:coherent_state}. Coherent states are always ``local'' in the sense that a multimode coherent state can always be written as a direct prodect of single-mode coherent states.
    
    \item \textit{Thermal state:} A (single-mode) thermal state $\rho_{\rm th}$ with $\bar{n}=\Tr(\hat{n}\rho_{\rm th})$ quanta can be written in the Fock basis as 
    \begin{equation}\label{eq:thermal_fock}
        \rho_{\rm th}=\frac{1}{\Bar{n}+1}\sum_{n=0}^\infty\left(\frac{\Bar{n}}{\Bar{n}+1}\right)^n\dyad{n}.
    \end{equation}
    If we have a harmonic bath of independent vibrational frequencies $\omega$ at inverse temperature $\beta$, then we describe the bath by the equilibrium state $\bigotimes_{\omega}\rho_{\rm th}(\omega)$ with Planckian spectrum $\Bar{n}_\omega=(\exp(\beta\omega)-1)^{-1}$.

    \item \textit{Squeezed vacuum:} A squeezed vacuum state, generated by acting with $\textbf{Sq}(r)$ on the vacuum [Eq.~\eqref{eq:sqz_transform}], can be written in the Fock basis as,
    \begin{equation}\label{eq:sqzvac_fock}
        \ket{\rm SV}\triangleq U_{\textbf{Sq}(r)}\ket{0}=\frac{1}{\sqrt{\cosh{r}}}\sum_{n=0}^\infty\frac{\sqrt{(2n)!}}{2^n n!}\left(\tanh{r}\right)^n \ket{2n}.
    \end{equation}
    Hence, photons are created in degenerate pairs via squeezing. Squeezed vacuum are currently employed for quantum-enhanced gravitational-wave astronomy~\cite{Caves1981LIGOsqueeze, Schnabel2010NatComm_QuMetrologyGW, LIGO2024Sci_squeeze}.
    
    \item \textit{Two-mode squeezed vacuum (TMSV):} A TMSV, generated via $\bm{S}(r)$ on the vacuum [Eq.~\eqref{eq:tms_transform}, is an entangled state between modes $A$ and $B$. One can think of a TMSV state as a purification of the thermal state~\eqref{eq:thermal_fock}, with $\Bar{n}=\sinh^2{r}$ number of quanta. In which case, it is easy to show that
    \begin{equation}\label{eq:tmsv_fock}
        \ket{\rm TMSV}\triangleq U_{\bm{S}(r)}\ket{0}=\frac{1}{\cosh{r}}\sum_{n=0}^{\infty}(\tanh{r})^n\ket{n}_A\ket{n}_B.
    \end{equation}
    The number of quanta in each mode are perfectly correlated.
\end{itemize}
\end{example}

\clearpage

\begin{nutshell}{}
\begin{itemize}
    \item A quantum harmonic oscillator with Hamiltonian $\hat{H}_{\rm qho}=\hbar\omega(\hat{q}^2+\hat{p}^2)/2$ has (\textit{i}) \emph{canonical operators} $\hat{q}$ and $\hat{p}$ that satisfy the CCR $\comm{\hat{q}}{\hat{p}}=i$ and (\textit{ii}) an infinite dimensional Hilbert space $\mathscr{H}$.
    \item We describe a network of $N$ oscillators (or modes) by the vector of canonical operators
    \begin{equation*}
        \hat{\bm r}=(\hat{q}_1,\hat{p}_1,\dots,\hat{q}_N,\hat{p}_N),
    \end{equation*}
    such that $\comm{\hat{\bm r}_k}{\hat{\bm r}_j}=i\bm\Omega_{ij}$ where $\bm\Omega=\bm I_N\otimes\bm\Omega_1$ is the $N$-mode \emph{symplectic form} and $\bm\Omega_1=\big(
    \begin{smallmatrix}
    0 & 1\\
    -1 & 0
    \end{smallmatrix}\big)$. 

    \item Displacement operators (or \emph{Weyl operators}) $D_{\bm\mu}=\exp(i\bm\mu^\top\bm\Omega\hat{\bm r})$ form an operator basis in $\mathscr{H}$ by which $f(\hat{\bm r})=(2\pi)^{-2N}\int\dd{\bm\mu}\Tr(fD_{\bm\mu}^\dagger)D_{\bm\mu}$ for any bounded operator-valued function $f$. They act as $D_{\bm\mu}^\dagger\hat{\bm r}D_{\bm \mu}=\hat{\bm r}-\bm\mu$.

    \item Open, Markovian Gaussian dynamics arises when system quadratures couple linearly to a Gaussian environment. This leads equivalently to Gaussian Lindblad master equations for states or to Heisenberg-Langevin equations (and input-output relations) for operators.

    \item Bilinear interactions induce linear relations, $\hat{\bm r}\rightarrow\bm S\hat{\bm r}$, where $\bm S\in{\rm Sp}(2N, \mathbb{R})$ is a \emph{symplectic transformation}, such that $\bm S\bm\Omega\bm S^\top=\bm\Omega$.
    
    \item The bilinear Hamiltonian $\hat{H}_G=\hat{\bm r}^\top\bm{\mathcal{H}}\hat{\bm r}/2$, with $\bm{\mathcal{H}}>0$, can be decomposed into its \emph{normal modes} by a symplectic transformation, $\hat{H}_G\overset{\bm S}{\longrightarrow}\sum_{i=1}^N\hbar\omega_i(\hat{q}_i^2+\hat{p}_i^2)/2$.
    
    \item A general \emph{Gaussian state} $\rho_G$ of $N$ oscillators is formally equivalent to a set of displaced thermal states that are coupled by a symplectic transformations:
    \begin{equation*}
        \rho_G=D_{\bm\mu}U_{\bm S}\left(\bigotimes_{i=1}^N\frac{\exp(-\beta\hat{H}_{\rm qho,i})}{\mathcal{Z}_i}\right)U_{\bm S}^\dagger D_{\bm\mu}^\dagger.
    \end{equation*}
\end{itemize}
\end{nutshell}

\clearpage

\part{It's All About the Moments}\label{lecture:2}

In Part~\ref{lecture:1}, we heuristically introduced the moniker \textit{Gaussian states}. Here we justify that name and describe how a quantum state is Gaussian iff its Wigner function is a  multivariate Gaussian distributions in phase space, fully characterized by the mean vector $\bm{\mu}$ and covariance matrix $\bm{\sigma}$ of the quadrature operators. We discuss evolution of Gaussian states through coherent (unitary) evolution and, more generally, Gaussian quantum channels---including noise and loss. Crucially, Gaussian evolution acts linearly on $\bm{\mu}$ and $\bm{\sigma}$, reducing dynamics to matrix algebra. We conclude with Gaussian quantum measurements, such as homodyne detection of quadratures, which likewise preserves Gaussianity, and close with brief remarks on excitation measurements (e.g., photon counting), which are intrinsically non-Gaussian.

\vspace{4em}
\begin{figure}[h]
    \centering
    \includegraphics[width=.49\linewidth]{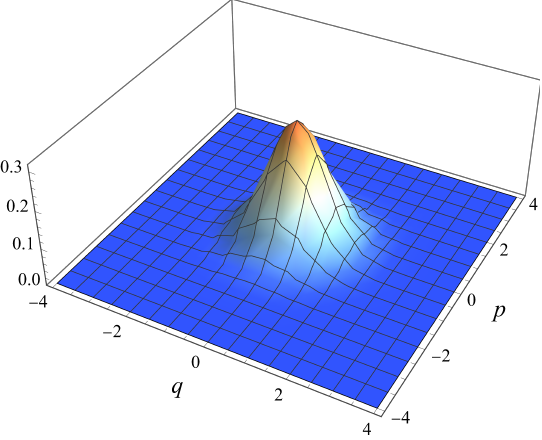}
    \includegraphics[width=.49\linewidth]{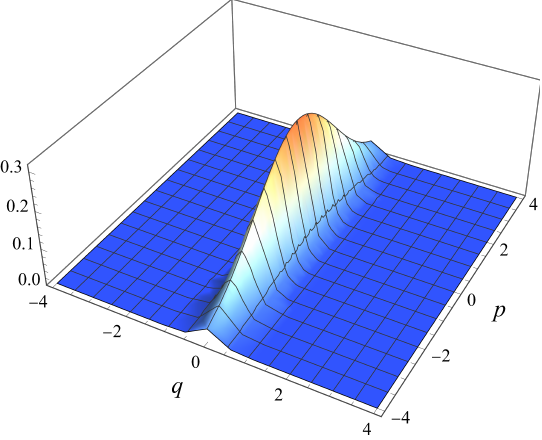}
    \caption{Wigner functions of Gaussian states. Left: vacuum state. Right: squeezed vacuum. Gaussian states are described by multivariate Gaussian distributions in phase space.}
    \label{fig:wigner_fn}
\end{figure}

\clearpage

\section{Mean and Covariance}

We define the mean and covariance matrix of a quantum state $\rho$ as
\begin{align}
    \bm\mu&\triangleq\Tr(\hat{\bm r}\rho),\label{eq:mean_general}\\
    \bm\sigma_{ij}&\triangleq\Tr(\acomm{\hat{\bm r}_i-\bm\mu_i}{\hat{\bm r}_j-\bm\mu_j}\rho).\label{eq:cov_general}
\end{align}
The mean $\bm\mu\in\mathbb{R}^{2N}$ is a $2N$-dimensional real vector, whereas the covariance $\bm\sigma\in\mathbb{R}^{2N}\times \mathbb{R}^{2N}$ is a $2N\times2N$ real, symmetric matrix. Typically higher-order moments (beyond these first and second) are necessary to specify the quantum state, analogous to general probability distributions, however for Gaussian states $\rho_G$, the mean and covariance completely determine the state. 

Consider a Gaussian state $\rho_G$, which takes the generic form in Eq.~\eqref{eq:rhoG}, with mean and covariance $(\bm\mu, \bm\sigma)$. We can decompose the covariance matrix as
\begin{equation}
   \bm\sigma=\bm S\left(\bigoplus_{j=1}^N\nu_j\bm I_2\right)\bm S^\top,\label{eq:cov-rhoG}
\end{equation}
where $\bm S$ denotes the symplectic matrix that diagonalizes $\bm\sigma$ and $\nu_j$ are the symplectic eigenvalues of $\bm\sigma$ that relate to the mode frequencies $\omega_j$ and inverse temperature $\beta$ via
\begin{equation}
    \nu_j=\coth(\beta\omega_j/2)\geq1.
\end{equation}
The symplectic eigenvalues determine the mean number of quanta in the modes via $\bar{n}_{j}=(\nu_j-1)/2=1/(e^{\beta\omega_j}-1)$. We remark that reduced Gaussian states can be found by simply extracting the relevant components of $\bm\mu$ and the corresponding diagonal blocks of $ \bm\sigma$.

Not all symmectric matrices serve as good covariance matrices, since the covariance matrix must correspond to a proper quantum state and adhere to Heisenberg's uncertainty principle. Any proper covariance matrix $\bm\sigma$ must satisfy the \textit{Robertson-Schr\"odinger uncertainty relation},
\begin{equation}\label{eq:robertson_uncertainty}
    \bm\sigma+i\bm\Omega\geq0.
\end{equation}
We will not prove this relation here (see Section 3.4 in Ref.~\cite{serafini2017book}) but just know that it originates from the CCR, which introduces the canonical form $\bm\Omega$ above, and the fact that $\rho>0$. Since $\bm u^\top\bm\Omega\bm u=0$ for any vector $\bm u$, a less restrictive condition $\bm\sigma>0$ immediately follows, which is a condition that classical probability distributions must satisfy. For single-mode covariance matrices, the Robertson-Schr\"odinger uncertainty relation implies that $\det\bm\sigma\geq 1$. For intance, given $\bm\sigma={\rm diag}(2\sigma^2_q,2\sigma^2_p)$, where $\sigma^2_q$ and $\sigma^2_p$ are the quadrature variances,\footnote{The factors of 2 derive from our definition of the covariance matrix in Eq.~\eqref{eq:cov_general}.} we recover Heisenberg's uncertainty principle, $\sigma^2_q\sigma^2_p\geq1/4$.

Below, we characterize common Gaussian states by their first and second moments. This characterization is formally equivalent to the Fock-basis expansion of these states in Part~\ref{lecture:1}, however the description in terms of moments is cleaner and highlights the simplicity of the phase-space methods.

\begin{example}{Gaussian states}
    \begin{itemize}
    \item \textit{Pure Gaussian states}: Any pure Gaussian state $\ket{\psi_G}=D_{\bm\mu}U_{\bm S}\ket{0}$ has a mean and covariance that can be written in terms of the displacement parameters $\bm\mu$ and the symplectic transformation $\bm S$, viz.,
    \begin{equation}
        \bm\mu[\psi_G]=\bm\mu \qq{and} \bm\sigma[\psi_G]=\bm S\bm S^\top.
    \end{equation}
    \item \textit{Coherent state}: The coherent state $D_{\bm\mu}\ket{0}$ is a pure Gaussian state with non-trivial mean and vacuum covariance matrix,
    \begin{equation}
        \bm\mu_{\rm coh}=\bm\mu \qq{and} \bm\sigma_{\rm coh}=\bm I.
    \end{equation} 
    A Fock basis representation of the coherent state is given in Eq.~\eqref{eq:coherent_state}.
    \item \textit{Thermal states}: A thermal states $\rho_{\rm th}$ with $\bar{n}$ quanta has
    \begin{equation}
        \bm\mu_{\rm th}=\bm 0 \qq{and} \bm\sigma_{\rm th}= (1+2\Bar{n})\bm I_2.
    \end{equation}
    A Fock basis representation of the thermal state is given in Eq.~\eqref{eq:thermal_fock}.
    
    \item \textit{Squeezed vacuum}: Acting with a squeezer [Eq.~\eqref{eq:sqz_transform}] on the vacuum state generates a single-mode squeezed vacuum, which has first and second moments,
    \begin{equation}
        \bm\mu_{\rm SV}=\bm 0 \qq{and} \bm\sigma_{\rm SV}=e^{2r\bm Z}.
    \end{equation}
    A Fock basis representation of the squeezed vacuum state is given in Eq.~\eqref{eq:sqzvac_fock}. Squeezed vacuum are currently employed for quantum-enhanced gravitational-wave astronomy~\cite{Caves1981LIGOsqueeze, Schnabel2010NatComm_QuMetrologyGW, LIGO2024Sci_squeeze}.
    
    \item \textit{Two-mode squeezed vacuum (TMSV)}: Consider a TMS transformation $\bm S_{G}$~\eqref{eq:tms_transform}, with gain $G=\cosh^2{r}$. Acting with a TMS on two vacuum states generates the entangled two-mode squeezed vacuum (TMSV) state with mean and covariance,
    \begin{equation}\label{eq:sigma_tmsv}
        \bm\mu_{\rm TMSV}=\bm 0 \qq{and} \bm\sigma_{\rm TMSV}=
        \begin{pmatrix}
           \left(2G-1\right)\bm I_2 & 2\sqrt{G(G-1)}\bm Z \\
           2\sqrt{G(G-1)}\bm Z & \left(2G-1\right)\bm I_2
        \end{pmatrix}.
    \end{equation}
    If we restrict to a single mode (i.e., diagonal blocks of $\bm\sigma_{\rm TMSV}$), the reduced state looks locally thermal with $\Bar{n}=G-1=\sinh^2{r}$ number of quanta. A Fock basis representation of the TMSV state is given in Eq.~\eqref{eq:tmsv_fock}.
\end{itemize}
\end{example}

\subsection*{Gaussian quantum evolution}

Under a unitary Gaussian transformation, corresponding to a displacement $\bm \nu$ and symplectic transformation $\bm S$, the mean and covariance transform simply. From Part~\ref{lecture:1}, we know that $\hat{\bm r}^{\rm out}=\bm S\hat{\bm r}^{\rm in}+\bm\nu$. Therefore:
\begin{align}
    \bm\mu^{\rm out}&=\bm S\bm\mu^{\rm in}+\bm \nu,\label{eq:mean_inout}\\
    \bm\sigma^{\rm out}&=\bm S\bm\sigma^{\rm in}\bm S^{\top}.\label{eq:covariance_inout}
\end{align}
This prescription applies to time evolution via $\hat{H}=\hat{\bm{r}}^\top \bm{\mathcal{H}}\hat{\bm{r}}/2$, in which case $\bm S= \exp(t\bm{\Omega}\bm{\mathcal{H}})$.

What about non-unitary evolution by the Gaussian quantum channel $\mathcal{E}_G$? In Part~\ref{lecture:1} [Eq.~\eqref{eq:dilation_gauss}], we formally introduced the unitary dilation of a Gaussian channel by considering an environment $E$ in a Gaussian state $\tau_E$ that couples to the system by means of a Gaussian unitary $U_{\bm S}$. We can push the analysis further. Write the interaction symplectic matrix in block form:
\begin{equation}\label{eq:S_partition}
    \bm S =
\begin{pmatrix}
    \bm A & \bm B \\
    \bm C & \bm D
\end{pmatrix}.
\end{equation}
The $2N\times 2N$ matrix $\bm A$ determines the internal evolution of the system, while the $2N\times 2M$ rectangular matrix $\bm B$ encodes the interaction between the $N$ system modes and $M$ environment modes. Now, given a Gaussian environment state ${\rho}_E$, characterized by the first and second moments $\bm \mu_E$ and $\bm\sigma_E$, the Gaussian quantum channel $\mathcal{E}_G:\mathscr{H}^{\otimes N}\rightarrow\mathscr{H}^{\otimes N}$ is completely determined by the displacement-noise vector $\bm d$, the scaling matrix $\bm X$, and the noise matrix $\bm Y$, given as:
\begin{align}
    \bm d &= \bm B\bm\mu_E,\\
    \bm X &= \bm A,\\
    \bm Y &= \bm B\bm\sigma_E\bm B^\top.
\end{align}
To see why this characterization is complete, consider the output operators for the system, $\hat{\bm r}^{\rm out}_S=\bm A\hat{\bm r}^{\rm in}_S+\bm B\hat{\bm r}^{\rm in}_E$, which have contributions from both the system and the environment. If we compute $N$th moments of the output operators, $\hat{\bm r}^{\rm out}_S$, this computation will carry terms like $\bm A^N\ev{(\hat{\bm r}^{\rm in}_S)^N}$, $\bm B^N\ev{(\hat{\bm r}^{\rm in}_E)^N}$, and cross terms of the form $\bm A^{K}\bm B^{N-K}\ev{(\hat{\bm r}^{\rm in}_S)^K}\ev{(\hat{\bm r}^{\rm in}_E)^{N-K}}$. Note the cross terms decouple since the initial system-environment state is uncorrelated [Eq.~\eqref{eq:dilation_gauss}]. Furthermore, since the environment state is Gaussian, any $\ell$th order moment $\ev{(\hat{\bm r}^{\rm in}_E)^{\ell}}$ can be written in terms of the first and second moments, $\bm\mu_E$ and $\bm\sigma_E$, via Wick's theorem. Finally, the scaling of $\bm B$ coupled with $\ev{(\hat{\bm r}^{\rm in}_E)^{N-K}}$ and that of $\bm A$ with $\ev{(\hat{\bm r}^{\rm in}_S)^K}$ implies that $\bm d$, $\bm X$, and $\bm Y$ are sufficient to describe the transformation induced by $\mathcal{E}_G$.

\begin{figure}
    \centering
    \includegraphics[width=.7\linewidth]{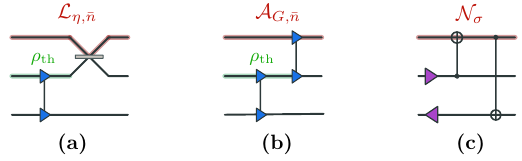}
    \caption{Unitary dilation of single-mode Gaussian channels: (a) thermal loss channel, (b) thermal amplifier channel, (c) additive Gaussian noise channel (a.k.a., random displacements). A purification of the environment thermal state $\rho_{\rm th}$ is also shown as a two-mode squeezed vacuum for (a, b).}
    \label{fig:dilation}
\end{figure}

We can thus characterize \textit{all} bosonic Gaussian channels by three objects $(\bm d,\bm X,\bm Y)$. Unitary operations have $\bm Y=\bm 0$ and $\bm X\in{\rm Sp}(2N,\mathbb{R})$ such that $\bm X\bm\Omega\bm X^\top=\bm\Omega$, with $\bm d$ corresponding to a (unitary) displacement. The evolution of a Gaussian system through a Gaussian channel is also quite simple. Given a quantum system with input moments $\bm\mu^{\rm in}$ and $\bm\sigma^{\rm in}$, a Gaussian channel $\mathcal{E}_G$ with characterization $(\bm d,\bm X,\bm Y)$ yields the input-output relations
\begin{equation}\label{eq:moments-inout}
    \bm\mu^{\rm out}=\bm X\bm^{\rm in}\bm\mu^{\rm in} + \bm d \qq{and}
    \bm\sigma^{\rm out}=\bm X\bm\sigma^{\rm in}\bm X^\top + \bm Y.
\end{equation}
Hence we have reduced Gaussian quantum evolution of a Gaussian state to simple matrix multiplication! 

Below we give example characterizations of canonical single-mode Gaussian channels and discuss (non-unique) unitary dilations. For visualization, we showcase symplectic diagrams of the dilations in Fig.~\ref{fig:dilation}.

\begin{example}{Characterizing single-mode Gaussian channels}
\begin{itemize}
    \item \textit{Thermal loss.} A thermal loss channel $\mathcal{L}_{\eta,\bar{n}}$ with transmittance $\eta$ and noisy quanta $\bar{n}$ has a characterization
    \begin{align}
    \bm d_{\mathcal{L}}&=\bm 0,\\
    \bm X_{\mathcal{L}}&=\sqrt{\eta}\bm I_2,\\
    \bm Y_{\mathcal{L}}&=(1-\eta)(1+2\bar{n})\bm I_2. \label{eq:app_thermal_loss_channel}
    \end{align}
    The thermal loss channel attenuates the input intensity by $\eta$ and adds $(1-\eta)(1/2+\bar{n})$ thermal noise to each quadrature. A unitary dilation of a thermal loss channel is a beamsplitter interaction between the system and an environment in a thermal state. [The thermal environment may be further extended to a pure two-mode squeezed vacuum; see Fig.~\ref{fig:dilation}(a).] For $\bar{n}=0$, the channel is a pure-loss channel which still adds vacuum noise to each quadrature.

    \item \textit{Thermal amplifier.} A thermal amplifier channel $\mathcal{A}_{G,\bar{n}}$ with gain $G$ and noisy quanta $\bar{n}$ has a characterization 
    \begin{align}
    \bm d_{\mathcal{A}}&=\bm 0,\\
    \bm X_{\mathcal{A}}&=\sqrt{G}\bm I_2,\\
    \bm Y_{\mathcal{A}}&=(G-1)(1+2\bar{n})\bm I_2.
    \end{align}
    The amplifier channel amplifies the input intensity by the gain $G$ and adds $(G-1)(1/2+\bar{n})$ noise to each quadrature. A unitary dilation is a two-mode squeezing interaction between the system and an environment mode occupying a thermal state. For $\bar{n}=0$, the channel is referred to as a quantum-limited amplifier---an important quantum technology~\cite{Caves1982:LinAmps}.
    
    \item \textit{Additive noise.} An additive Gaussian noise (AGN) channel $\mathcal{N}_{\sigma}$ with variance $\sigma^2$ has a characterization $\bm d_{\mathcal{N}}=\bm0$, $\bm X_{\mathcal{N}}=\bm0$, and $\bm Y_{\mathcal{N}}=2\sigma^2\bm I_2$. This corresponds to stochastic displacements, i.e., a random walk in phase space. One unitary dilation is found by coupling the system to two separate environment modes via SUM gates.
    \end{itemize}
    \end{example}

We will come to realizations of these channels when we discuss semi-classical black holes in Part~\ref{lecture:4}. We will see that late-time interaction with a non-rotating black hole permits a description as a thermal-loss channel, with the transmittance ($\eta$) characterizing the greybody factor and the additive thermal noise corresponding to thermal Hawking radiation emitted by the black hole. In the superradiant regime, late-time interaction with a rotating black hole has a description as a thermal amplifier channel, where thermal Hawking radiation seeds superradiant wave amplification.



\section{Characteristic and Wigner Functions}

A useful description of a bosonic quantum state is in terms of either its characteristic function or Wigner function. For Gaussian quantum states, these functions correspond to multi-variate Gaussian distributions, which justifies the moniker Gaussian quantum state.

Physical quantum states are bounded operators on the Hilbert space. We can therefore expand physical states in the Weyl operator basis per Eq.~\eqref{eq:weyl_transform}:
\begin{equation}\label{eq:characteristic_fn}
    \rho=\frac{1}{(2\pi)^{2N}}\int_{\bm r\in\mathbb{R}^{2N}}\dd{\bm r}\chi(\bm r)D_{\bm r},
\end{equation}
where $\chi(\bm r)\triangleq\Tr(\rho D_{\bm r}^\dagger)$. The quantity $\chi(\bm r)$ is known as the \emph{characteristic function} of $\rho$. Given $\Tr(\rho)=1$, $\rho^\dagger=\rho$, and $\Tr(D_{\bm r})=(2\pi)^N\delta^{2N}(\bm r)$ [Eq.~\eqref{eq:weyl_opbasis}], we find that $\chi(\bm 0)=1$ and $\chi^*(-\bm r)=\chi(\bm r)$. Consider a Gaussian unitary transformation, $\rho\rightarrow D_{\bm\mu} U_{\bm S}\rho U_{\bm S}^\dagger D_{\bm\mu}$. By Eq.~\eqref{eq:weyl_Stransform} and symplectic invariance of the measure $\dd(\bm S\bm r)=\dd{\bm r}$, it follows that
\begin{equation}\label{eq:characteristic_transform}
    \chi(\bm r)\, \longrightarrow \, e^{-i\bm\mu^\top\bm\Omega\bm r}\chi(\bm S^{-1}\bm r).
\end{equation}

If we perform a Fourier transform on the characteristic function, then we obtain the \emph{Wigner function}:
\begin{equation}
    W(\bm x)\triangleq\frac{1}{(2\pi)^{2N}} \int_{\bm r\in\mathbb{R}^{2N}}\dd{\bm r}{\rm e}^{i\bm x^\top\bm\Omega\bm r}\chi(\bm r).
    \label{eq:Wigner_func}
\end{equation}
The Wigner function is a quasi-probability distribution in phase space, with ``quasi'' referring to the fact that it is generically non-positive. On the other hand, cuts (or marginals) of the Wigner function do produce genuine probability distributions; for instance, given a $N$-mode quantum state $\rho$, one can show that $\int\dd^N{q}W(q_1,q_2,\dots,q_N;p_1,p_2,\dots,p_N)=\expval{\rho}{p_1,p_2,\dots,p_N}$, where the integral is over all $q$ quadratures. This is just the probability to measure the quantum system with momenta $p_1,\dots,p_N$.

The defining feature of a Gaussian state is that the characteristic and Wigner functions of a Gaussian state are multi-variate Gaussian distributions with mean $\bm\mu$ and covariance $\bm\sigma$. Therefore, we can completely describe Gaussian quantum states via Gaussian probability distributions in phase space. We show this explicitly by calculating the characteristic function of a Gaussian state using its Gibbs form~\eqref{eq:rhoG}. In fact, we need only calculate the characteristic function of a thermal state and then use Eq.~\eqref{eq:characteristic_transform} to find the characteristic function for a general Gaussian state.

For simplicity, we calculate the characteristic function $\chi_{\rm th}$ for a single-mode thermal state $\rho_{\rm th}$, which we can straightforwardly extend to a direct product of thermal state $\bigotimes_i\rho_{{\rm th},i}$. We make use of the following relation:\footnote{Interpretation: We can generate a thermal state by randomly displacing the vacuum.}
\begin{equation}
    \rho_{\rm th}=\frac{1}{(2\pi\bar{n})}\int_{\bm r\in\mathbb{R}^2}\dd{\bm r}e^{-\frac{\abs{\bm r}^2}{2\bar{n}}} D_{\bm r}\dyad{0}D_{\bm r}^\dagger.
\end{equation}
Whence, to calculate $\chi_{\rm th}(\bm s)=\Tr(D_{\bm s}^\dagger\rho_{\rm th})$, we must compute the characteristic function of the coherent state $\ket{\bm{r}}=D_{\bm r}\ket{0}$. We find:
\begin{align}
    \chi_{\bm r}(\bm s)&=\Tr{\dyad{\bm{r}}D_{\bm s}^\dagger}\\ 
    &=\expval{D_{\bm r}^\dagger D_{\bm s}^\dagger D_{\bm r}}{0}\nonumber\\
    &=e^{i\bm s^\top\bm\Omega\bm r}\expval{D_{\bm s}^\dagger}{0}\nonumber\\
    &= e^{i\bm s^\top\bm\Omega\bm r} e^{-\frac{1}{4}\bm s^2}.
\end{align}
The third equality follows from the composition rule~\eqref{eq:weyl_comp}. The final equality follows from the fact that $\bra{0}D_{\bm s}^\dagger=\bra{\bm{s}}$, and the vacuum component $\innerproduct{\bm{s}}{0}$ is an exponential function per Eq.~\eqref{eq:coherent_state}. Thus:
\begin{align}
    \chi_{\rm th}(\bm s)&=\frac{1}{(2\pi\bar{n})}\int_{\bm r\in\mathbb{R}^2}\dd{\bm r}e^{-\frac{\abs{\bm r}^2}{2\bar{n}}}\chi_{\bm r}(\bm s)\nonumber\\
    &=\frac{e^{-\frac{1}{4}\bm s^2}}{(2\pi\bar{n})}
    \int_{\bm r\in\mathbb{R}^2}\dd{\bm r}e^{-\frac{\abs{\bm r}^2}{2\bar{n}}}e^{i\bm s^\top\bm\Omega\bm r} \nonumber\\
    &=\exp\left(-\frac{1}{4}\bm s^\top\bm s(1+2\bar{n})\right). \label{eq:chi_th_1mode}
\end{align}
To arrive at the final equality, we employ the Fourier transform of a multi-variate Gaussian distribution.

\begin{example}{Fourier transform of a multivariate Gaussian}
Consider a $2N\times2N$ symmetric matrix $\bm V>0$ and a vector $\bm b\in\mathbb{R}^{2N}$, then:
\begin{equation}\label{eq:multivariate_fourier_transform}
    \int_{\mathbb{R}^{2N}}\dd{\bm r}{\rm e}^{-\bm r^\top\bm V\bm r + \bm b^\top\bm r}=\frac{\pi^{N}}{\sqrt{\det\bm V}}{\rm e}^{\frac{1}{4}\bm b^\top\bm V^{-1}\bm b}.
\end{equation}
The proof is quite simple. Consider a symplectic transformation $\bm S:\bm r\rightarrow\bm S\bm r$ that diagonalizes $\bm V$, i.e. $\bm S\bm V\bm S^\top=\bigoplus V_i\bm I_2$. Note that $\dd{\bm r}\rightarrow\dd{(\bm S\bm r)}=\dd{\bm r}$. Then, the $2N$-dimensional integral reduces to a product of $2N$ one-dimensional Gaussian integrals.
\end{example}

We extend now to $N$ modes. Recall that $\bm\sigma_{\rm th}=(1+2\Bar{n})\bm I_2$ is the covariance matrix for a single-mode thermal state; hence, the characteristic function for the thermal state [Eq.~\eqref{eq:chi_th_1mode}] is completely determined by $\bm\sigma_{\rm th}$. For an uncorrelated $N$ mode thermal state, the covariance matrix is the direct sum $\bigoplus_{i=1}^N\bm\sigma_{{\rm th} ,i}$. Therefore, the characterstic function of an uncorrelated $N$-mode thermal state is the product $\chi_{\rm th}(\bm s)\triangleq\prod_{i=1}^N\chi_{{\rm th},i}(\bm s_i)$. 

Recalling the Gibbs form of a generic Gaussian state~\eqref{eq:rhoG} and by Eq.~\eqref{eq:characteristic_transform}, we deduce the characteristic function for a general Gaussian state $\rho_G$ with mean $\bm\mu$ and covariance $\bm\sigma$ to be $\chi_G(\bm r)=e^{-i\bm\mu^\top\bm\Omega\bm r}\chi_{\rm th}(\bm S^{-1}\bm r)$. Given that $\bm\sigma=\bm S(\bigoplus_{i=1}^N\bm\sigma_{{\rm th} ,i})\bm S^\top$, which we derive from the input-output relation~\eqref{eq:covariance_inout}, and using $\bm S^{-\top}\bm S^{-1}=\bm\Omega^\top\bm S\bm S^\top\bm\Omega$, we compute 
\begin{equation}\label{eq:characteristic_gaussian}
    \chi_G(\bm r)=\exp\left(-i\bm\mu^\top\bm\Omega\bm r\right)\exp\left(-\frac{1}{4}\bm r^\top\bm\Omega^\top\bm\sigma\bm\Omega\bm r\right).
\end{equation}
Taking the Fourier transform [Eq.~\eqref{eq:Wigner_func}] and using Eq.~\eqref{eq:multivariate_fourier_transform}, we finally arrive at the Wigner function for a Gaussian state:
\begin{equation}\label{eq:wigner_gaussian}
    W_G(\bm x)=\frac{1}{\pi^N\sqrt{\det\bm\sigma}}\exp\left[-(\bm x-\bm\mu)^\top\bm\sigma^{-1}(\bm x-\bm\mu)\right].
\end{equation}
Since $W_G(\bm x) \geq 0$ and $\int\dd{\bm{x}} W_G(\bm{x})=1$, the Wigner function defines a bona fide multivariate Gaussian probability distribution in phase space. In Fig.~\ref{fig:wigner_fn}, we plot the Wigner function for the vacuum state (with $\bm\mu=\bm 0$ and $\bm\sigma_{\rm vac}=\bm I_2$) and squeezed vacuum state (with $\bm\mu_{\rm SV}=\bm 0$ and $\bm\sigma_{\rm SV}=e^{2r\bm Z}$).

\section{Quantum Measurements}

Quantum measurements of a bosonic system can be broadly divided into \emph{Gaussian} and \emph{non-Gaussian} categories, depending on whether the measurement outcomes depend linearly or nonlinearly on the quadratures. Gaussian measurements---such as homodyne and heterodyne detection---yield measurement results that are linear combinations of the canonical variables $(\hat{q},\hat{p})$. Consequently, their fundamental noise floor is set by the vacuum fluctuations of the ground state. When the quantity of interest depends on nonlinear combinations of the field, most notably the excitation number $\hat{n} = \hat{a}^\dagger \hat{a}$, Gaussian measurements become intrinsically limited: if the mean excitation is well below a single photon (e.g., $\langle \hat{n} \rangle \ll 1$), the signal is effectively buried beneath vacuum noise. In this scenario, direct excitation measurements can dramatically improve sensitivity by bypassing the vacuum-noise floor. Modern photonic platforms routinely employ simplified photon detectors that can identify the \emph{presence} of an excitation (a ``click'') without fully resolving the excitation number. These techniques are now standard tools in microwave, optical, and trapped-ion architectures for sensing and quantum-information-processing tasks.

In this section, we discuss Gaussian measurements, photon counting, and click detection, as well as how one might perform correlation measurements in an two-mode bosonic system.

\subsection*{Gaussian Measurements}

\begin{figure}
    \centering
    \includegraphics[width=.75\linewidth]{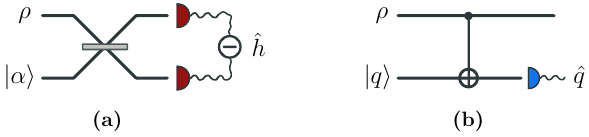}
    \caption{Homodyne measurement. (a) Mix the system mode with a local oscillator (strong classical field with $\abs{\alpha}^2\gg\Tr{\hat{n}\rho}$) on a beamsplitter followed by an intensity-difference measurement. This direct measurement is generally destructive. (b) Ancilla-assisted measurement where we couple the system to an ancilla by a SUM-gate and follow with homodyne measurement of the ancilla. This indirect measurement is non-destructive.}
    \label{fig:measurement}
\end{figure}

An ideal Gaussian measurement is a projection onto a pure Gaussian state $\ket{\psi_G}$. We can think of this at the level of a positive operator-valued measure (POVM) via resolution of the identity:
\begin{equation}\label{eq:generaldyne}
    \frac{1}{(2\pi)^N}\int_{\bm\mu\in\mathbb{R}^{2N}}\dd{\bm r}D_{\bm r}U_{\bm S}\dyad{0}U_{\bm S}^\dagger D_{\bm r}^\dagger = \hat{I},
\end{equation}
which generalizes Eq.~\eqref{eq:alpha_completeness}. The (parameterised) Gaussian state is $\ket{\psi_G(\bm r,\bm S)}=D_{\bm r}U_{\bm S}\ket{0}$. The POVM is then $\{E_{\bm r,\bm S}\}$ with $E_{\bm r,\bm S}=\dyad{\psi_G}/(2\pi)^N$. For a Gaussian state $\rho$ with mean $\bm\mu$ and covariance $\bm\sigma$, we can immediately calculate the probability to project onto $\ket{\psi_G(\bm\mu,\bm S)}$:
\begin{equation}
    \Tr(E_{\bm r,\bm S}\rho)=\frac{1}{\pi^N\sqrt{\det(\bm\sigma+\bm S\bm S^\top)}}\exp\left[-(\bm r-\bm\mu)^\top\left(\bm\sigma+\bm S\bm S^\top\right)^{-1}(\bm r-\bm\mu)\right].
\end{equation}
This follows by: (1) $\expval{\rho}{\psi} =\int_{\bm r\in\mathbb{R}^{2N}}\dd{\bm r}\chi_\rho(\bm{r})\chi_\psi^*(\bm{r})/(2\pi)^N$ per Eq.~\eqref{eq:characteristic_fn}; (2) using the characteristic functions for Gaussian states~Eq.~\eqref{eq:characteristic_gaussian}; and (3) employing the Fourier transform~\eqref{eq:multivariate_fourier_transform}.

A common measurement is a so-called \emph{homodyne measurement} where the state $\rho$ is projected onto a position (or momentum) eigenstate $\ket{q}$ (or $\ket{p}$). Formally, this corresponds to taking $U_{\bm S}$ as a squeezing operator in the limit of infinite squeezing. Practically, if we want to perform a homodyne measurement in, e.g., the optical domain,\footnote{In principle, this scheme works for any bosonic system. Though in practice, things become can become complicated depending on the actual physical setup. For instance, homodyne detection of microwave fields is a bit more involved---requiring a chain of amplifiers, frequency mixers, and a low-pass filter.} we can mix the system with a local oscillator (a.k.a., strong coherent field with $\abs{\alpha}^2\gg\Tr{\hat{n}\rho}$) on a beam-splitter and measure intensity differences of the output fields via photodetection; see Fig.~\ref{fig:measurement}(a) for an illustration and the example below for details. 

\begin{example}{Homodyne detection scheme}
    We argue that the setup in Fig.~\ref{fig:measurement}(a) realizes a homodyne measurement. Consider the joint state $\rho\otimes\dyad{\alpha}$ described by mode operators $\hat{a}$ and $\hat{b}$, respectively, and consider photodetectors $D_a$ and $D_b$ which are situated at the outputs of a 50:50 beamsplitter. Define the intensity difference operator $\hat{h}=\hat{a}^{(\rm out)\,\dagger}\hat{a}^{(\rm out)}-\hat{b}^{(\rm out)\,\dagger}\hat{b}^{(\rm out)}$, which can be measured by the photodetectors $D_a$ and $D_b$, where
\begin{align}
    \hat{a}^{(\rm out)}&=\frac{1}{\sqrt{2}}(\hat{a}^{(\rm in)}+\hat{b}^{(\rm in)}),\\
     \hat{b}^{(\rm out)}&=\frac{1}{\sqrt{2}}(\hat{b}^{(\rm in)}-\hat{a}^{(\rm in)}),
\end{align}
are the output modes of the beamsplitter. Writing $\hat{h}$ in the ``in'' basis, we have $\hat{h}= \hat{a}^{(\rm in)\,\dagger}\hat{b}^{(\rm in)}+\hat{b}^{(\rm in)\,\dagger}\hat{a}^{(\rm in)}$. It is then easy to show that 
\begin{equation}
    \Tr{\hat{h}(\rho\otimes\dyad{\alpha})}=\sqrt{2}\abs{\alpha}\Tr{\hat{q}_\phi^{(\rm in)}\rho},
\end{equation}
where $\hat{q}_\phi^{(\rm in)}=(\hat{a}^{(\rm in)}e^{-i\phi}+\hat{a}^{(\rm in)\,\dagger}e^{i\phi})/\sqrt{2}$ and $\phi=\arg(\alpha)$ is the phase of the local oscillator. Thus the physical scheme realizes the first moments exactly. To estimate the second moments, we consider $\Tr{\hat{h}^2(\rho\otimes\dyad{\alpha})}$. After some algebra, we obtain:
\begin{equation}
\Tr{\hat{h}^2(\rho\otimes\dyad{\alpha})}=2\abs{\alpha}^2\left(\Tr{\hat{q}_\phi^{(\rm in)\,2}\rho}+\frac{\Tr{\hat{n}\rho}}{\abs{\alpha}^2}\right).
\end{equation}
As $\abs{\alpha}^2/\Tr(\hat{n}\rho)\rightarrow\infty$, the second moments are realized. Similar relations hold for higher-order moments.
\end{example}

\begin{figure}
    \centering
    \includegraphics[width=.5\linewidth]{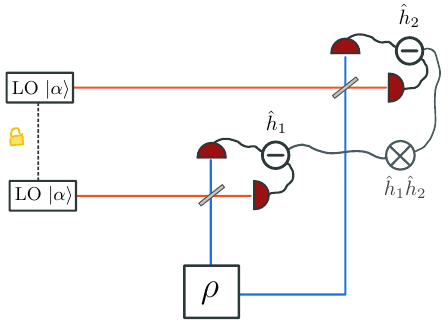}
    \caption{Correlation measurements via joint homodyne. Outputs $\hat{h}_1$ and $\hat{h}_2$ of two homodyne measurements are multiplied to estimate the two-mode correlations, e.g., $\expval{\hat{q}_1\hat{q}_2}$.}
    \label{fig:multimode_measurement}
\end{figure}

The discussion above was restricted to a single mode. What about multimode measurements? We will not delve too much into the details here but simply mention that one possible setup to measure two-mode correlations (encoded in the covariance matrix $\bm\sigma$) is a joint homodyne scheme, as depicted in Fig.~\ref{fig:multimode_measurement}. In this setup, the outputs of two homodyne measurements are multiplied to infer quadrature correlations. For a multimode Gaussian state, measuring all possible two-mode correlations is sufficient to characterize the state. I do note, however, that this does not imply optimal estimation of interesting parameters (e.g., temperature, mode-mode scattering rates etc.) encoded in the state.

The practical schemes above are destructive measurement processes due to direct photodetection of the quantum state under consideration. To perform a non-destructive homodyne measurements of the signal state, we can couple the signal modes to ancillary modes (prepared in position eigenstates) by the SUM-gate [Eq.~\eqref{eq:sum_gate}] and then perform homodyne measurements on the ancillary modes; see Fig.~\ref{fig:measurement}(b). To illustrate this, consider a single-mode quantum state $\ket{\Psi}$ and observe:
\begin{align}
\underbrace{e^{-i\hat{q}\otimes\hat{p}}}_{\text{SUM gate}}\ket{\Psi}\otimes\ket{q^\prime}&=e^{-i\hat{q}\otimes\hat{p}}\int\dd{Q}\Psi_Q\ket{Q}\otimes\ket{q^\prime}\\
    &=\int\dd{Q}\Psi_Q\ket{Q}\otimes e^{-iQ\hat{p}}\ket{q^\prime}\\
    &=\int\dd{Q}\Psi_Q\ket{Q}\otimes\ket{q^\prime+Q},
\end{align}
We have completely correlated the position of the signal with the ancilla and, therefore, can infer the position of the signal mode ($Q$) by homodyne measurements of the ancillary mode (with outcomes $s=q'+Q$). This ancilla-assisted measurement protocol is non-destructive on the signal mode and, furthermore, allows us to perform a ``mid-circuit'' homodyne measurement in the middle of the signal-mode evolution.

\subsection*{Counting Quanta}

Photon counting provides a genuine non-Gaussian measurement that resolves the discrete excitations of an oscillator (e.g., photons, phonons etc.). These measurements are particularly useful in the few-photon regime, where excitation statistics carry significant information and where linear detection may be suboptimal for certain tasks.\footnote{Full photon counting corresponds to the POVM $\{E_n\}$ with $E_n =\dyad{n}$, resolving the state in the Fock basis. This becomes experimentally challenging as the occupation number grows.} While full number resolution grants access to the Fock-basis representation of the state, click (or ``bucket'') detection---which merely distinguishes the presence or absence of an excitation---can be highly informative. In the case of Gaussian states, the click probabilities remain straightforward to compute. In regimes where the field population is well below a single photon on average, click detection effectively approximates true photon counting while being far simpler to implement.

\paragraph*{Click detection.}
Suppose we have a photodetector that only ``clicks" when at least one photon arrives. We describe this device with measurement operators $E_0=\dyad{0}$ (no click) and $E_1 = I-E_0$ (click). We see that a ``click'' does not strictly tell us how many photons arrive, only that \textit{at least one} photon arrived. This is okay so long as the photon count is low. Click detectors are also much easier to build and implement in the lab than full photon counting.

The no-click probability for a single-mode
Gaussian state $\rho_G$ is given by
\begin{equation}
    p_0[\rho_G]
    = \expval{\rho_G}{0}
    = \frac{2}{\sqrt{\det(\bm\sigma+\bm{I}_2)}}\exp\!\left[
        -\bm\mu^\top(\bm\sigma+\bm{I}_2)^{-1}\bm\mu
    \right] ,
\end{equation}
where $\bm{\mu}$ and $\bm{\sigma}$ denote its first and second moments. For a
thermal state with $\bm{\sigma}_{\rm th}=(1+2\bar{n})\bm{I}_2$, this yields
$p_0[\rho_{\rm th}]=1/(\bar{n}+1)$, matching the $n=0$ component of
Eq.~\eqref{eq:thermal_fock}; thus
$p_1[\rho_{\rm th}]=\bar{n}/(\bar{n}+1)$. In the small-population limit,
$p_0\approx 1$ and $p_1\approx \bar{n}$, so a click event estimates the mean excitation number when $\bar{n}\ll 1$.

\paragraph*{Correlation measurements.}
In principle, beamsplitters and photon counting provides an alternative way to perform \emph{phase-insensitive} correlation measurements and, thus, reconstruct (part of) the quadrature covariance. Consider interfering two modes on a balanced 50:50 beamsplitter, with output modes
\begin{equation}
    \hat{a}_\pm = \frac{\hat{a}_1 \pm \hat{a}_2}{\sqrt{2}},
    \qquad 
    \hat{n}_\pm = \hat{a}_\pm^\dagger \hat{a}_\pm .
\end{equation}
A straightforward expansion shows
\begin{equation}
    \hat{n}_+ - \hat{n}_-
    = \hat{a}_1^\dagger \hat{a}_2 
    + \hat{a}_2^\dagger \hat{a}_1 =\hat{q}_1 \hat{q}_2 + \hat{p}_1 \hat{p}_2
\end{equation}
Thus, the photon-number difference on the outputs of the beamsplitter measures the quadrature-insensitive interference between the modes---probing correlations that do not depend on the optical phase of the fields.

In terms of the standard first-order coherence function $G^{(1)}_{12} = \langle \hat{a}_1^\dagger \hat{a}_2 \rangle$, the photon-difference observable
satisfies $\expval*{\hat{n}_+ - \hat{n}_-}=2\Re[G^{(1)}_{12}]$. This emphasizes in another language that excitation measurements of this type are sensitive only to \emph{phase-insensitive} interference between the modes. A Mach-Zehnder interferometer with a tunable phase can measure both the real and imaginary parts of $G^{(1)}_{12}$. Generally though, \emph{phase-sensitive} correlations associated with $\expval{\hat{a}_1 \hat{a}_2}$ requires more sophisticated measurement settings.\footnote{%
A passive (Mach-Zehnder) interferometer implements a number-conserving $\rm{SU}(2)$ transformation, which only mixes normal coherences such as $\expval*{\hat{a}_1^\dagger \hat{a}_2}$. In contrast, an $\rm{SU}(1,1)$ interferometer replaces the beamsplitters with parametric amplifiers (two-mode squeezers). This allows phase-sensitive correlations to be converted into measurable intensity signals, in principle.
}

\clearpage

\begin{nutshell}{}
    \begin{itemize}
        \item The mean and covariance matrix of a quantum state $\rho$ are defined as $\bm\mu=\Tr(\hat{\bm r}\rho)$ and $\bm\sigma_{ij}=\Tr(\acomm{\hat{\bm r}_i-\bm\mu_i}{\hat{\bm r}_j-\bm\mu_j}\rho)$.

        \item The covariance matrix $\bm\sigma$ of any quantum state $\rho$ must satisfy the Robertson-Schr\"odinger uncertainty relation,
        \begin{equation*}
            \bm\sigma +i\bm\Omega\geq0.
        \end{equation*}
        
        \item Given a quantum state $\rho$ with covariance matrix $\bm\sigma$, there exists a symplectic transformation $\bm S$ such that $\bm S\bm\sigma\bm S^\top=\bigoplus_j\nu_j\bm I_2$, where $\nu_j\geq1$ are the symplectic eigenvalues of $\rho$.

        \item A general Gaussian channel $\mathcal{E}_G$ can be described by a displacement $\bm d$, a scaling matrix $\bm X$, and a noise matrix $\bm Y$. Given $\rho^{\rm out}=\mathcal{E}_G(\rho^{\rm in})$, the input-output relation for the first and second moments is
        \begin{equation*}
            \bm\mu^{\rm out}=\bm X\bm\mu^{\rm in} + \bm d \qq{and} \bm\sigma^{\rm out}=\bm X\bm\sigma^{\rm in}\bm X + \bm Y.
        \end{equation*}
        If $\mathcal{E}_G$ is unitary, then $\bm Y=0$ and $\bm X$ is a symplectic transformation.

        \item The Wigner function of a Gaussian state $\rho_G$ is a multi-variate Gaussian distribution,
        \begin{equation*}
            W_G(\bm x)=\frac{1}{\pi^N\sqrt{\det\bm\sigma_G}}\exp\left[-\left(\bm x-\bm\mu_G)^\top\bm\sigma_G^{-1}(\bm x-\bm\mu_G\right)\right],
        \end{equation*}
        where $\bm\mu_G$ and $\bm\sigma_G$ are, respectively, the first and second moments of $\rho_G$ (which are sufficient to specify the state $\rho_G$).
        
        \item A Gaussian measurement of a quantum state $\rho$ is a projection onto Gaussian state $\ket{\psi_G}=D_{\bm\mu} U_{\bm S}\ket{0}$ with probability $\expval{\rho}{\psi_G}/(2\pi)^N$, which follows from the completeness relation
        \begin{equation*}
            \frac{1}{(2\pi)^N}\int_{\bm\mu\in\mathbb{R}^{2N}}\dd{\bm\mu}D_{\bm\mu}U_{\bm S}\dyad{0}U_{\bm S}^\dagger D_{\bm\mu}^\dagger=\hat{I}.
        \end{equation*}

         \item Homodyne measurements are projections onto position (momentum) eigenstates. Practically, homodyne measurements of a quantum state $\rho$ can be achieved by mixing the system with a classical field (a.k.a., local oscillator) at a beamsplitter and measuring intensities of the outputs via photodetection.

         \item Photon counting (viz., $E_n=\dyad{n}$) is a non-Gaussian measurement that resolves the excitation number. In practice, full number-resolution is challenging, and a common alternative is \emph{click detection}, which discriminates between absence and presence of excitations (viz., $E_0=\dyad{0}$ and $E_1=I-E_0$). Click detection is powerful in the low-photon regime, where it provides a robust and experimentally available means to access signals buried beneath vacuum noise.

        
    \end{itemize}
   
\end{nutshell}

\clearpage

\part{Entangled}\label{lecture:3}

In Part~\ref{lecture:2} we developed the moment-based description of Gaussian states and their evolution under Gaussian transformations. Here we examine how quantum correlations are structured and quantified in multimode Gaussian states. We begin with the entropy of Gaussian states, which ultimately equates to a thermal-type entropy, illustrating the principle of \emph{Gaussian extremality}: At fixed second moments, Gaussian states maximize entropy (Jaynes' Principle). We then briefly discuss a modewise entanglement decomposition that reveals how correlations in general multimode Gaussian states can be concentrated into pairs of two-mode–squeezed vacua. Since entropy alone cannot distinguish classical from quantum correlations in mixed states, we turn to separability criteria based on the positivity of partial transpose (PPT) and the associated logarithmic negativity---a readily computable entanglement measure for Gaussian states. We quantitatively highlight how decoherence, such as initial thermal fluctuations, can swamp quantum correlations. Finally, we discuss experimentally friendly separability tests derived from the PPT condition, which take the form of Cauchy–Schwarz inequalities applied to covariance data. These tools position us to analyze quantum correlations in QFCS and analogue-gravity systems.

\vspace{4em}
\begin{figure}[h]
    \centering
    \includegraphics[width=.8\linewidth]{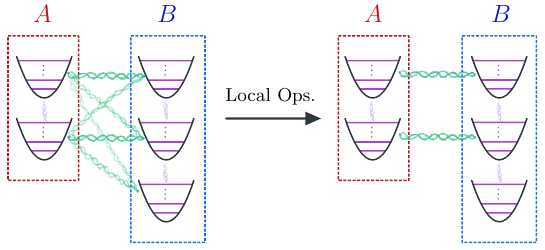}
    \caption{Modewise entanglement structure. A system of $N_A+N_B$ oscillators is partitioned into subsystems $A$ and $B$; here $N_A=2$ and $N_B=3$; here $N_A=2$ and $N_B=3$. Quantum correlations between $A$ and $B$ are genuinely multimode, but for special classes of Gaussian states (e.g., pure states and isotropically noisy states), local operations concentrate correlations into (noisy) two-mode-squeezed vacua (plus uncorrelated modes for $N_B>N_A$).}
    \label{fig:entangled}
\end{figure}

\clearpage


\section{Preliminary Remarks}

A defining feature of quantum mechanics is the existence of nonlocal correlations (\emph{entanglement}) between subsystems. Entanglement serves as a key resource for quantum communication, metrology, and computation. Yet quantifying entanglement, and distinguishing it from classical correlations, is subtle and often computationally challenging. It is therefore essential to establish a clear framework for characterizing entanglement in CV setting relevant to these notes.

Operationally, entanglement is defined in contrast to \emph{separability}. A
bipartite state $\rho_{AB}$ is separable if it can be expressed as a convex
mixture of product states,
\begin{equation}
    \rho_{AB}^{\rm{sep}}
    = \sum_x p_{AB}(x)\,
    \rho_A^{(x)} \!\otimes \rho_B^{(x)},
    \label{eq:separable_state}
\end{equation}
where $p_{AB}(x)$ denotes a classical probability distribution and $\rho_A^{(x)}$ and $\rho_B^{(x)}$ denote local density operators. This state exhibits only classical correlations: That is, all statistical dependence between $A$ and $B$ can be attributed to shared randomness in $p_{AB}(x)$. A state is \emph{entangled} iff it cannot be written in the form \eqref{eq:separable_state}.

A familiar example is the Bell state
\begin{equation}
    \ket{\Phi^+}_{AB}
    = \frac{1}{\sqrt{2}}\big(
        \ket{\mathsf{0},\mathsf{0}}+\ket{\mathsf{1},\mathsf{1}}
      \big),
    \label{eq:bell_state}
\end{equation}
representing maximal correlations between two qubits---for instance, a pair of single photons encoded in polarization. Measuring photon $A$ to be horizontally polarized ($\mathsf{0}$) guarantees the same outcome for photon $B$, and likewise for vertical polarization ($\mathsf{1}$).

The CV analogue is provided by the two-mode squeezed vacuum [Eq.~\eqref{eq:tmsv_fock}]. In the limit of large squeezing $(r\to\infty)$, the two-mode squeezed vacuum converges to the famous Einstein-Podolsky-Rosen (EPR) state~\cite{EPR1935}.


\section{Entropy, Extremality, and Modewise Entanglement}

\subsection*{Gaussian Entropy}
For a pure bipartite state $\ket{\Psi}_{AB}$, the von Neumann entropy across the $A,B$ partition is a faithful measure of entanglement, quantified in units of
\emph{entangled bits} (ebits):
\begin{equation}
    S(\rho_A)=-\Tr(\rho_A\log\rho_A),
\end{equation}
with $\log$ taken base 2. By the Schmidt decomposition,
$S(\rho_A) = S(\rho_B)$. As example, the Bell state corresponds to precisely 1 ebit by definition. 

In general, computing the entropy of a mixed state requires diagonalizing its reduced density operator, which is computationally taxing. For Gaussian states, however, the entropy admits a closed analytic form. Using the Gibbs representation of a Gaussian state
[Eq.~\eqref{eq:rhoG}] and the fact that entropy is invariant under unitary
transformations, we find:
\begin{align}
    S(\rho_G)&=S\left(D_{\bm\mu} U_{\bm S}\left(\bigotimes_{j=1}^N\rho_{{\rm th} ,j}\right)U_{\bm S}^\dagger D_{\bm\mu}^\dagger\right)\\
    &=S\left(\bigotimes_{j=1}^N\rho_{{\rm th} ,j}\right)\\
    &=\sum_{j=1}^N S(\rho_{{\rm th} ,j}).
\end{align}
where additivity, $S(\rho_A\otimes\rho_B)=S(\rho_A)+S(\rho_B)$,\footnote{More generally, $S(\rho_{AB}) \le S(\rho_A) + S(\rho_B)$, with equality for product states.} implies that the entropy of a Gaussian state is simply the sum of the individual entropies of the $N$ thermal modes.

A single-mode thermal state $\rho_{{\rm th},j}$ with mean occupation $\bar{n}_j$ has entropy
\begin{equation}
    s_{\rm th}(x)
    = (x+1)\log(x+1) - x\log x,
\end{equation}
where $s_{\rm th}$ denotes the thermal entropy function. This relation follows from the Fock expansion of the thermal state [Eq.~\eqref{eq:thermal_fock}] or, more physically, from the thermodynamics of a quantum harmonic oscillator.

Using the relation $\bar{n}_j = \nu_j - \tfrac{1}{2}$, where $\nu_j$ are the symplectic eigenvalues of the state's covariance matrix, we arrive at the general expression
\begin{equation}\label{eq:gaussian_entropy}
S(\rho_G)=\sum_{j=1}^N s_{\rm th}\left(\nu_j-\frac{1}{2}\right).
\end{equation}
Remarkably, the entropy depends only on the symplectic eigenvalues---that is, only on the \emph{second moments} of the state---and is completely insensitive to its first moments.

\begin{example}{Blackbody entropy flux}
    Consider a stationary photon flux from a blackbody at temperature $T$ impinging on a photodetector. We model the field arriving at the detector a one-dimensional Markovian continuum with annihilation operators $\hat{a}(t)$ satisfying
    \begin{equation}
        [\hat{a}(t),\hat{a}^\dagger(t')] = \delta(t-t'),
    \end{equation}
    so that $\hat{a}(t)$ has units of $\sqrt{\rm{photons/s}}$. Moving to
    frequency space,
    \begin{equation}
        \hat{a}(t)
        = \int \frac{\dd\omega}{2\pi}\,e^{i\omega t}\hat{a}(\omega),
        \qquad
        [\hat{a}(\omega),\hat{a}^\dagger(\omega')] = 2\pi\delta(\omega-\omega'),
    \end{equation}
    and $\hat{a}(\omega)$ has units of $\sqrt{\rm{photons/Hz}}$. The photon-number spectral density $n(\omega)$ is defined by
    \begin{equation}
        \langle \hat{a}^\dagger(\omega)\hat{a}(\omega') \rangle
        = 2\pi\,n(\omega)\,\delta(\omega-\omega'),
    \end{equation}
    so that $n(\omega)\,\dd\omega$ gives the average number of photons per second in the frequency interval $\dd\omega$.

    For thermal radiation,
    \begin{equation}
        n_{\rm th}(\omega;T)
        = \frac{1}{e^{\hbar\omega/k_{\rm{B}}T}-1},
    \end{equation}
    Since each frequency mode is a
    single-mode thermal Gaussian state with entropy
    $s_{\rm th}(x)$ [Eq.~\eqref{eq:gaussian_entropy}], the entropy flux carried
    by the thermal photon field is
    \begin{equation}
        \dot{S}(T)
        = \int_0^\infty \frac{\dd\omega}{2\pi}\,
        s_{\rm th}\!\big(n_{\rm th}(\omega;T)\big),
    \end{equation}
    In the high-frequency (Wien) regime, $n_{\rm th}(\omega)\ll 1$ and the entropy contribution is negligible; in the low-frequency (Rayleigh--Jeans) regime, $n_{\rm th}(\omega)\gg 1$ and the entropy flux grows. In this one-dimensional model, the integral diverges at $\omega\to 0$ because all frequencies are assumed equally available to the detector. Real radiation fields always impose a mode structure or filtering---such as the $\omega^2$ electromagnetic density of states in free space or graybody transmission factors in black-hole evaporation---that suppress low-frequency modes and render the total entropy flux finite.
\end{example}

\subsection*{Gaussian extremality}
Gaussian states are special for many reasons. A particularly important one is
that they \emph{extremize} many information-theoretic quantities (under constraints). In particular~\cite{wolf2006extremality}, given any state $\rho$
and a Gaussian state $\rho_G$ with the same first and second moments,
\begin{equation}
    S(\rho) \le S(\rho_G).
    \label{eq:gauss_extreme}
\end{equation}
In other words, Gaussian states \emph{maximize} entropy under fixed covariance. This property connects subtly to Jaynes’ maximum-entropy principle~\cite{Jaynes1957}: thermal (Gibbs) states maximize entropy for a fixed mean energy.

We make a few observations. Displacements do not affect entropy, so first
moments are irrelevant. Using symplectic diagonalization,
$S(\rho) = S(U_{\bm S}\rho U_{\bm S}^\dagger)$ with
$\bm S\bm\sigma\bm S^\top = \bigoplus_j \nu_j \bm I_2$, where $\nu_j$ are the
symplectic eigenvalues of $\rho$ (and of $\rho_G$ by assumption). For Gaussian
states,
\begin{equation*}
    S(\rho_G)
    = \sum_{j=1}^N s_{\rm th}\!\left(\nu_j - \tfrac12\right),
\end{equation*}
as established in Eq.~\eqref{eq:gaussian_entropy}. Thus
Eq.~\eqref{eq:gauss_extreme} becomes
\begin{equation}
    S(\rho)
    \le \sum_{j=1}^N s_{\rm th}\!\left(\nu_j - \tfrac12\right),
\end{equation}
highlighting that the extremality statement can be framed entirely in terms of
the fixed symplectic eigenvalues $\nu_j$, or equivalently, the corresponding
thermal occupations $\bar{n}_j = \nu_j - \tfrac12$.

To illustrate this connection with Jaynes’ principle, we provide a variational
derivation for a single mode (the multimode case follows by imposing $N$
independent constraints). Among all states with fixed mean photon number
$\bar{n}$, those diagonal in the Fock basis maximize entropy; thus we consider
$\rho = \sum_n q_n \dyad{n}$. Introducing the Lagrangian
\begin{equation}
    \mathscr{L}
    = -\sum_n q_n \log q_n
    - \lambda_1\!\left(\sum_n q_n - 1\right)
    - \lambda_2\!\left(\sum_n n q_n - \bar{n}\right)
\end{equation}
and extremize with respect to $q_n$ to obtain
\begin{equation}
    \partial_{q_n}\mathscr{L} = 0
    \;\;\Rightarrow\;\;
    q_n \propto e^{-\lambda_2 n}.
\end{equation}
With $\bar{n} = 1/(e^{\lambda_2}-1)$, this is precisely the thermal state
$\rho_{\rm th}$ of Eq.~\eqref{eq:thermal_fock}. Therefore Gaussian states maximize entropy at fixed second moments, establishing Gaussian extremality.

\subsection*{Modewise entanglement}

A striking feature of multimode Gaussian states is that their entanglement is fundamentally \emph{pairwise}. For pure Gaussian states, all correlations between subsystems $A$ and $B$ can be converted, by local symplectic operations, into a tensor product of TMSV states. That is, pure Gaussian entanglement admits a perfect mode pairing. This generalizes to certain mixed states with \emph{isotropic noise}, where each mode experiences the same thermal background (cf., Ref.~\cite{reznik2003modewise} and Section 7.3 of Ref.~\cite{serafini2017book} for more details). See Fig.~\ref{fig:entangled} for an illustration.

To state this formally, consider a bipartition $AB$ of $K=N+M$ modes. Let $\bm{Y}_{AB}=y\,\bm{S}\bm{S}^\top > 0$ be a positive matrix describing the second moments of the system (e.g., its covariance matrix), where $y>0$ and $\bm{S}\in\rm{Sp}(2K,\mathbb{R})$. Then:

\begin{theorem}[Modewise Entanglement~{\cite{reznik2003modewise}}]
There exist local symplectic transformations $\bm{S}_A\in\rm{Sp}(2N,\mathbb{R})$ and $\bm{S}_B\in\rm{Sp}(2M,\mathbb{R})$ such that
\begin{equation}
\label{eq:modewise_entanglement}
\big(\bm{S}_A \oplus \bm{S}_B\big)\,
\bm{Y}_{AB}\,
\big(\bm{S}_A^\top \oplus \bm{S}_B^\top\big)
=
y \left(
\bigoplus_{i=1}^{N}
\bm{S}_{G_i}\bm{S}_{G_i}^\top
\right)
\oplus \bm{I}_{2(M-N)},
\end{equation}
where each $\bm{S}_{G_i}$ is a two-mode–squeezing transformation between modes $A_i$ and $B_i$ [see Eq.~\eqref{eq:tms_transform}], and $\bm{I}_{2(M-N)}$ acts on uncorrelated modes.
\end{theorem}

As a concrete example, let $\rho_{AB}$ be a Gaussian state with covariance matrix $\bm{Y}_{AB} = (1+2\bar{n})\,\bm{S}\bm{S}^\top$, corresponding to isotropic noise $\bar{n}$ on all modes. By locally applying $\bm{S}_A$ and $\bm{S}_B$, the state can be converted into $N$ noisy TMSV states (one across each mode pair) plus $M-N$ uncorrelated thermal states. This structure is the continuous-variable analogue of the Schmidt decomposition for finite-dimensional systems.

Although we phrased this as a statement about states, the result really applies to the underlying symplectic structure: That is, it is a property of the matrix $\bm{S}\bm{S}^\top$, which fundamentally describes how correlations are distributed in phase space.

\section{PPT Criterion and the Logarithmic Negativity}

We introduce a well-known and oft-used separability test: the \emph{positivity of partial transpose (PPT) criterion}, also known as the Peres-Horodecki criterion~\cite{peres96,simon2000criterion}. It provides a powerful diagnostic of entanglement and serves as the foundation for a computable entanglement measure, the \textit{logarithmic negativity}.

A valid quantum state must satisfy:
\begin{itemize}
    \item[(\textit{i})] $\Tr\rho=1 \quad$ (unit trace)
    \item[(\textit{ii})] $\rho^\dagger = \rho \quad$ (self-adjoint)
    \item[(\textit{iii})] $\ev{\rho}{\psi} \geq 0 \ \  \forall \ \psi \quad$ (non-negativity)
\end{itemize}
In particular, the non-negativity condition means that the eigenvalues of $\rho$ are positivity. The PPT criterion exploits non-negativity in the following way: If a quantum state $\rho_{AB}$ is separable, then the partially transposed state, $\rho_{AB}^{\top_B}$, is non-negative.\footnote{The transposition could just as well be done on the $A$ side.} This condition is \emph{necessary} for separability, which is easy to show. 

Consider a separable quantum state $\rho_{AB}$ which can be written as in Eq.~\eqref{eq:separable_state}. Then, partial transposition with respect to the system $B$ results in
\begin{equation}
    \widetilde{\rho}_{AB}\triangleq\rho_{AB}^{\top_B}=\sum_x p_{AB}(x) \rho^{(x)}_A\otimes\left(\rho^{(x)}_B\right)^\top,\label{eq:PT_separable}
\end{equation}
Observe that $(\rho^{(x)}_B)^\top$ is a bona fide quantum state that satisfies properties (i-iii) above. Thus, the partially transposed state, $\widetilde{\rho}_{AB}$, is non-negative. This shows that the PPT criterion is necessary for separability, and as a consequence, violation of the PPT criterion is sufficient for entanglement. However, this condition is only necessary \emph{and} sufficient for separability in particular cases. Remarkably, for an $(N+1)$-mode Gaussian bosonic system, violation of the PPT criterion is both necessary and sufficient for any 1 versus $N$ mode bipartition.  

However, the PPT criterion only tells us whether a state is entangled; it does not quantify how much. it does now tell us how entangled a state may be. This motivates the \emph{logarithmic negativity}, an entanglement measure that is efficiently computable and operationally meaningful~\cite{vidal02,plenio05}.

For a bipartite state $\rho_{AB}$,
\begin{equation}
    E_N(\rho_{AB})
    \triangleq \log_2 \| \rho_{AB}^{\top_B} \|_1,
    \label{eq-log_neg}
\end{equation}
where $\|A\|_1 = \Tr\{\sqrt{A^\dagger A}\}$ is the trace norm. Since $\Tr\{\rho_{AB}^{\top_B}\} = 1$, positivity of $\rho_{AB}^{\top_B}$ implies that all eigenvalues are non-negative and thus $\norm{\rho_{AB}^{\top_B}}_1 = 1$, giving $E_N = 0$. If \(\rho_{AB}^{\top_B}\) has negative eigenvalues (PPT violation), then the trace norm must exceed 1, yielding $E_N > 0$.

Let $\rho_{AB}$ be an $(N+M)$-mode Gaussian state with covariance matrix
$\bm{\sigma}_{AB}$. Partial transposition corresponds to conjugation by
\begin{equation}
    \widetilde{\bm\sigma}_{AB}
    = \bm{T} \bm{\sigma}_{AB} \bm{T}, \qq{where} \bm{T} = \bm{I}_{2N} \oplus
    \Big(\!\bigoplus_{j=1}^M \bm{Z}\Big)
\end{equation}
and $\bm Z = \rm{diag}(1,-1)$ flips the sign of each momentum quadrature in subsystem $B$. The covariance matrix of $\rho_{AB}^{\top_B}$ is
\begin{equation}
    \widetilde{\bm\sigma}_{AB}
    = \bm{T} \bm{\sigma}_{AB} \bm{T}.
    \label{PT}
\end{equation}
Let $\{\tilde{\nu}_j\}$ be the symplectic eigenvalues of $\widetilde{\bm\sigma}_{AB}$. Then~\cite{serafini2017book}
\begin{equation}
    E_N(\rho_{AB})
    = \sum_{j=1}^{N+M}
      \max\!\bigl[ 0,\ -\log_2(\tilde{\nu}_j) \bigr].
\end{equation}
Thus a sufficient entanglement condition is
\[
\min_j \tilde{\nu}_j < 1.
\]
We refer the reader to Ch. 7.1 of Ref.~\cite{serafini2017book} for further details.

\begin{figure}
    \centering
    \includegraphics[width=.55\linewidth]{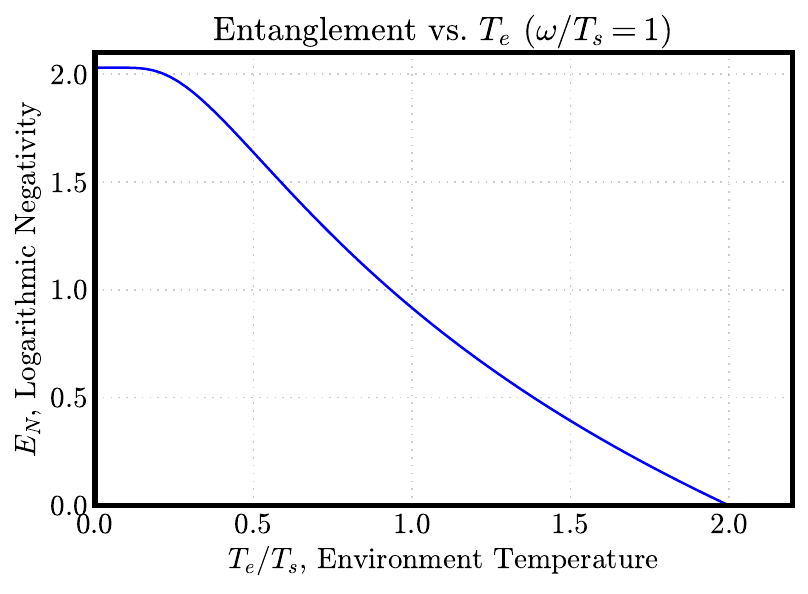}
    \caption{Log-negativity versus temperature $T_e$ for a thermally seeded TMSV state with $\bar{n}_s=1/[\exp(\omega/T_s)-1]\approx.58$ squeezed quanta ($\omega/T_s=1$) and $\bar{n}_e=1/[\exp(\omega/T_e)-1]$ noisy thermal quanta. Entanglement vanishes at $T_e=2T_s$ for any fixed frequency $\omega$.}
    \label{fig:logneg_noisyTMSV}
\end{figure}

\begin{example}{Log-Negativity of Two-Mode S;queezed Vacuum}
    We compute the log-negativity of a TMSV state
    $\ket{\rm{TMSV}}$ with squeezing parameter $r$. Using $\sinh^2 r \triangleq \bar{n}_s$ as the mean photon number per mode, its
    covariance matrix is
    \begin{equation}
        \bm\sigma_{\rm{TMSV}}
        =
        \begin{pmatrix}
            (1+2\bar{n}_s)\bm I_2
            &
            2\sqrt{\bar{n}_s(\bar{n}_s+1)}\,\bm Z
            \\
            2\sqrt{\bar{n}_s(\bar{n}_s+1)}\,\bm Z
            &
            (1+2\bar{n}_s)\bm I_2
        \end{pmatrix}.
        \label{eq:sigma_tmsv_r}
    \end{equation}
    Partial transposition flips the sign of $\hat{p}$ on one mode, turning $\bm Z \to \bm I_2$ in the off-diagonal blocks:
    \begin{equation}
        \widetilde{\bm\sigma}_{\rm{TMSV}}
        =
        \begin{pmatrix}
            (1+2\bar{n}_s)\bm I_2
            &
            2\sqrt{\bar{n}_s(\bar{n}_s+1)}\,\bm I_2 \\
            2\sqrt{\bar{n}_s(\bar{n}_s+1)}\,\bm I_2
            &
            (1+2\bar{n}_s)\bm I_2
        \end{pmatrix}.
        \label{eq:sigma_tmsv_PT}
    \end{equation}
    This matrix has two symplectic eigenvalues: one is $\tilde{\nu}_+=1$ (no contribution), while the nontrivial one is $\tilde{\nu}_-=e^{-2r}$. Thus,
    \begin{equation}
        E_N(\ket{\rm{TMSV}})
        =
        \max\!\left[0, -\log_2 \tilde{\nu}_-\right]
        =
        \frac{2r}{\ln 2}.
    \end{equation}
    The log-negativity therefore grows linearly with the squeezing strength $r$.
\end{example}

We now investigate how entanglement degrades in the presence of thermal noise, which is relevant to physical scenarios such as Hawking radiation immersed in the cosmic microwave background or analogue-gravity systems dissipatively coupled to a thermal environment.

\begin{example}{Thermally Seeded TMSV}
    Let a TMSV state with $\bar{n}_s=\sinh^2 r$ be subjected to isotropic thermal
    noise corresponding to $\bar{n}_e$ excitations per mode.
    The resulting covariance matrix is a simple rescaling:
    \begin{equation}
        \bm\sigma = (1+2\bar{n}_e)\,\bm\sigma_{\rm{TMSV}}.
    \end{equation}
    Since symplectic eigenvalues scale identically, the non-trivial PT symplectic eigenvalue becomes
    \begin{equation}
        \tilde{\nu}_- = (1+2\bar{n}_e)\,e^{-2r}.
    \end{equation}
    A necessary and sufficient condition for Gaussian entanglement is
    $\tilde{\nu}_- < 1$, implying
    \begin{equation}
        e^{2r} > 1 + 2\bar{n}_e
        \;\;\Longleftrightarrow\;\;
        \bar{n}_s + \sqrt{\bar{n}_s(\bar{n}_s + 1)}
        > \bar{n}_e.
        \label{eq:noisyTMSV-ent}
    \end{equation}

    Let us parametrize the squeezing and noise populations via Bose-Einstein distributions,
    $\bar{n}_{i} = 1/(e^{\omega/T_i}-1)$ with $i\in\{s,e\}$. In that case, Eq.~\eqref{eq:noisyTMSV-ent} reduces to a simple thermality condition:
    \begin{equation}
        T_e < 2T_s.
    \end{equation}
    Thus, entanglement survives only if the environment temperature is at most twice the effective squeezing temperature. This threshold behavior is shown in Fig.~\ref{fig:logneg_noisyTMSV} and is relevant to systems where correlated modes interact locally and dissipatively with a thermal bath, while being driven by a two-mode-squeezing process (e.g., analogue Hawking process).
\end{example}


\section{Separability via Cauchy-Schwarz Witnesses}

Full entanglement quantification generally requires complete state tomography, which is impractical for many platforms. Even in the Gaussian regime, reconstructing all $2N^2 + N$ covariance parameters (plus $N$ first moments) quickly becomes costly: for $N=2$, one already needs 12 independent observables without assuming further structure. It is therefore valuable to have operational \emph{entanglement witnesses} that certify entanglement from only a small set of experimentally accessible moments.

Here we restrict attention to the Gaussian regime, where second moments contain all relevant entanglement information. We focus on witnesses derived from the PPT condition for two-mode states, which can be tested via quadratic moments; see, e.g., Ref.~\cite{simon2000criterion} and applications to analogue gravity in Refs.~\cite{busch14,steinhauer2015entanglement}. For higher-order moment criteria in non-Gaussian settings, see Ref.~\cite{vogel2005inseparability} and Sec.~7.6 of Ref.~\cite{serafini2017book}.\footnote{We also mention the celebrated Duan criterion~\cite{duan2000entCriterion}, which is necessary but not sufficient for Gaussian two-mode entanglement and is therefore weaker than the PPT condition.}

\subsection*{PPT-based condition}
Let the covariance matrix of a two-mode Gaussian state $\rho_{AB}$ be
\begin{equation}
    \bm\sigma_{AB}
    =
    \begin{pmatrix}
        \bm A & \bm C \\
        \bm C^\top & \bm B
    \end{pmatrix},
\end{equation}
with $\bm A$ and $\bm B$ describing the local second moments of subsystems $A$ and $B$, while $\bm C$ captures correlations. Simon~\cite{simon2000criterion} derived a necessary and sufficient condition for separability of Gaussian two-mode states:
\begin{multline}
    \mathcal{P}_-
    \triangleq
    \det\bm A\,\det\bm B
    + \left(1-\abs{\det\bm C}\right)^2
    - \Tr(\bm A\bm\Omega_1 \bm C\bm\Omega_1 \bm B\bm\Omega_1 \bm C^\top\bm\Omega_1)
    \\
    - \det\bm A - \det\bm B
    \;\ge 0.
    \label{eq:pminus}
\end{multline}
This follows from the Robertson-Schr\"odinger uncertainty relation [Eq.~\eqref{eq:robertson_uncertainty}] applied to the partially transposed covariance matrix, $\widetilde{\bm\sigma}_{AB}+i\bm\Omega\geq 0 \iff \mathcal{P}_-\ge 0$. Thus, $\mathcal{P}_-<0$ unequivocally signals entanglement for any two-mode Gaussian state.

Despite its strength, using $\mathcal{P}_-$ experimentally is nearly as difficult as full tomography, as it depends on every covariance entry. Moreover, $\mathcal{P}_-$ is not a quantitative measure: its negativity certifies entanglement but does not track ``how much’’ of it is present.

\subsection*{Cauchy-Schwarz-type witness}
A simpler witness was obtained in Ref.~\cite{busch14} by relaxing the PPT condition:
\begin{equation}
    \Delta
    \triangleq
    \expval*{\hat{a}^\dagger\hat{a}}\,
    \expval*{\hat{b}^\dagger\hat{b}}
    -
    \abs*{\expval*{\hat{a}\hat{b}}}^2
    \ge 0,
    \label{eq:cs_inequality}
\end{equation}
which is a Cauchy–Schwarz constraint on normally ordered correlations. A violation $\Delta < 0$ implies entanglement. For certain physically relevant Gaussian states, e.g., noisy TMSV states satisfying
$\expval*{\hat{a}\hat{a}} = \expval*{\hat{b}\hat{b}} = \expval*{\hat{a}^\dagger\hat{b}} = 0$, this witness is \emph{equivalent} to $\mathcal{P}_-<0$, and thus to the PPT violation.

\begin{figure}[h]
    \centering
    \includegraphics[width=.55\linewidth]{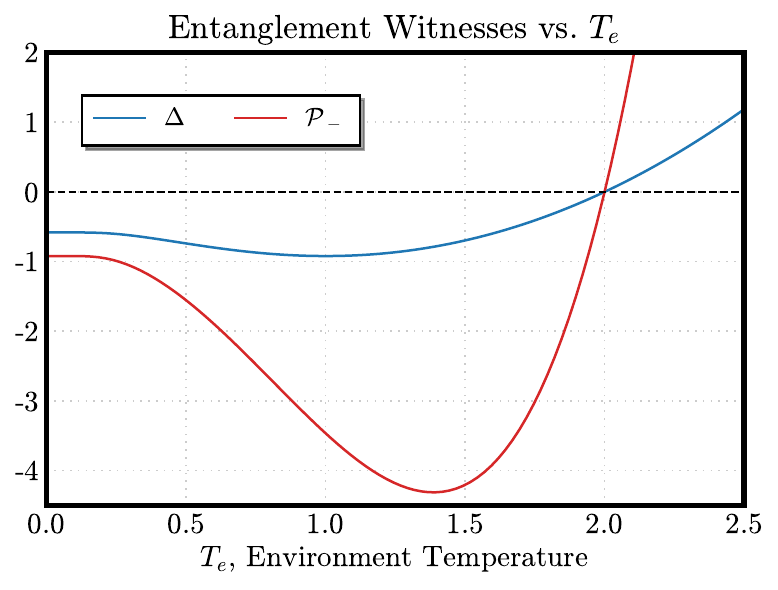}
    \caption{
        Comparison of PPT witness $\mathcal{P}_-$ and the Cauchy–Schwarz witness $\Delta$
        for a noisy TMSV state with fixed squeezing ($\omega/T_s = 1$) versus environmental
        temperature $T_e$.
    }
    \label{fig:witness_compare}
\end{figure}

\subsection*{Limitations and interpretation}
Witnesses like $\Delta$ are experimentally attractive but must be used with care:
\begin{itemize}
\itemsep0.25em
    \item $\Delta<0$ signals entanglement, but a more negative value does \emph{not} indicate stronger entanglement.
    \item Even in Gaussian two-mode systems, $\Delta\ge 0$ does not guarantee separability unless additional structure is known.
    \item Witnesses do not track the degradation of entanglement under noise, loss, or thermal fluctuations. So though experimentally attractive, its theoretical limitation must be known.
\end{itemize}

These caveats are visible in Fig.~\ref{fig:witness_compare}, where one observes that both $\Delta$ and $\mathcal{P}_-$ can become \emph{more} negative at higher noise, even as the log-negativity (Fig.~\ref{fig:logneg_noisyTMSV}) decreases monotonically. If interpreted naively, $\Delta$ suggests that increasing thermal background enhances entanglemen---contradicting physical intuition and the PPT criterion. Nevertheless, these witnesses remain indispensable for entanglement verification in resource-constrained experiments, including currently outstanding attempts to detect Hawking-pair entanglement in analogue gravity settings.

\clearpage

\begin{nutshell}{}
\begin{itemize}
    \item For pure bipartite states, the von Neumann entropy of a subsystem quantifies entanglement.

    \item The entropy of a Gaussian state is fully determined by its symplectic eigenvalues:
    \begin{equation*}
        S(\rho_G)
        = \sum_{j=1}^N
        s_{\rm th}\!\left(\nu_j - \frac{1}{2}\right),
        \qquad
        s_{\rm th}(x)
        = (x+1)\log(x+1)-x\log x,
    \end{equation*}
    with $s_{\rm th}(x)$ the thermal entropy function.

    \item Gaussian states \emph{maximize} entropy at fixed first and second moments:
    \begin{equation*}
        S(\rho)\le S(\rho_G),
    \end{equation*}
    a manifestation of Jaynes’ maximum-entropy principle.

    \item Pure (and isotropically noisy) multimode Gaussian states admit a modewise decomposition into $N$ (noisy) two-mode squeezed vacuum pairs plus uncorrelated modes---i.e., Gaussian correlations are fundamentally pairwise.

    \item For bipartite states, violation of positivity under partial transpose (PPT) is a sufficient condition for entanglement. For two-mode Gaussian states, it is also necessary.

    \item The \emph{logarithmic negativity},
    \begin{equation*}
        E_N(\rho_{AB}) = \log \|\rho^{\top_B}_{AB}\|_1,
    \end{equation*}
    is an entanglement monotone and is efficiently computed for Gaussian states from the symplectic spectrum of the partially transposed covariance matrix.

    \item Entanglement witnesses based on Cauchy–Schwarz–type inequalities require only a limited set of observables. They can certify (or witness) entanglement without full tomography, but they do not quantify it.
\end{itemize}
\end{nutshell}

\clearpage

\part{Gaussian Physics on the Horizon}\label{lecture:4} 

\begin{figure}[h]
    \centering
    \includegraphics[width=.7\linewidth]{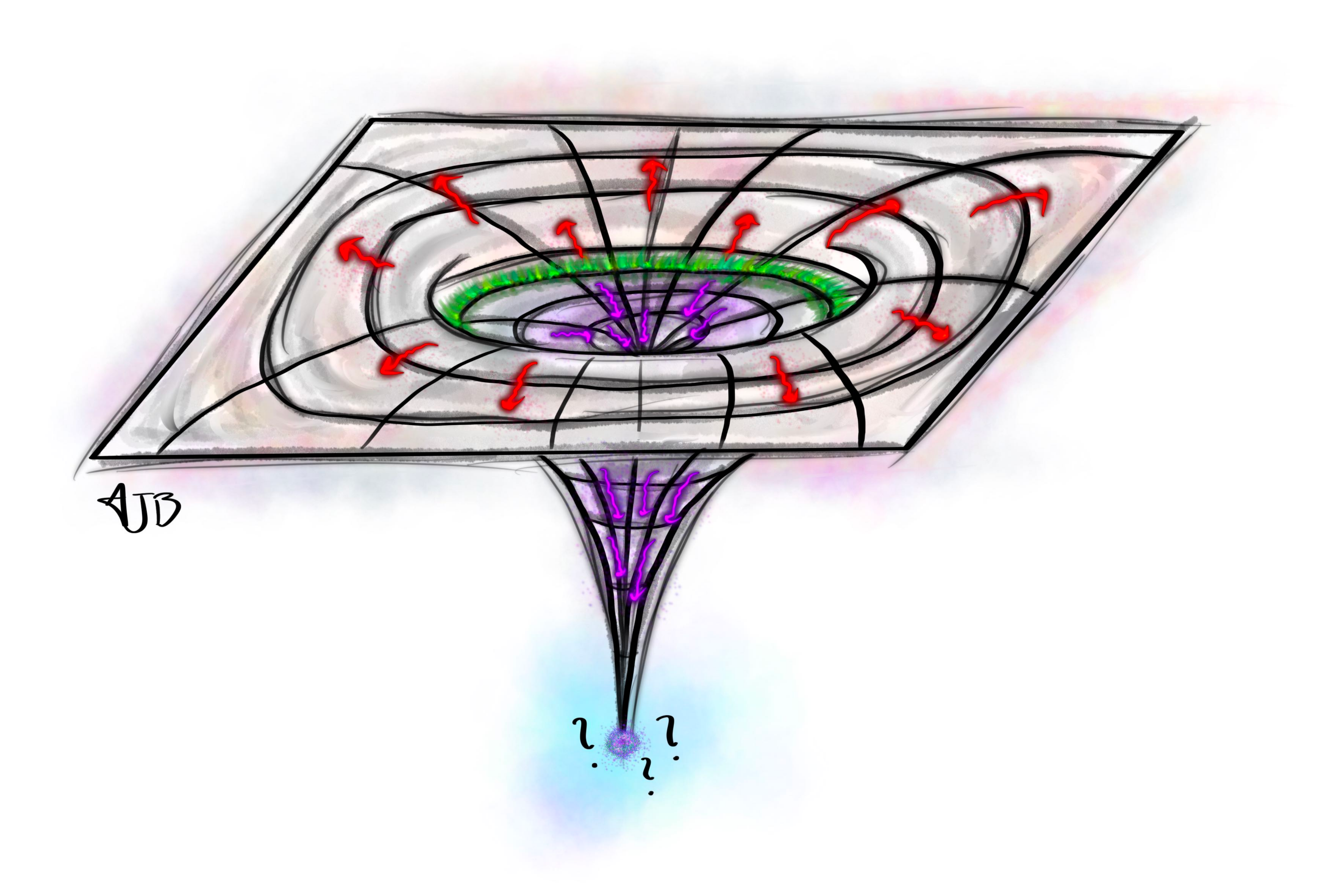}
    \caption{Black hole evaporation. Hawking radiation (red wiggles) escape the gravitational pull of the black hole, while entangled Hawking partners (purple wiggles) fall towards the singularity.}
    \label{fig:bh_evaporation}
\end{figure}

Semi-classical gravity treats spacetime as a classical, smooth geometry while quantizing the fields that propagate upon it. Within this framework, black holes are not perfectly black: quantum fluctuations near the event horizon generate thermal Hawking radiation---a process known as the \emph{Hawking effect}. Outgoing quanta are entangled with partner modes that fall behind the horizon [Fig.~\ref{fig:bh_evaporation}], linking black-hole physics to quantum information. Rotation introduces a closely related process, \emph{quantum superradiance}, in which certain incident waves are amplified through ergoregion-induced mode mixing.

Laboratory \emph{analogue-gravity} platforms---such as optical dielectric media with controlled refractive-index perturbations, or quantum fluids of light formed by exciton–polaritons in semiconductor microcavities---emulate black-hole and white-hole horizons, as well as ergoregions, and reproduce essential features of the Hawking effect and superradiance. Effective spacetime geometries emerge from the propagation of low-energy excitations, providing a controlled setting to study particle creation and observe spacetime-induced entanglement.

In Parts~(\ref{lecture:1}--\ref{lecture:3}) we developed the phase-space formalism for Gaussian bosonic systems. Here we apply those tools to semi-classical gravity and to analogue-gravity models. A central theme is that the essential semi-classical processes (Hawking effect and superradiance) are, at their core, \emph{Gaussian bosonic transformations}. In what follows, we introduce the QFCS origin of these effects and then discuss representative optical and polaritonic analogue models. Our aim is not to analyze these platforms in detail, but rather to illustrate how the universal features of semi-classical gravitational phenomena emerge and how the Gaussian phase-space language provides a clean way to describe them. We point to representative literature and include simple toy models to highlight the core ideas and how practical hurdles, such as thermal backgrounds, residual interactions (e.g., phonon–polariton scattering), and imperfect detection, negatively affect important quantum signatures.

\vspace{-1em}
\clearpage


\section{Digression: Mode Reduction for Quantum Fields}

Up to this point, we have worked with bosonic systems possessing a \emph{finite} number of modes. Quantum field theory (QFT), by contrast, contains an \emph{infinite} number of degrees of freedom: roughly speaking, one oscillator for each momentum mode (or spatial point). Before applying the phase-space tools of Parts~(\ref{lecture:1}--\ref{lecture:3}) to gravitational settings, we must understand how this infinity of modes plausibly reduces to finite, tractable subsystems.

\subsection*{Free fields: infinitely many harmonic oscillators}

Consider a free, massless scalar quantum field $\hat\varphi$ in Minkowski spacetime, satisfying the Klein--Gordon equation
\begin{equation}
  (\partial_t^2 - \nabla^2)\,\hat\varphi(\mathbf x,t)=0.
\end{equation}
Because the KG equation is linear, the field can be expanded in a complete basis of positive-frequency solutions. Using standard continuum normalization,\footnote{Throughout we use
$[\hat a_{\mathbf k},\hat a_{\mathbf p}^\dagger]=(2\pi)^3\delta^{(3)}(\mathbf
k-\mathbf p)$.}
\begin{equation}
  \hat\varphi(\mathbf x,t)
  = \int\!\frac{d^3k}{(2\pi)^3}\;
    \frac{1}{\sqrt{2\omega_{\mathbf k}}}
    \!\left[
      \hat a_{\mathbf k}\,e^{-i(\omega_{\mathbf k}t-\mathbf k\!\cdot\!\mathbf x)}
      + 
      \hat a_{\mathbf k}^\dagger\,e^{+i(\omega_{\mathbf k}t-\mathbf k\!\cdot\!\mathbf x)}
    \right],
\end{equation}
where $\omega_{\mathbf k}=|\mathbf k|$. The creation and annihilation operators define the usual Poincar\'e-invariant (Minkowski) vacuum, $\hat a_{\mathbf k}\ket{0}=0$, and thus generate a Fock basis. The corresponding Hamiltonian (after discarding the vacuum energy) is
\begin{equation}
  \hat H
  = \int\!d^3k\;\omega_{\mathbf k}\,
    \hat a_{\mathbf k}^\dagger\hat a_{\mathbf k},
\end{equation}
making explicit that the free field is an \emph{infinite} collection of decoupled harmonic oscillators labeled by $\mathbf k$. 

In the setting of QFCS, it is not the infinitude of modes that causes conceptual difficulty, but rather the fact that interactions, curved backgrounds, and scattering may generically mix infinitely many modes unless additional structure is present.

\subsection*{From an infinity of modes to finitely many}

In practice---both in theory and in laboratory quantum optics---physicists regularly work with quantum fields using only a \emph{finite} subset of degrees of freedom. Two mechanisms routinely achieve this:

\paragraph*{(a) Symmetry reduction.}
If the background spacetime or physical system has symmetries, then conserved quantities (such as energy, momentum, or angular momentum) label independent mode sectors. The full Hilbert space decomposes into invariant subspaces
\begin{equation}
  \mathcal H
  \;\cong\;
  \bigoplus_{J} \mathcal H_{J},
\end{equation}
where $J$ denotes a collective set of quantum numbers. Crucially, the \emph{dynamics} is block-diagonal in $J$: modes with distinct quantum numbers do not mix. Within each sector, however, a \emph{finite} number of asymptotic modes (ingoing/outgoing on each side of a scattering region) can mix through, e.g., a Gaussian transformation. Thus each $J$-sector can be treated as a finite-mode Gaussian system.


\paragraph*{(b) Filtering and smearing.}
Operationally, experiments only measure a restricted subset of modes.
Detectors have finite bandwidth, finite temporal resolution, angular
acceptance, and spatial mode selectivity. Mathematically, this corresponds to
working with smeared field operators $\hat\varphi(f)$ or with a selected set of
mode operators $\hat a_{i}=\int d^3k\,f_i(\mathbf k)\hat a_{\mathbf k}$. Restricting attention to a finite collection of such modes is operationally equivalent to tracing out or ignoring all remaining degrees of freedom. This is the sense in which laboratory QFT is routinely reduced to tractable finite-mode systems.

\begin{figure}
    \centering
    \includegraphics[width=.8\linewidth]{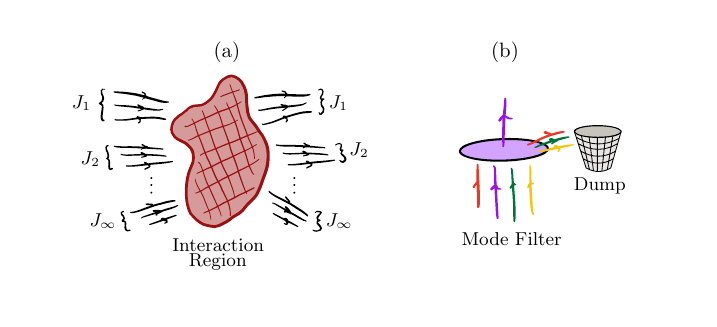}
    \caption{High-level depiction of mode reduction. (a) \emph{Symmetry reduction:} conserved quantum numbers (collectively labeled by $J$) partition the field into invariant mode sectors. Only modes sharing the same $J$ mix under the dynamics, allowing each sector to be treated as a finite-dimensional Gaussian process. (b) \emph{Filtering/smearing:} an experiment or detector selects a finite set of accessible modes (e.g., via bandwidth, spatial mode matching, or temporal resolution), while remaining degrees of freedom are effectively traced out.}
    \label{fig:inf_to_finite_dof}
\end{figure}

We illustrate these two mechanisms conceptually in Fig.~\ref{fig:inf_to_finite_dof}.

\section{Black Holes: Hawking Radiation, Superradiance, and Beyond}
\label{section:bhs}

In this section we turn to the spontaneous creation of quanta during the gravitational collapse of a massive body---the phenomenon known as the (astrophysical) \emph{Hawking effect}, first derived in Refs.~\cite{Hawking74BHexplosions,Hawking1975} (see also~\cite{Hawking1976breakdown,Unruh76notes}). Quantum fluctuations near the forming horizon give rise to outgoing radiation with a thermal spectrum, entangled with partner modes that fall into the black hole. Rotation leads to a closely related process, \emph{superradiance}~\cite{starobinskii1973superrad,unruh1974superrad}, in which certain incident waves are amplified by mode mixing in the ergoregion. Both effects are intrinsically quantum, and both generate correlations between exterior radiation and degrees of freedom hidden behind the horizon.

We first describe these semi-classical gravitational effects at a field-theoretic level. The dynamics of a quantum field on a curved spacetime leads to Bogoliubov mixing between ``in'' and ``out'' mode sectors. This mixing is what produces spontaneous particle creation during collapse (the Hawking effect) and wave amplification in the presence of an ergoregion (superradiance). Crucially, the resulting input-output relations are \emph{Gaussian}: each mode sector evolves through a linear map that preserves the canonical commutation relations. As a result, both the Hawking process and superradiance can be viewed as Gaussian quantum channels. This Gaussian perspective provides a streamlined route for understanding phenomenological features of black-hole evaporation, including the entanglement structure between interior and exterior modes~\cite{Agullo2024:EntRotBH}, the semi-classical mass loss of the black hole, and the emergence of steady-state thermal behavior when the geometry is slowly varying~\cite{Frolov1998bible}. We end with speculative comments connecting black holes and quantum information.

Throughout, we focus on a massless scalar field $\varphi$ (spin $s=0$) for simplicity. Analogous conclusions hold for higher-spin bosonic fields, such as photons ($s=1$) and gravitons ($s=2$). For a detailed and comprehensive treatment, the reader may consult Ref.~\cite{Frolov1998bible}.

\begin{figure}
    \centering
    \includegraphics[width=.4\linewidth]{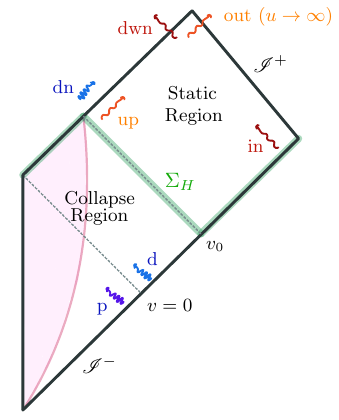}
    \caption{Penrose diagram of gravitational collapse forming a black hole. High-frequency progenitor modes ``p'' and ``d'' in the collapse region ($v<v_0$) seed the Hawking process; modes localized near the last escaping ray $v=0$ provide the dominant contribution. The resulting ``up'' mode emerging into the static region ($v>v_0$) carries thermal Hawking radiation, while its entangled partner mode ``dn'' lies behind the event horizon. Outgoing up-modes encounter the Schwarzschild/Kerr potential barrier: They may transmit to $\mathscr{I}^+$ as late-time Hawking quanta ($u\!\to\!\infty$), or be reflected back toward the horizon as ``dwn'' modes. For rotating black holes, modes satisfying $\omega - m\Omega_H < 0$ undergo superradiant amplification: scattering off the black-hole potential extracts rotational energy in the ergoregion, yielding a reflection coefficient greater than unity.}
    \label{fig:penrose_diagram}
\end{figure}

\subsection*{Scattering of quantum fields by black holes}

Consider a massive body that collapses to form a black hole. After the collapse settles, the exterior geometry becomes stationary and, if non-rotating, is described by the Schwarzschild solution. A Penrose diagram illustrating the in- and out-regions $\mathscr I^\pm$ is shown in Fig.~\ref{fig:penrose_diagram}. With well-defined asymptotic regions, we pose a scattering problem: given initial data $\varphi|_{\mathscr I^-}$, what is the late-time field $\varphi|_{\mathscr I^+}$ measured by distant observers?

For a non-rotating black hole, the late-time metric takes the familiar form
\begin{equation}
    ds^{2}
    = -\!\left(1-\frac{r_s}{r}\right)dt^{2}
      +\!\left(1-\frac{r_s}{r}\right)^{-1}dr^{2}
      + r^{2}d\Omega^{2},
\end{equation}
where $r_s = 2M$ is the Schwarzschild radius (in $G=c=1$ units). The stationary and spherically symmetric geometry admits conserved quantities $J = \{\omega,\ell,m\}$ (frequency measured at infinity, angular momentum, and its $z$-component). These labels define natural mode bases $\{\varphi^{\rm in/out}_{J}\}$ on $\mathscr I^\pm$ and imply that the scattering map decomposes into independent sectors,
\begin{equation}
    \mathcal{N}
    \;\approx\;
    \bigoplus_{J}\, \mathcal{N}_{J},
\end{equation}
so that each $J$-sector evolves independently. Thus the full problem reduces to understanding a single channel $\mathcal N_J$ at a time.

For a free bosonic field (or any field with at most bilinear interactions), each sector $J$ behaves as a \emph{Gaussian bosonic channel}: the field operators transform linearly, and tracing out the unobserved degrees of freedom (e.g., the portion of the wave falling behind the horizon) yields a single-mode Gaussian channel that maps ingoing modes on $\mathscr{I}^-$ to outgoing modes on $\mathscr I^+$.

\begin{figure}
    \centering
    \includegraphics[width=.6\linewidth]{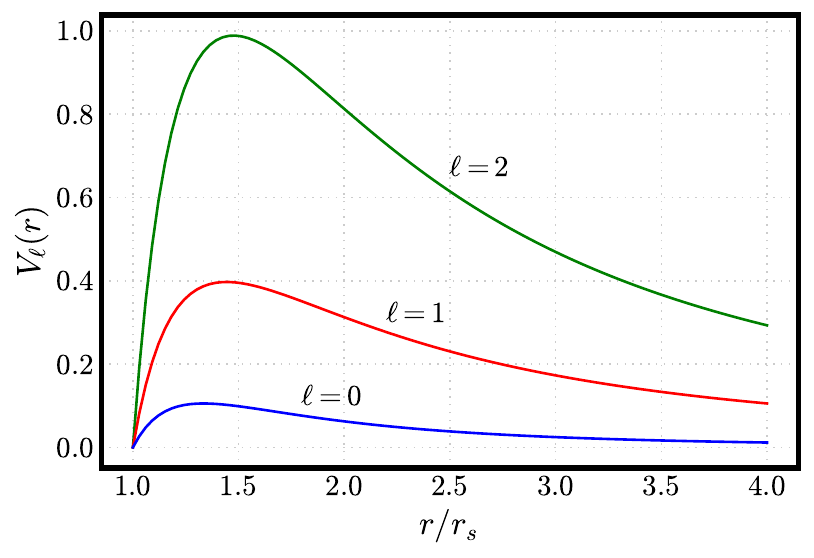}
    \caption{Plot of the potential $V_{\ell}(r)$ versus the Schwarzschild radial coordinate $r$ ($r_s=1$).}
    \label{fig:Vell}
\end{figure}

\paragraph*{Asymptotic basis and the Regge--Wheeler equation.}
Separating variables in Schwarzschild coordinates leads to
\begin{equation}
    \varphi_{J}
    = \frac{1}{\sqrt{4\pi\omega}}
      \frac{R_{\ell}(r)}{r}\,
      Y_{\ell m}(\theta,\phi)\,
      e^{-i\omega t},
\end{equation}
with the radial function obeying the Regge-Wheeler equation in the tortoise coordinate $r_*$,
\begin{equation}
    \frac{d^{2}R_\ell}{dr_*^{2}}
    + \bigl(\omega^{2} - V_\ell(r)\bigr)R_\ell = 0,
    \qquad
    \frac{dr_*}{dr}
    = \!\left(1-\frac{r_s}{r}\right)^{-1},
\end{equation}
\begin{equation}
    V_\ell(r)
    = \left(1-\frac{r_s}{r}\right)
      \!\left(\frac{\ell(\ell+1)}{r^{2}} + \frac{r_s}{r^{3}}\right).
\end{equation}
The potential vanishes at the horizon ($r\!\to\!2M$) and at infinity, with a peak near the photon sphere ($r\!\approx\!3M$). Thus, asymptotically $R_\ell \sim e^{\pm i\omega r_*}$, giving the standard positive-frequency in/out modes
\begin{equation}\label{eq:inout_basis}
    \varphi^{\rm in}_{J}\Big|_{\mathscr I^-}
    \approx
    \frac{e^{-i\omega v}}{\sqrt{4\pi\omega}\, r}Y_{\ell m}(\theta, \phi),
    \qquad
    \varphi^{\rm out}_{J}\Big|_{\mathscr I^+}
    \approx
    \frac{e^{-i\omega u}}{\sqrt{4\pi\omega}\, r}Y_{\ell m}(\theta, \phi),
\end{equation}
with ingoing and outgoing Eddington–Finkelstein coordinates
$v = t+r_*$ and $u=t-r_*$.

For Schwarzschild, the channel $\mathcal N_J$ corresponds to ordinary scattering: part of an ingoing wave falls into the black hole, and the remainder is reflected to $\mathscr I^+$. No amplification occurs because Schwarzschild has no ergoregion.

\paragraph*{Greybody factors and the Gaussian channel.}
Let $\Gamma_J$ be the transmission probability through the potential barrier (so $1-\Gamma_J$ is the reflection probability). The outgoing mode at $\mathscr I^+$ is a linear combination of the reflected ``in'' mode and an ``up'' mode emerging from near the horizon. At the level of Gaussian channels, this means that $\mathcal N_J$ is an \emph{attenuator channel} with the up-mode serving as the environment. In the standard notation of Part~\ref{lecture:2}:
\begin{align}
    \bm X_J &= \sqrt{1-\Gamma_J}\,\bm I_2,\\
    \bm Y_J &= \Gamma_J\,\bm\sigma^{\rm up}_J,\\
    \bm d_J &= \sqrt{\Gamma_J}\,\bm\mu^{\rm up}_J,
\end{align}
where $(\bm\mu^{\rm up}_J,\bm\sigma^{\rm up}_J)$ are the first and second moments of the up-mode. 

\paragraph*{The origin of the Hawking effect.}
The up-mode is innocuous in the static exterior region:
near the horizon it behaves as a positive-frequency outgoing mode,
\begin{equation}\label{eq:upmode}
    \varphi^{\rm up}_J\Big|_{\Sigma_H}
    \approx
    \frac{e^{-i\omega u}}{\sqrt{4\pi\omega}\,r}\,Y_{\ell m}(\theta, \phi)
    \qquad (u\to\infty).
\end{equation}
The nontrivial physics emerges when we propagate this mode backward through the
collapsing region. Using the near-horizon relation
$u \sim -\kappa^{-1}\ln(-\kappa v)$ (with $\kappa=1/4M$ the surface gravity), one finds for $v<0$,
\begin{equation}
    \varphi^{\rm up}_J\Big|_{\mathscr I^-}
    \approx
    \frac{e^{\,i(\omega/\kappa)\ln(-\kappa v)}}{\sqrt{4\pi\omega}\,r}\,
    Y_{\ell m}(\theta, \phi)
    \qquad (v<0),
\end{equation}
which oscillates with exponentially increasing frequency as $v\to 0^-$.\footnote{The point $v \to 0^-$ defines the last ray just before horizon formation.} The modes relevant to the Hawking effect ($v\to 0^-$) have very high energy, which supports the assumption that, during backwards propagation, we can ignore interactions with the collapsing matter (geometric optics approximation).

Decomposing this mode into positive- and negative-frequency asymptotic modes on $\mathscr I^-$ yields a Bogoliubov transformation (see example calculation below). At the level of mode operators,
\begin{equation}\label{eq:up_operator}
    \hat a^{\rm up}_J
    = c_J \hat a^{\rm p}_J + s_J \hat a^{\rm d\,\dagger}_J,
    \qquad
    \frac{s_J}{c_J} = e^{-\omega/2T_H},
\end{equation}
with $c_J^2 - s_J^2 = 1$ and Hawking temperature $T_H = \kappa/2\pi$.  If the ``p'' and ``d'' progenitor modes are initially in vacuum, the mean number of outgoing quanta is
\begin{equation}
    \bar n_{H}
    = s_J^2
    = \frac{1}{e^{\omega/T_H}-1},
\end{equation}
the familiar thermal (Bose–Einstein) spectrum. In this case the up-mode has zero mean and isotropic thermal fluctuations,
\begin{equation}
    \bm\mu^{\rm up}_J = 0,
    \qquad
    \bm\sigma^{\rm up}_J = \bigl(1+2\bar n_H\bigr)\,\bm I_2.
\end{equation}
The negative-frequency component associated with $s_J$ is the seed of both Hawking radiation and the entanglement between exterior Hawking quanta and their interior partners.

\begin{example}{Mode decomposition of $\varphi_J^{\rm up}$}\label{example:phiup}

We sketch how the ``up'' mode $\varphi^{\rm up}_J$ carries thermal quanta with respect to the natural positive/negative-frequency basis on $\mathscr{I}^-$.  The asymptotic form of the ``up'' mode on $\mathscr{I}^-$ is given by Eq.~\eqref{eq:upmode}.  Writing $\varphi = \Phi/r$ on asymptotic 2-spheres, we have
\[
\Phi^{\rm up}_J(v) \;=\; 
\frac{1}{\sqrt{4\pi\omega}}\,
(-\kappa v)^{\,i\omega/\kappa}\, Y_{\ell m}(\theta,\phi)
\qquad (v<0),
\]
where we use $e^{i\omega\ln(-\kappa v)/\kappa}=(-v\kappa)^{i\omega/\kappa}$. Likewise, the ``in'' basis on $\mathscr{I}^-$ is
\[
\Phi^{\rm in}_J(v) \;=\; 
\frac{1}{\sqrt{4\pi\omega}}\, e^{-i\omega v}\,
Y_{\ell m}(\theta,\phi),
\quad 
\Phi^{\rm in\,*}_J(v) = \frac{1}{\sqrt{4\pi\omega}} e^{+i\omega v} Y_{\ell m}(\theta,\phi).
\]

The “up’’ mode can be expanded in the asymptotic basis through the Bogoliubov relation  
\[
\varphi_J^{\rm up}
=\sum_{J'}
\Big(\alpha_{JJ'}\,\varphi^{\rm in}_{J'} 
    + \beta_{JJ'}\,\varphi^{\rm in\,*}_{J'}\Big),
\]
where the Bogoliubov coefficients follow from the Klein–Gordon (KG) product on
$\mathscr{I}^-$:
\[
\alpha_{JJ'} = (\varphi_{J'}^{\rm in},\varphi^{\rm up}_J)_{\mathscr{I}^-},
\qquad
\beta_{JJ'} = -(\varphi_{J'}^{\rm in\,*},\varphi^{\rm up}_J)_{\mathscr{I}^-}.
\]

Using the KG product on a null surface, we obtain
\begin{align}
\alpha_{JJ'}
&=
i\!\int_{-\infty}^0\!\!dv 
\int d\Omega\,
\Big(
\Phi_{J'}^{\rm in\,*}\,\partial_v \Phi_J^{\rm up}
-
\Phi_J^{\rm up}\,\partial_v \Phi_{J'}^{\rm in\,*}
\Big)
\nonumber\\[2mm]
&=
\sqrt{\frac{\omega'}{\omega}}\,
\Bigg(\int d\Omega\, Y_{\ell m} Y_{\ell' m'}^* \Bigg)
\left[
\frac{1}{2\pi}
\int_{-\infty}^0 
dv\, (-v)^{\,i\omega/\kappa}\, e^{+i\omega' v}
\right].
\end{align}
Orthogonality of spherical harmonics gives 
\[
\int d\Omega\, Y_{\ell m} Y_{\ell' m'}^*
= \delta_{\ell\ell'}\delta_{m m'}.
\]

Similarly, for $\beta_{JJ'}$,
\begin{align}
\beta_{JJ'} 
&=
-\,i\!\int_{-\infty}^0\!\! dv
\int d\Omega\,
\Big(
\Phi_{J'}^{\rm in}\,\partial_v \Phi_J^{\rm up}
-
\Phi_J^{\rm up}\,\partial_v \Phi_{J'}^{\rm in}
\Big)
\nonumber\\[2mm]
&=
\sqrt{\frac{\omega'}{\omega}}\,
\Bigg(\int d\Omega\, Y_{\ell m} Y_{\ell' m'} \Bigg)
\left[
\frac{1}{2\pi}
\int_{-\infty}^0 
dv\, (-v)^{\,i\omega/\kappa}\, e^{-i\omega' v}
\right].
\end{align}
Using the identity
\[
Y_{\ell m} = (-1)^m Y_{\ell,-m}^*,
\]
we obtain the angular selection rule
\[
\int d\Omega\, Y_{\ell m}Y_{\ell' m'}
= (-1)^m\, \delta_{\ell\ell'}\,\delta_{m,-m'}.
\]

The remaining integrals are of the standard form
\[
\int_{-\infty}^0 dv\, (-v)^{ia}\,e^{\pm i\omega' v}
= (\pm i\omega')^{-1-ia}\Gamma(1+ia),
\qquad a=\omega/\kappa,
\]
where the branch of $(-v)^{ia}$ is fixed by analytic continuation.  Taking the ratio of magnitudes yields
\[
\frac{|\beta_{JJ'}|}{|\alpha_{JJ'}|}
= e^{-\pi\omega/\kappa}.
\]
Identifying
\[
\frac{s_J}{c_J} = e^{-\pi\omega/\kappa},
\qquad c_J^2 - s_J^2 = 1,
\]
we recover the operator relation
\[
\hat a_J^{\rm up}
= c_J\,\hat a_J^{\rm p}
  + s_J\,\hat a_J^{\rm d\,\dagger},
\]
which is Eq.~\eqref{eq:up_operator}. Thus the negative–frequency component (the $\beta$ coefficient) is suppressed by the Boltzmann factor $e^{-\omega/T_H}$ with $T_H=\kappa/2\pi$, and if the ``p'' and ``d'' progenitor modes are in vacuum, the outgoing Hawking particle number is $\bar n_H = s_J^2 = 1/(e^{\omega/T_H}-1)$.

\end{example}

\subsection*{Gaussian, bosonic channel description}

Relation~\eqref{eq:up_operator} together with the ratio
$s_J/c_J = \exp(-\omega/2T_H)$ immediately implies that the ``up'' mode $\varphi^{\rm up}_J$ occupies a thermal state $\rho_{{\rm th},J}$ with mean occupation number $\bar n_H = s_J^2= 1/(e^{\omega/T_H}-1)$. Since a single-mode thermal state has zero first moments and covariance $\bm\sigma_{\rm th}=(1+2\bar n_H)\bm I_2$, we obtain a complete Gaussian specification of the late-time Schwarzschild scattering channel $\mathcal{N}_J:\mathscr{H}^{\rm in}_J \to \mathscr{H}^{\rm out}_J$.  The ``in'' mode is attenuated by a factor $\sqrt{1-\Gamma_J}$ by the potential barrier, while the remaining fraction is replaced by the thermal ``up'' mode.  Thus
\begin{equation}
    \bm d_{\mathcal{N}_J}=0,\qquad
    \bm X_{\mathcal{N}_J}=\sqrt{1-\Gamma_J}\,\bm I_2,\qquad
    \bm Y_{\mathcal{N}_J}=\Gamma_J(1+2\bar n_H)\,\bm I_2,
\end{equation}
which is precisely a thermal-loss channel
$\mathcal{L}_{1-\Gamma_J,\bar n_H}$.

\medskip

For rotating (Kerr) black holes the story is similar, with one crucial
difference: \emph{superradiance}.  In Kerr spacetime the natural near-horizon
frequency is $\varpi=\omega - m\Omega_H$ (the Killing frequency associated with
$\partial_t+\Omega_H\partial_\phi$).\footnote{The quantity
$\Omega_H=a/(2M\left(M+\sqrt{M^2-a^2}\right))$ roughly denotes the angular velocity of the event horizon, where $a=L/M$ and $L$ is the black hole's spin.} Modes with $\omega>0$ but
\begin{equation}
    \varpi = \omega - m\Omega_H < 0
\end{equation}
have \emph{negative Klein-Gordon norm} at the horizon.  Mixing between the ingoing ``in'' mode and the negative-norm ``up'' mode at the potential barrier results in \emph{stimulated particle creation}---viz., wave amplification.

For superradiant modes, the channel $\mathcal{N}_J$ is a thermal-amplifier channel $\mathcal{A}_{G_J,\bar n_H}$ with specification
\begin{equation}
    \bm d_{\mathcal{N}_J}=0,\qquad
    \bm X_{\mathcal{N}_J}=\sqrt{G_J}\,\bm I_2,\qquad
    \bm Y_{\mathcal{N}_J}=(G_J-1)(1+2\bar n_H)\,\bm I_2,
\end{equation}
where $G_J\ge 1$ is the superradiant gain and
\begin{equation}
    \bar n_H
    = \frac{1}{e^{|\varpi|/T_H}-1},
\end{equation}
the Hawking occupation associated with the horizon generator. For $\varpi>0$ (non-superradiant sector), the channel reverts to a thermal-loss channel as in the Schwarzschild case, except that the greybody factor $\Gamma_J$ now depends on $m$ due to broken spherical symmetry.

Putting everything together, the single-mode quantum channel describing the interaction between a distant observer and a rotating black hole formed through gravitational collapse is
\begin{equation}
\mathcal{N}_J =
    \begin{cases}
        \mathcal{L}_{1-\Gamma_J,\;\bar n_H}, & \varpi>0, \\[4pt]
        \mathcal{A}_{G_J,\;\bar n_H}, & \varpi<0,
    \end{cases}
\qquad
\bar n_H=\frac{1}{e^{|\varpi|/T_H}-1},
\qquad
\varpi=\omega-m\Omega_H,
\end{equation}
where $T_H=\kappa/2\pi$ is the Hawking temperature of the rotating black hole
($T_H=1/8\pi M$ for Schwarzschild).  

\begin{figure}[t]
    \centering
    \includegraphics[width=.85\linewidth]{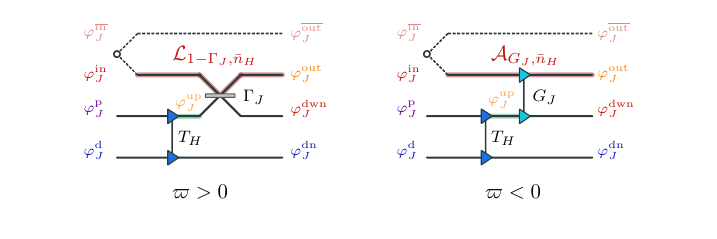}
    \caption{Symplectic diagrams for the input-output
    channel of a rotating black hole in the non superradiant ($\varpi>0$) and superradiant ($\varpi<0$) regimes.}
    \label{fig:bh_circuits}
\end{figure}

Figure~\ref{fig:bh_circuits} illustrates the symplectic input-output scattering diagrams of the Hawking and superradiant processes; the color scheme matches the spacetime picture in Fig.~\ref{fig:penrose_diagram}.  These circuits are precisely the unitary dilations of the thermal-loss and thermal-amplifier channels reviewed in Fig.~\ref{fig:dilation} of Part~\ref{lecture:2}, augmented here by a complementary ``$\overline{\text{\i n}}$'' mode that purifies the ``in'' mode.  If the black hole is probed by thermal radiation at temperature $T_{\rm in}$, the joint state of ``in'' and ``$\overline{\text{\i n}}$'' is a TMSV (thermofield-double) state.  Including this complementary mode allows one to track how entanglement between exterior Hawking quanta and the black-hole interior depends on the mixedness of the incoming radiation.  For simplicity, we assume that the complementary system evolves trivially (e.g., it could be stored in an idealized quantum memory for later joint measurements).  

\subsection*{``In'' and ``out'' Gaussian states}

In the preceding discussion we described how quantum fields scatter in the geometry of a collapsing black hole using mode functions and their associated annihilation and creation operators.  We now determine the quantum states of the ingoing and outgoing radiation.  Throughout we occasionally use the terms ``spontaneous'' and ``stimulated'' emission to distinguish whether the ingoing modes carry excitations.

The mapping $\mathcal{N}_{J} : \mathscr{H}^{\rm in}_{J} \to \mathscr{H}^{\rm out}_{J}$ is a single-mode Gaussian channel.  Consequently, \emph{Gaussian input states remain Gaussian at the output}, and the evolution is completely specified by the input means and covariance matrices, as well as the channel parameters.  For this reason we restrict attention to Gaussian initial data on $\mathscr{I}^-$.  This assumption is already forced upon us for the high-frequency progenitor modes (``p'' and ``d''), whose frequencies make them effectively impossible to populate; their initial states are vacuum to excellent approximation, a fact essential to the reduction of $\mathcal{N}_{J}$ to a Gaussian channel.

Let the uncoupled ingoing radiation be in a Gaussian state with block-diagonal mean and covariance 
\[
\bm\mu^{\rm in}=\bigoplus_J \bm\mu^{\rm in}_{J}, 
\qquad 
\bm\sigma^{\rm in}=\bigoplus_J \bm\sigma^{\rm in}_{J}.
\]
Each sector $J$ may be mixed, and one may include an orthogonal ``$\overline{\text{\i n}}$'' mode to purify it, as in Fig.~\ref{fig:bh_circuits}. Because the dynamics factorize across $J$, the outgoing radiation also decomposes as $\bm\mu^{\rm out}=\bigoplus_J \bm\mu^{\rm out}_{J}$ and $\bm\sigma^{\rm out}=\bigoplus_J \bm\sigma^{\rm out}_{J}$.

Using the fact that $\mathcal{N}_{J}$ is a thermal-loss channel when 
$\varpi=\omega-m\Omega_H>0$ and a thermal-amplifier channel when $\varpi<0$, 
the transformation rules for the first and second moments follow immediately:
\begin{align}
\bm\mu^{\rm out}_{J}
&=
\begin{cases}
\sqrt{1-\Gamma_J}\,\bm\mu^{\rm in}_{J}, & \varpi>0,\\[4pt]
\sqrt{G_J}\,\bm\mu^{\rm in}_{J}, & \varpi<0,
\end{cases}
\label{eq:kerr_inout_mean}
\\[1em]
\bm\sigma^{\rm out}_{J}
&=
\begin{cases}
(1-\Gamma_J)\bm\sigma^{\rm in}_J 
+ \Gamma_J(1+2\bar n_H)\bm I_2, 
& \varpi>0,
\\[4pt]
G_J\,\bm\sigma^{\rm in}_J
+ (G_J-1)(1+2\bar n_H)\bm I_2,
& \varpi<0,
\end{cases}
\label{eq:kerr_inout_cov}
\end{align}
where 
\[
\bar{n}_H = \frac{1}{e^{|\varpi|/T_H}-1}
\]
is the Hawking occupation number in the relevant sector.

\paragraph*{Spontaneous Hawking emission.}
The \emph{spontaneous} Hawking process corresponds to taking vacuum initial data in every mode on $\mathscr{I}^-$, i.e. 
\[
\bm\mu^{\rm in}_{J}=0,
\qquad
\bm\sigma^{\rm in}_{J}=\bm I_2.
\]
Substituting these values into Eqs.~\eqref{eq:kerr_inout_mean}-\eqref{eq:kerr_inout_cov} yields
\begin{equation}
\bm\sigma^{\rm out}_{J}
=
\begin{cases}
(1+2\Gamma_J\bar n_H)\,\bm I_2, & \varpi>0,
\\[4pt]
\bigl[1+2(G_J-1)(\bar n_H+1)\bigr]\bm I_2,
& \varpi<0,
\end{cases}
\label{eq:spont_sigma_out}
\end{equation}
with $\bm\mu^{\rm out}_{J}=0$. Writing $\bm\sigma^{\rm out}_{J}=(1+2\bar n^{\rm out}_{J})\bm I_{2}$, the 
mean number of spontaneously emitted quanta per mode $J$ is
\begin{equation}
\bar n^{\rm out}_{J}
=
\begin{cases}
\Gamma_J\,\bar n_H, & \varpi>0,\\[4pt]
(G_J-1)(\bar n_H+1), & \varpi<0.
\end{cases}
\label{eq:spont_nout}
\end{equation}
Even as $\bar n_H\!\to\!0$ (i.e., no Hawking contribution), a rotating black hole spontaneously emits superradiant quanta:  for $\varpi<0$, one always has $\bar n^{\rm out}_J>0$ because $G_J>1$.

\paragraph*{Stimulated emission.}
We use the term \emph{stimulated Hawking emission} broadly to describe processes in which some ingoing modes on $\mathscr{I}^-$ carries excitations.  One may adopt a stricter definition---namely, that only excitations in the progenitor modes (``p'' or ``d'') truly \emph{seed} the Hawking process, which conversely may be possible in analogue-gravity experiments. However, in astrophysical settings the progenitor modes have enormous frequencies and are effectively impossible to populate. Thus, for astrophysical black holes, the only practical source of stimulation comes from populating the accessible ``in'' modes.  Although this does not seed pair creation in the collapse region, it does lead to additional output radiation through scattering with the Hawking flux, and it plays an essential role in processes such as feeding, spinning up, or establishing thermal equilibrium with the black hole.

\subsection*{Semi-classical evolution}

Until now we have only considered evolution of quantum fields propagating on a fixed, stationary black-hole spacetime. However, due to spontaneous Hawking emission and superradiance, the black hole itself must evolve: its mass $M$ and angular momentum $L$ decrease in time as radiation escapes. On short dynamical time-scales $\sim M$ (in $G=c=1$ units) the geometry is effectively stationary, but on much longer time-scales the cumulative effect of the outgoing flux drives a quasi-static evolution. The goal of this subsection is to give a compact description of this semi-classical evolution; see also the numerical work of Page~\cite{page1976PRDI,page1976PRDII} and
Section~10.5 of Ref.~\cite{Frolov1998bible}.

\begin{figure}
    \centering
    \includegraphics[width=.49\linewidth]{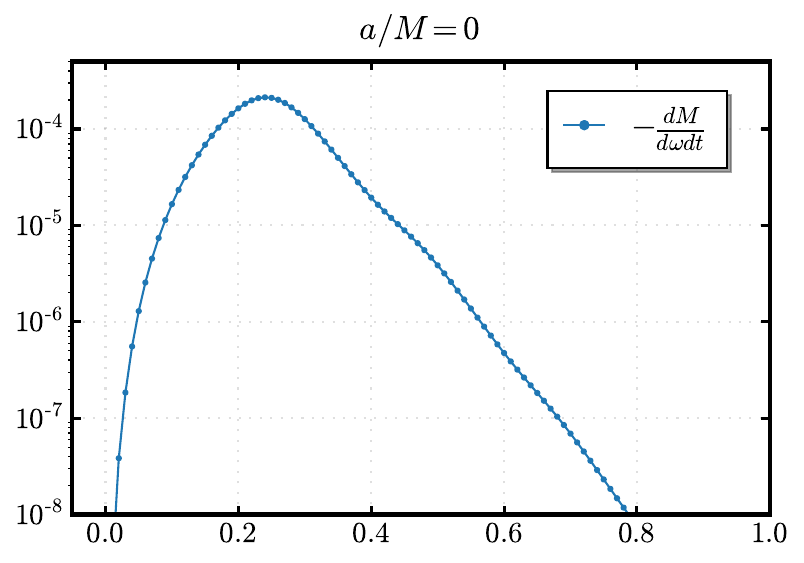}
    \includegraphics[width=.49\linewidth]{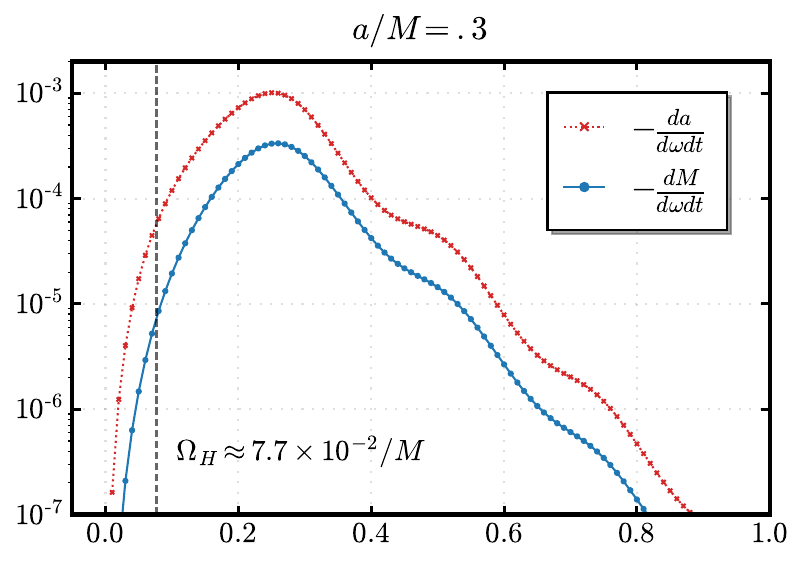}
    \includegraphics[width=.49\linewidth]{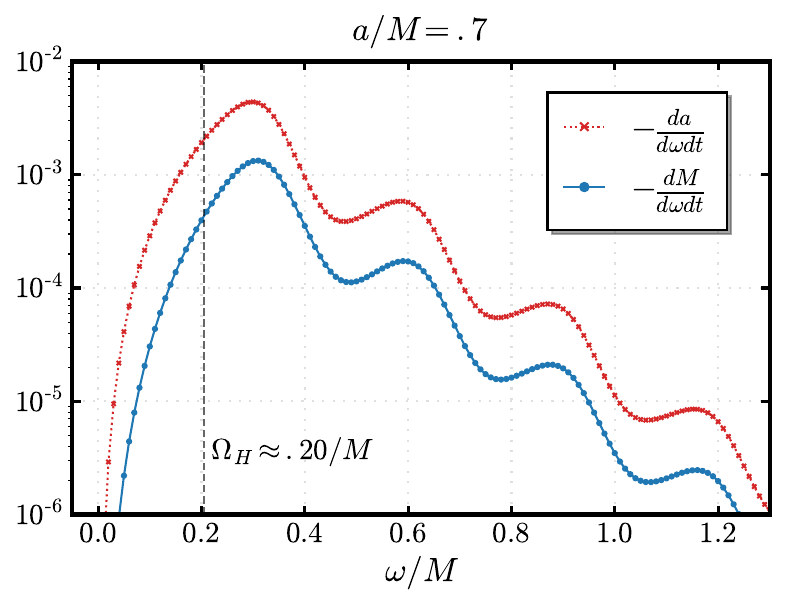}
    \includegraphics[width=.49\linewidth]{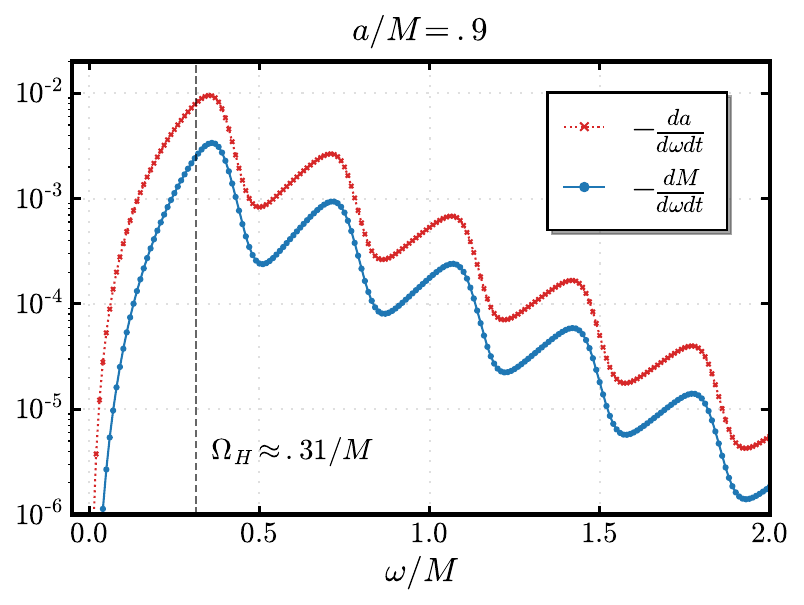}
    \caption{
    Power spectra (blue circles) and angular-momentum spectra (red crosses) for spontaneously emitted photons from a Kerr black hole with $M=1$ and spin parameter $a=L/M$. Frequencies satisfying $\omega<m\Omega_H$ (to the left of $\Omega_H$ on each panel) lie in the superradiant regime and exhibit amplification, while the $\omega>m\Omega_H$ region is dominated by Hawking emission. Numerical input data courtesy of Adri\`a Delhom.}
    \label{fig:dMdL_spectra}
\end{figure}

For each mode $J=(\omega,\ell,m)$ we denote by $\bar{n}^{\rm in}_{\omega\ell m}$ and $\bar{n}^{\rm out}_{\omega\ell m}$ the mean occupation numbers of ingoing and outgoing quanta measured at $\mathscr{I}^-$ and $\mathscr{I}^+$, respectively.\footnote{Here $\bar{n}^{\rm out}_{\omega\ell m}$ includes the effects of greybody factors and, in the rotating case, superradiant amplification, as derived previously in the Gaussian-channel description.} The ingoing and outgoing power spectral densities (energies per unit time per unit frequency) as measured by asymptotic observers are
\begin{equation}
  \mathcal{P}^{\rm in}(\omega)
  \;\triangleq\;
  \frac{1}{2\pi}\sum_{\ell,m}\omega\,\bar{n}^{\rm in}_{\omega\ell m},
  \qquad
  \mathcal{P}^{\rm out}(\omega)
  \;\triangleq\;
  \frac{1}{2\pi}\sum_{\ell,m}\omega\,\bar{n}^{\rm out}_{\omega\ell m},
\end{equation}
so that the total ingoing and outgoing powers are
\begin{equation}
  \mathcal{P}^{\rm in}
  = \int_0^\infty \dd{\omega}\,\mathcal{P}^{\rm in}(\omega),
  \qquad
  \mathcal{P}^{\rm out}
  = \int_0^\infty \dd{\omega}\,\mathcal{P}^{\rm out}(\omega).
\end{equation}
Energy conservation then implies that the net flux $\mathcal{P}^{\rm out}-\mathcal{P}^{\rm in}$ dictates the quasi-static change in the black-hole mass:
\begin{equation}
  -\dv{M}{t}
  \;=\;
  \frac{1}{2\pi}\int_0^\infty\dd{\omega}
  \sum_{\ell,m}\omega\,
  \Big(\bar{n}^{\rm out}_{\omega\ell m}
      -\bar{n}^{\rm in}_{\omega\ell m}\Big).
  \label{eq:dMdt}
\end{equation}
Similarly, each quantum carries angular momentum $m$ along the spin axis, so the net angular-momentum flux is
\begin{equation}
  -\dv{L}{t}
  \;=\;
  \frac{1}{2\pi}\int_0^\infty\dd{\omega}
  \sum_{\ell,m} m\,
  \Big(\bar{n}^{\rm out}_{\omega\ell m}
      -\bar{n}^{\rm in}_{\omega\ell m}\Big).
  \label{eq:dLdt}
\end{equation}
For a rotating (Kerr) black hole, superradiant modes with $\varpi = \omega - m\Omega_H < 0$ are preferentially amplified and carry away positive $m$ at relatively low energy cost, so the dimensionless spin $a/M$ decreases more rapidly than the mass itself: the hole spins down efficiently and is driven away from extremality before Hawking evaporation significantly reduces $M$. For illustrative purposes, I plot the power and angular-momentum spectra of photons spontaneously emitted from a rotating black hole in Fig.~\ref{fig:dMdL_spectra}.

Evolution of $M(t)$ and $L(t)$ fully specifies the macroscopic state of the black hole (``black holes have no hair''), but it is instructive to connect this to its thermodynamic properties. Bekenstein argued that black holes are thermodynamic systems with an entropy $S_{\rm BH}$---the \emph{Bekenstein--Hawking entropy}---proportional to the horizon area $\mathscr{A}$, while Hawking's calculation of thermal radiation at temperature $T_H$ corroborates this picture. The Bekenstein--Hawking entropy is
\begin{equation}
  S_{\rm BH} = \frac{\mathscr{A}}{4},
\end{equation}
where for a Kerr black hole of mass $M$ and spin parameter $a=L/M$ (in mass units) the horizon area is
\begin{equation}
  \mathscr{A}
  = 4\pi\Big[\big(M+\sqrt{M^2-a^2}\big)^2 + a^2\Big].
\end{equation}
For a non-rotating Schwarzschild black hole ($a=0$), $S_{\rm BH}=4\pi M^2 \sim 10^{77}(M/M_\odot)^2$ in ordinary units. It remains an open question what microscopic degrees of freedom $S_{\rm BH}$ counts, but the standard interpretation is that it measures the logarithm of the number of microstates compatible with the macroscopic parameters $(M,L)$.

\begin{example}{Tidbits about $S_{\rm BH}$ and Hawking's area theorem}
\begin{itemize}
  \item In classical GR, the Bekenstein--Hawking entropy satisfies $\delta S_{\rm BH}\geq 0$ (Second Law of Black-Hole Mechanics). This follows from Hawking's area theorem for classical black holes~\cite{hawking1971AreaThm}. Recent gravitational-wave observations have provided consistency checks of the area theorem for astrophysical mergers~\cite{isi2021BHareaTest}.

  \item Hawking's area theorem constrains, for example, the maximum efficiency of gravitational-wave emission in black-hole mergers. Consider two initially non-rotating black holes of mass $M$ that merge to form a final black hole of mass $M_f=2M-\delta M$, where $\delta M$ is the mass radiated away in gravitational waves. The initial area scales as $\mathscr{A}_0 \propto 2M^2$, while the final area scales as $\mathscr{A}_f \propto (2M-\delta M)^2$. Imposing $\mathscr{A}_f \geq \mathscr{A}_0$ yields $\delta M \leq (2-\sqrt{2})M$, so the efficiency $\delta M/2M \leq (2-\sqrt{2})/2 \approx 0.29$. For the GW150914 event, the inferred efficiency was only $\sim 4\%$.

  \item Classically, the time-reverse of a merger (a single hole ``fissioning'' into two) is forbidden by the area theorem as this requires $\delta \mathscr{A}<0$.

  \item In Hawking's original derivation, the area theorem relies on the weak energy condition (positive energy density) and classical GR. In the semi-classical regime the weak energy condition is violated near the horizon because of the negative-energy partner modes that accompany Hawking radiation~\cite{Frolov1998bible}. A Schwarzschild black hole in isolation loses mass through Hawking emission; since $\delta S_{\rm BH}/S_{\rm BH} = 2\,\delta M/M$ and $\delta M<0$, this implies $\delta S_{\rm BH}<0$, motivating a generalized second law for the sum of black-hole and radiation entropies.
\end{itemize}
\end{example}

We have so far discussed \emph{macroscopic} black-hole
degrees of freedom ($M$ and $L$), ignoring the microscopic radiation degrees of freedom of the quantum fields inside and outside the horizon. These can be characterized using the entropy and log-negativity tools developed in Part~\ref{lecture:3}.

Suppose the ingoing radiation in mode $J=(\omega,\ell,m)$ is in a (possibly mixed) Gaussian state. We can always purify the ingoing system by introducing a complementary mode ``$\overline{\text{\i n}}$'' in a reference Hilbert space, such that the joint state on
$({\rm in}\cup\overline{\text{\i n}})$ is pure and
$s_J^{\rm rad}[{\rm in}\cup\overline{\text{\i n}}]=0$.\footnote{Here and below
we use a lowercase $s$ to emphasize that these are entropy \emph{densities}
(e.g., per unit time per mode), whereas capital $S$ will denote integrated
entropies.} The global unitary describing black-hole scattering then acts
nontrivially on the physical modes (interior plus ingoing), while the
reference mode $\overline{\text{\i n}}$ evolves trivially $\overline{\text{\i n}} \to \overline{\rm{out}}$ by identity (it can be thought of as an ideal quantum memory storing the purification). After scattering, the global state on
$({\rm int}\cup{\rm out}\cup\overline{\text{out}})$ remains pure, and by
unitarity one has
\begin{equation}
  s_J^{\rm rad}[{\rm int}]
  \;=\;
  s_J^{\rm rad}[{\rm out}\cup\overline{\text{out}}\,],
\end{equation}
even though $s_J^{\rm rad}[{\rm out}]$ and $s_J^{\rm rad}[{\rm int}]$ need not be equal when the ingoing state is mixed.

To quantify the radiation entropy per unit time (an \emph{entropy flux}) as seen by an asymptotic observer with access to a subsystem $A$ (for example, $A={\rm out}$ or $A={\rm int}$), we sum over all modes,
\begin{equation}
  \dv{S^{\rm rad}[A]}{t}
  \;=\;
  \frac{1}{2\pi}\int_0^\infty\dd{\omega}
  \sum_{\ell,m} s^{\rm rad}_{\omega\ell m}[A].
  \label{eq:entropy_flux}
\end{equation}
In the simple spontaneous-emission case, the ingoing modes are in vacuum (so $s^{\rm rad}_J[{\rm in}]=0$) and the outgoing and interior radiation form a pure bipartite Gaussian state, so $s^{\rm rad}_J[{\rm out}]=s^{\rm rad}_J[{\rm int}]$.

Using Eq.~\eqref{eq:gaussian_entropy} from Part~\ref{lecture:3}, the entropy density of the outgoing radiation in mode $J$ can be expressed in terms of the mean outgoing occupation number $\bar{n}^{\rm out}_J$ as
\begin{equation}
  \qq{Spontaneous emission:}\quad
  s^{\rm rad}_J[{\rm out}]
  =
  \big(\bar{n}^{\rm out}_{J}+1\big)\log\big(\bar{n}^{\rm out}_{J}+1\big)
  -\bar{n}^{\rm out}_{J}\log\big(\bar{n}^{\rm out}_{J}\big),
  \label{eq:srad_out_spont}
\end{equation}
where $\bar{n}^{\rm out}_J$ is given by Eqs.~\eqref{eq:spont_nout}. This expression applies equally in the superradiant and non-superradiant regimes, since in both cases the outgoing state for each mode is  effectively thermal and characterized only by a mean occupation number.

\begin{figure}
    \centering
    \includegraphics[width=.49\linewidth]{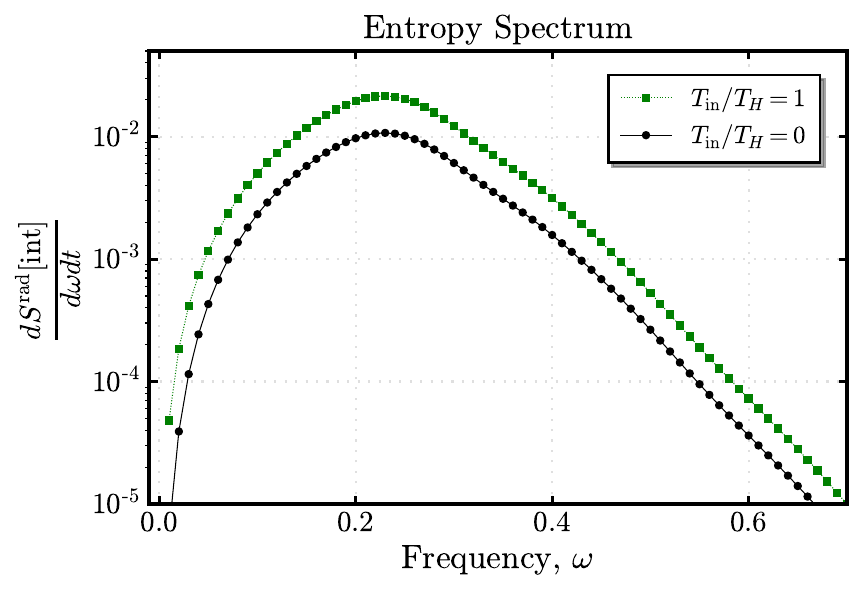}
    \includegraphics[width=.49\linewidth]{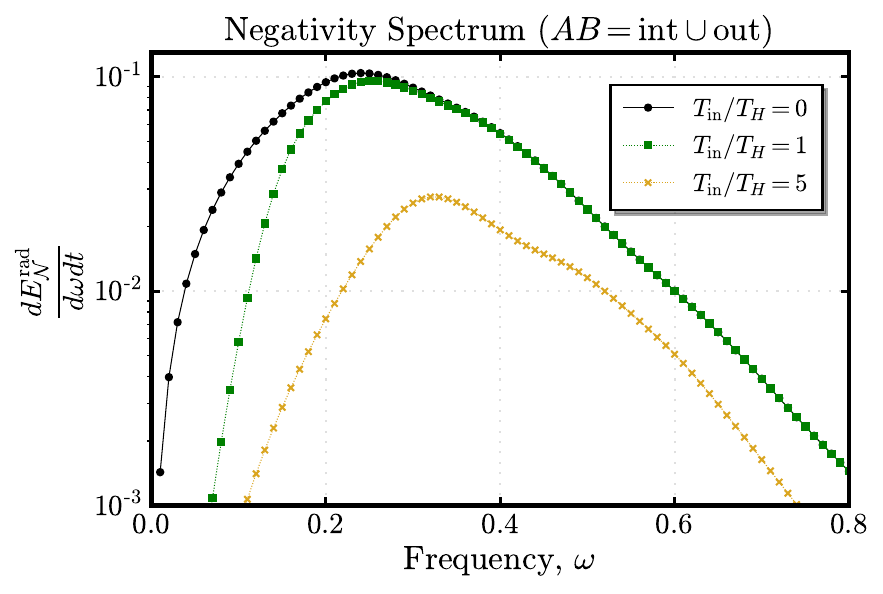}
    \caption{
    Entanglement and entropy spectra for radiation (photons and gravitons) emitted by a non-rotating black hole interacting with ingoing thermal radiation at temperature $T_{\rm in}$. \textbf{Left:} Entropy of the interior radiation $S^{\rm rad}[\mathrm{int}]=S^{\rm rad}[\mathrm{out}\cup\overline{\mathrm{out}}]$ assuming full access to the purifying bath degrees of freedom. This entropy grows with $T_{\rm in}$ as hotter ingoing radiation injects additional entropy into the system. \textbf{Right:} Log-negativity $E_{N}^{\rm rad}[\mathrm{int},{\rm out}]$ between interior modes and the outgoing radiation accessible at $\mathscr{I}^+$. This entanglement decreases with increasing $T_{\rm in}$ because the ingoing thermal flux acts as noise that decoheres the Hawking pairs and suppresses quantum correlations across the horizon (a contrived situation astrophysically, but natural to explore in tunable analogue systems). Numerical input data courtesy of Adri\`a Delhom.
    }
    \label{fig:negativity_entropy_Tin}
\end{figure}

Finally, we can also characterize the rate at which entanglement is generated between different subsystems using the logarithmic negativity; see Ref.~\cite{Agullo2024:EntRotBH} for details. For a bipartition $AB$ of the radiation degrees of freedom, the log-negativity in mode $J$ defines a ``negativity density'' $e_{N,J}^{\rm rad}[AB]$ (measured in ebits per unit time per mode). The associated \emph{negativity flux} is
\begin{equation}
  \dv{E_{N}^{\rm rad}[AB]}{t}
  \;=\;
  \frac{1}{2\pi}\int_0^\infty\dd{\omega}
  \sum_{\ell,m} e_{N,\omega\ell m}^{\rm rad}[AB].
  \label{eq:negativity_flux}
\end{equation}
We often take $A={\rm int}$ and $B={\rm out}$ so that $E_{N}^{\rm rad}[{\rm int},{\rm out}]$ quantifies the entanglement between the black hole and the outgoing radiation that reaches the asymptotic region.

To illustrate the contrast between entropy and log-negativity, Fig.~\ref{fig:negativity_entropy_Tin} shows the entropy spectrum (left) and the negativity spectrum (right) for a non-rotating black hole in contact with a thermal bath at temperature $T_{\rm in}$, including both photon and graviton channels. When we assume full control over the thermal bath--and hence over the purification of the ingoing modes---the relevant quantity is the entropy of the radiation interior to the black hole, $S^{\rm rad}[\mathrm{int}]=S^{\rm rad}[\mathrm{out}\cup\overline{\mathrm{out}}]$, which increases monotonically with $T_{\rm in}$ as hotter radiation pumps entropy into the system.  

If, however, the purifying bath degrees of freedom are not accessible (as is typical in analogue settings), then the entropy of the outgoing radiation alone does not measure entanglement with the interior. Instead, the appropriate quantity is the log-negativity $E_N^{\rm rad}[\mathrm{int},{\rm out}]$, which quantifies the distillable correlations between the black-hole interior modes and the accessible radiation that escapes to infinity. As shown in Fig.~\ref{fig:negativity_entropy_Tin}, this entanglement decreases with increasing $T_{\rm in}$: the ingoing thermal flux acts as noise that decoheres the Hawking pairs and suppresses quantum correlations across the horizon.

This behavior anticipates the tradeoff governing black-hole evaporation: Hawking emission tends to increase the entropy of the hole and reduce its mass, while external radiation can accelerate the process. These competing effects underlie the dynamical and thermodynamic considerations discussed next when we examine evaporation and the conditions for macroscopic equilibrium.


\subsection*{Evaporation and equilibrium}

If left in isolation, a black hole will lose mass and angular momentum due to the Hawking effect and superradiance. Pushing the semi-classical analysis to its limits, this implies that the hole will eventually dissipate, leaving behind thermal Hawking radiation as its legacy. On the other hand, we can slow down evaporation---or even arrest it---by feeding the black hole. Under suitable conditions, the hole can be brought into a macroscopic equilibrium. We now quantify these ideas.

In what follows we restrict attention to non-rotating black holes. This is not merely for simplicity: Fig.~\ref{fig:dMdL_spectra} shows that a rotating hole loses angular momentum faster than mass.\footnote{Were this not the case, a black hole could be driven to extremality and a naked singularity would appear, contradicting \href{https://en.wikipedia.org/wiki/Cosmic_censorship_hypothesis}{cosmic censorship}.} Hence, given enough time in isolation, a rotating black hole evolves toward a Schwarzschild state.

\begin{figure}
    \centering
    \includegraphics[width=.55\linewidth]{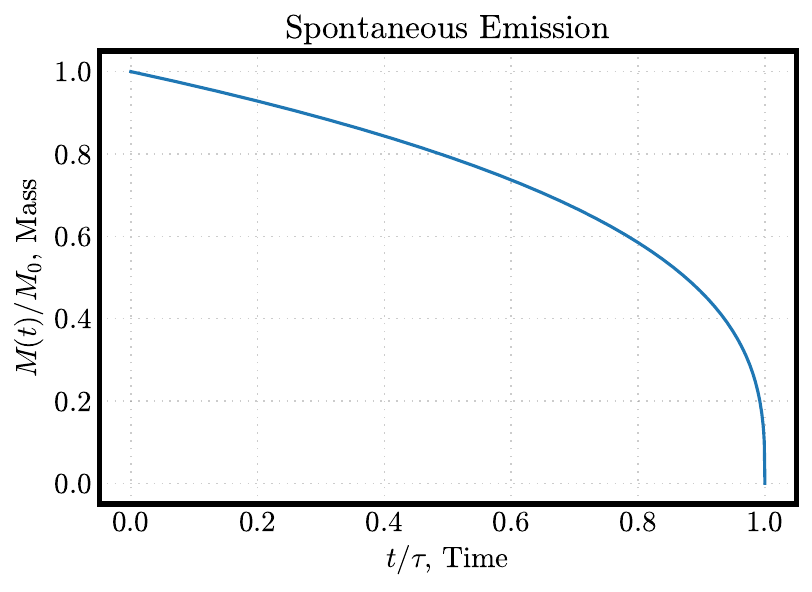}
    \caption{Semi-classical evolution of a (non-rotating) black hole's mass over time. Evaporation estimated to occur at $\tau\approx \gamma M_0^3$ with $\gamma=1/384\pi$ (better estimates have $\gamma\approx 1/8895$~\cite{page2013JCAP}).}
    \label{fig:bhMvtime}
\end{figure}

\paragraph*{Black hole evaporation.}

Here we estimate the lifetime of an isolated Schwarzschild black hole. Ignoring greybody factors and higher angular-momentum modes, we model the outgoing radiation as that of an ideal blackbody at temperature
\(T_H\). From the semiclassical mass-loss equation,
\[
-\frac{dM}{dt}
   \;=\;
   \int_0^\infty d\omega \sum_{\ell,m}
   \omega \big(\bar n^{\rm out}_{\omega\ell m}
               - \bar n^{\rm in}_{\omega\ell m}\big),
\]
and taking \(\bar n^{\rm in}=0\), we obtain the scaling
\(-dM/dt \sim T_H^2\). Using
\(\int_0^\infty dx\, x/(e^{\beta x}-1) \propto 1/\beta^2\) and
\(T_H = 1/8\pi M\), one finds that the black-hole lifetime scales as \(\tau \sim M^3\). Keeping track of constants gives \(M(t) \approx (M_0^3 - t/384\pi)^{1/3}\) (Fig.~\ref{fig:bhMvtime}), so that \(\tau \approx 384\pi M_0^3\).\footnote{More refined calculations including greybody factors give
\(\tau \approx 8895\, M_0^3\)~\cite{page2013JCAP}.}
Numerically,
\[
\tau \sim 10^{67}\ {\rm yr}\,(M/M_\odot)^3.
\]

Thus, a \href{https://en.wikipedia.org/wiki/Primordial_black_hole}{primordial black hole} evaporating today must have \(M \lesssim 10^{-19}M_\odot\sim 10^{11}\,{\rm kg}\), roughly the total mass of humanity (about 8 billion people at 80 kg each). Most of the energy is released in the final moments, when the Hawking temperature becomes very large. A final burst of high-energy thermal radiation is expected, although no such events have yet been observed.

\paragraph*{Macroscopic equilibrium.}

From Eqs.~\eqref{eq:dMdt} and~\eqref{eq:dLdt}, the macroscopic state of the black hole is stationary when there is no net flux of either energy or angular momentum, such that
\[
\bar n^{\rm out}_{\omega\ell m}
   = 
   \bar n^{\rm in}_{\omega\ell m}
   \qquad\forall\, (\omega,\ell,m).
\]
Using the Gaussian input–output rule \eqref{eq:kerr_inout_cov}, this occurs precisely when the ingoing radiation is thermal at the Hawking temperature,
\[
\bar n^{\rm in}_{\omega\ell m} = \bar n_H(\omega).
\]
Thus, a black hole achieves macroscopic equilibrium only when fed by a thermal bath at temperature \(T_{\rm in}=T_H\). Thermodynamically, macroscopic equilibrium means that the entropy fluxes balance,
\[
\frac{dS^{\rm rad}[{\rm out}]}{dt}
   =
   \frac{dS^{\rm rad}[{\rm in}]}{dt},
\]
so that on average the black hole neither heats nor cools.

\begin{figure}
    \centering
    \includegraphics[width=.7\linewidth]{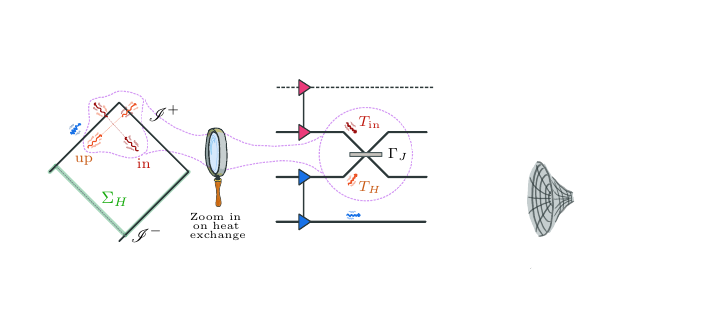}
    \caption{Unitary depiction of heat exchange between ingoing thermal radiation (at temperature $T_{\rm in}$) and Hawking radiation (at temperature $T_H$) for a Schwarzschild black hole. Macroscopic equilibrium occurs at $T_{\rm in}=T_H$. This can be viewed as two sides swapping their halves of two thermofield–double states.}
    \label{fig:bh_equil}
\end{figure}

\paragraph*{Microscopic stasis (or lack thereof).}

Although a Schwarzschild black hole in contact with a bath at \(T_{\rm in}=T_H\) appears macroscopically static, its microscopic degrees of freedom are not in stasis. The outgoing radiation is mixed because the observer at \(\mathscr{I}^+\) does not have access to the purifying ``$\overline{\rm out}$'' modes. Consequently,
\[
\delta S^{\rm rad}[{\rm out}\cup\overline{\rm out}]
   =
   \delta S^{\rm rad}[{\rm int}]
   \neq 0,
\]
meaning that the interior becomes increasingly entangled with the rest of the Universe even when all macroscopic fluxes vanish.

For channels with \(\Gamma_{\omega\ell m}\approx 1\), the interior modes form a two-mode Gaussian state with degenerate symplectic eigenvalue
\[
\nu^{\rm int}_{\omega\ell m}
   =
   \sqrt{
      1 + 4\Gamma_{\omega\ell m}\,\bar n_H
        + 4\Gamma_{\omega\ell m}\,\bar n_H^2 } ,
\]
leading to a radiation entropy on the order of twice the thermal entropy of the ingoing bath. For modes with \(\Gamma_{\omega\ell m}\approx 0\), the exterior state remains nearly pure because the ingoing radiation reflects from the potential barrier. I illustrate these behaviors in Fig.~\ref{fig:negativity_entropy_Tin}.

\subsection*{Black holes and information}

Up to this point we have worked strictly within the semi-classical framework: quantum fields evolving on a fixed spacetime.  We close with a brief and cautious step beyond this regime to outline the basic problem in information loss in black holes, leading to the ``Page curve" and why it is often invoked in discussions of black hole information. Our treatment here is intentionally minimal and conceptual; see Refs.~\cite{marolf2017BHinfoRvw,almheiri2021RMP,ashtekar2020BHevapLQG} for recent reviews and Refs.~\cite{PageCurve93PRL, zurek1982PRL, page2005rvwNJP,page2013JCAP} seminal analyses.

A useful starting point is a purely quantum-information-theoretic observation. Consider a finite quantum system $B$ interacting unitarily with a radiation system $R$.  The entanglement entropy $S(B)$ cannot grow indefinitely: it is bounded by $\log\abs{\mathscr{H}_B}$.  As information flows from $B$ to $R$, the entropy of the radiation $S(R)$ initially rises, reaches a maximum when $S(R)\approx \log\abs{\mathscr{H}_B}$, and then decreases as information returns to $R$. This characteristic ``rise and fall’’ is known as the \emph{Page curve}~\cite{PageCurve93PRL}.

To apply this intuition to black holes, one introduces an extra assumption not contained in semi-classical gravity: From the perspective of an exterior observer, a black hole behaves as a quantum system with $\abs{\mathscr{H}_{\rm BH}}\sim e^{S_{\rm BH}}$ internal states. This “Central Dogma’’~\cite{almheiri2021RMP} asserts that the Bekenstein–Hawking entropy counts all microstates of the black hole.  If so, the entropy of the Hawking radiation cannot exceed $S_{\rm BH}$, and one expects a Page curve during evaporation.

\begin{figure}
    \centering
    \includegraphics[width=.49\linewidth]{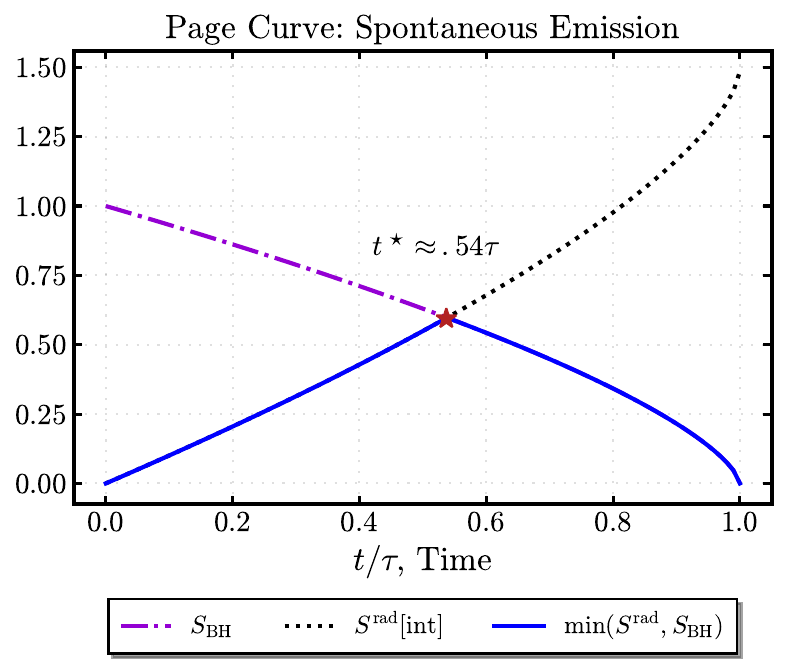}
    \includegraphics[width=.49\linewidth]{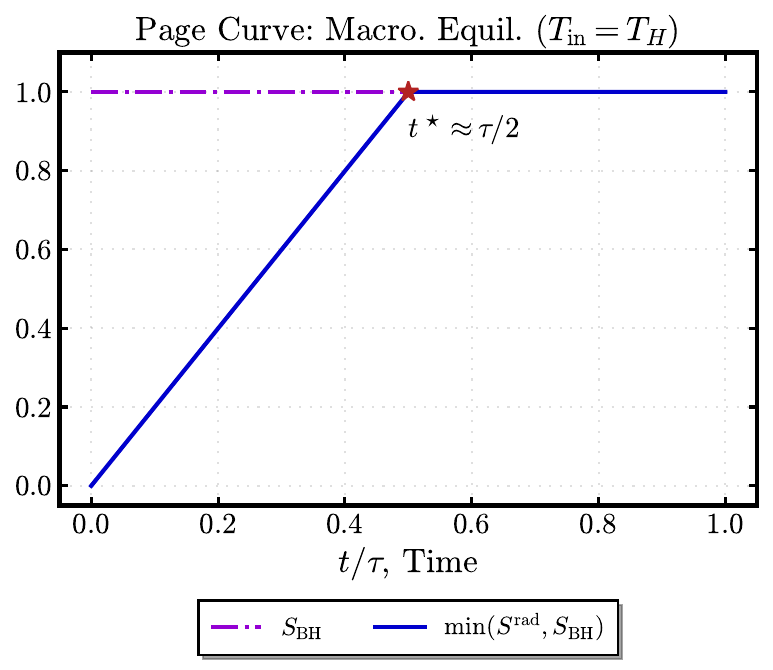}
    \caption{Schematic Page curves for a non-rotating black hole. Left: an isolated hole, where the radiation entropy $S^{\rm rad}[{\rm int}]$ rises and later falls during evaporation. Right: a hole held at macroscopic equilibrium ($T_{\rm in}=T_H$), for which $S^{\rm rad}[{\rm int}]$ grows linearly until it reaches the Bekenstein–Hawking entropy $S_{\rm BH}$ at the Page time $t^\star$.  The turnover is \textit{imposed} by the assumption that the black hole has $\sim e^{S_{\rm BH}}$ internal degrees of freedom. Data normalized by initial Bekenstein-Hawking entropy.}
    \label{fig:page_curves}
\end{figure}

This leads to a tension already visible in our semi-classical framework.  Consider a non-rotating black hole held in macroscopic equilibrium by thermal radiation at $T_{\rm in}=T_H$.  Its mass, and hence $S_{\rm BH}$, remain constant, yet the entanglement between the black hole and its exterior, viz., $S^{\rm rad}[{\rm int}]=S^{\rm rad}[{\rm out}\cup\overline{\rm out}]$, increases steadily in time.  Under the central dogma, this growth must halt once $S^{\rm rad}[{\rm int}]$ reaches $S_{\rm BH}$ at the \emph{Page time} $t^\star$. After which further entanglement would be forbidden and the radiation must effectively ``reflect’’ information back outward~\cite{HaydenPreskill2007}. 

Figure~\ref{fig:page_curves} illustrates this point. For an isolated black hole the entropy of the interior radiation rises and then falls as the black hole evaporates; in macroscopic equilibrium it rises linearly until it reaches $S_{\rm BH}$ and is then forced to saturate.  In either case, the turnover is not predicted by semi-classical physics but arises entirely from the postulate that the black hole has only $e^{S_{\rm BH}}$ internal degrees of freedom.

Whether this assumption is correct remains an open and actively debated question (see \href{https://relativity.phys.lsu.edu/ilqgs/panel113021.mp4}{here} for a particularly passionate one). From the semi-classical perspective, nothing enforces saturation at $S_{\rm BH}$; one could imagine a larger, dynamical internal Hilbert space. We therefore view the Page curve as a conceptual construct that helps frame what the black hole information problem begs of a deeper quantum description.

\section{Analogue Gravity}\label{lecture:6}

The discussion of black-hole evaporation invites a sharper question: which features of the Hawking effect depend on gravity, and which are universal consequences of quantum fields in the presence of horizons (space-time, geometrical objects)? A key conceptual insight dates back to Unruh’s seminal 1981 work on ``sonic'' black holes~\cite{unruh81analogue}, where Unruh realized that the Hawking effect is largely \emph{kinematic}. In particular, horizon-induced mode conversion and the associated mixing of positive- and negative-mode components are sufficient to produce thermal Hawking radiation, largely independent of microscopic short-distance physics of the platform~\cite{CorleyJacobson1996, Visser2003essential, Unruh2005:Universality}. Similar reasoning applies to superradiance in the presence of an ergoregion. Since then, this perspective has blossomed across many physical platforms and arenas, establishing analogue gravity as a framework for realizing---and stress-testing---semi-classical gravitational predictions in laboratory systems~\cite{jacquet2020nextGen, Jacquet2020:PolaritonAnalogues, Almeida2023:AnalogueGravHistory, Barcel2005:LivRvwAgrav}.

\begin{figure}
    \centering
    \includegraphics[width=.7\linewidth]{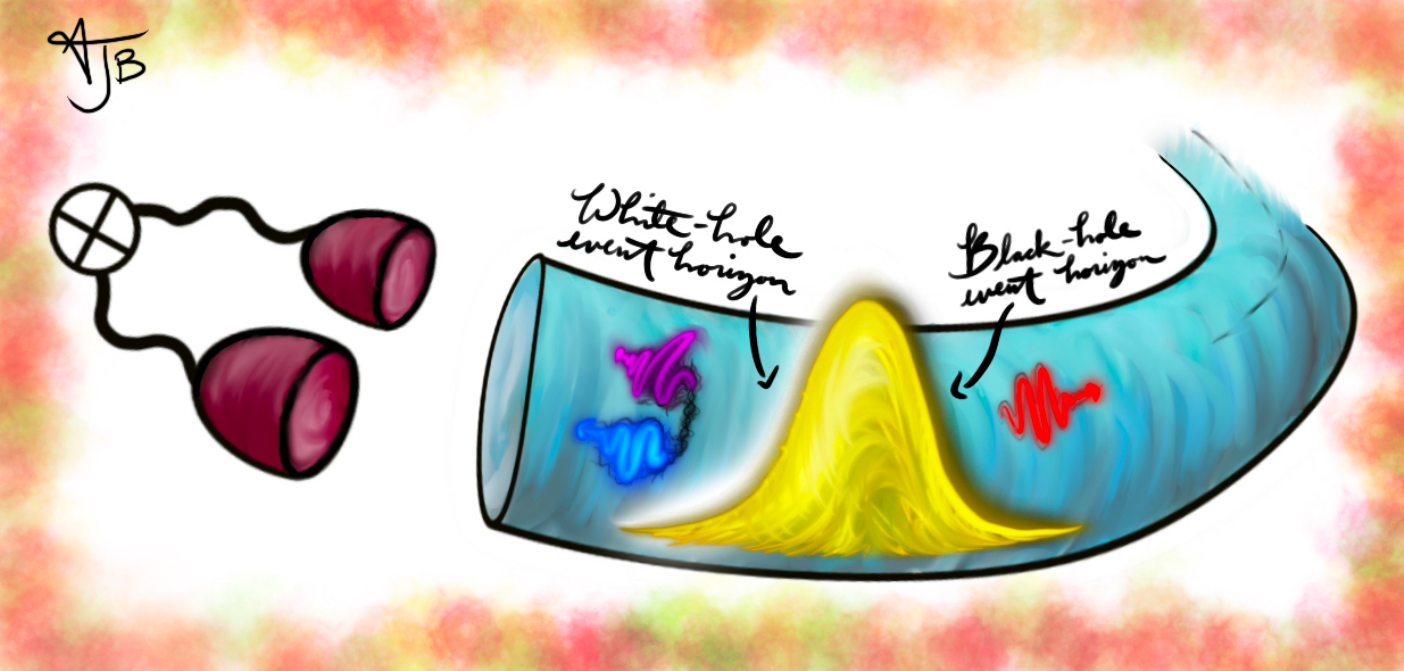}
    \caption{Optical white-black hole analogue. A strong pulse in a dielectric medium locally perturbs the refractive index, creating an effective flow profile for weak probe modes. Its trailing edge forms an analogue white-hole horizon, while the leading edge forms an analogue black-hole horizon. Horizon-induced mode conversion leads to Hawking pair creation at the black-hole horizon (red wave packet), with the partner mode propagating through the interior and seeding correlated emission from the white-hole horizon (purple and blue wave packets). Precision measurements access these correlations, providing a laboratory setting to probe the quantum physics of horizons.}\label{fig:wbh_analog}
\end{figure}

In this section, we discuss the universality of the Hawking effect. We then turn to a concrete optical realization: a traveling refractive-index perturbation in a dielectric medium that creates an effective flow profile supporting a white-black hole pair (Fig.~\ref{fig:wbh_analog}). Using a theoretical dielectric model, we summarize representative numerical results from the literature (based on my own biased explorations) demonstrating spontaneous particle creation at the horizons and the emergence of correlated Hawking pairs.

Finally, we discuss the possibility of analogue horizons and ergoregions in driven-dissipative polaritonic platforms. Analytical, numerical, and experimental work has established parameter regimes in which horizon- and ergoregion-like structures can form, based on the linearized dynamics of low-energy excitations propagating atop of the mean field. We do not pursue that program here. Rather, we adopt a complementary perspective motivated by an outstanding question in the field: How can we \emph{observe} entanglement between Hawking quanta and their partner modes under realistic laboratory conditions? Accordingly, we take space-time--induced pair creation as a working assumption and focus on how decoherence degrades the resulting entanglement. Using an intentionally crude open-system toy model, we designate radiative relaxation as an effective measurement channel and examine how phonon-induced diffusion and thermal fluctuations degrade (or possibly eliminate) the entanglement signatures we hope to measure.

\subsection*{Universality of the Hawking effect}

Remarkably, the Hawking effect is not a niche astrophysical quantum phenomenon but a broadly \textit{universal} one, arising from the curious interplay between quantum fields and event horizons. This universality is often summarized by the statement that ``Hawking radiation is kinematics''~\cite{Visser2003essential}: once an effective horizon is present, thermal particle creation follows almost independently of the microscopic physics at play. This perspective is reinforced by the wide range of systems in which analogue Hawking radiation appears, including classical fluids, Bose-Einstein condensates, polaritonic superfluids, and dielectric optical media. In this section, guided by Refs.~\cite{unruh81analogue,unruh1995DumbHole,CorleyJacobson1996,Visser2003essential,Unruh2005:Universality}, we lay out the universal ingredients of the Hawking effect and sketch how it arises in a generic physical setting.

\begin{quote}
    The essential ingredient of the Hawking effect is the presence of an effective horizon that leads to logarithmic phase pile-up of outgoing, near-horizon modes.
\end{quote}

That is all there is to it. We encountered this mechanism explicitly in Part~\ref{lecture:3} when analyzing the near-horizon ``up'' modes of a semi-classical black hole. This leads to a basic question: What then gives rise to an effective horizon?

A simple and widely applicable framework is provided by the river model of spacetime~\cite{hamilton2008river}. In this picture, we consider a background flow with velocity $\mathfrak{u}$ and a local propagation speed $\mathfrak{c}$, which may represent the speed of light or the fastest characteristic velocity in the medium (e.g., the speed of sound). In Painlev\'e-Gullstrand coordinates, the effective metric takes the form
\begin{equation}
    \dd{s}^2
    = - (\mathfrak{c}^2-\mathfrak{u}^2)\dd{t}^2
    - 2\mathfrak{u}\cdot\dd{\bm x}\dd{t}
    + \dd{\bm x}^2.
\end{equation}
For simplicity, we restrict the flow to a single spatial direction $x$, so that $\mathfrak{u}\cdot\dd{\bm x}=\mathfrak{u}\dd{x}$. An apparent horizon forms when $\abs{\mathfrak{u}}=\abs{\mathfrak{c}}$: the horizon is black-hole--like for $\mathfrak{u}\to-\mathfrak{c}$ and white-hole--like for $\mathfrak{u}\to+\mathfrak{c}$. We assume a black-hole horizon located near $x=x_H$, with $\mathfrak{c}$ and $\mathfrak{u}$ asymptotically constant far from the horizon.

To analyze wave propagation, consider a field $\varphi$ in the geometric-optics approximation,
\begin{equation}
    \varphi(x,t)\approx \mathcal{A}(x,t)e^{-i\phi(x,t)},
\end{equation}
where $\mathcal{A}$ is a slowly varying envelope and $\phi(x,t)=\omega t-\int^x\dd{x^\prime}k(x^\prime)$ is a rapidly varying phase. The phase satisfies the null-ray equation
\begin{equation}
    \mathfrak{g}^{\mu\nu}(\partial_\mu\phi)(\partial_\nu\phi)=0,
\end{equation}
with inverse metric
\begin{equation}
    \mathfrak{g}^{\mu\nu}=\frac{1}{\mathfrak{c}^2}
    \begin{pmatrix}
        -1 & -\mathfrak{u} \\
        -\mathfrak{u} & \mathfrak{c}^2-\mathfrak{u}^2
    \end{pmatrix}.
\end{equation}
This yields the local dispersion relation
\begin{equation}
    (\omega-\mathfrak{u}k)^2=\mathfrak{c}^2k^2,
\end{equation}
or equivalently
\begin{equation}
    \omega-\mathfrak{u}k=\sigma\mathfrak{c}k,
\end{equation}
with $\sigma=\pm1$ labeling outgoing ($\sigma=+1$) and ingoing ($\sigma=-1$) rays.

For ingoing rays,
\begin{equation}
    k_{\rm in}=-\frac{\omega}{\mathfrak{c}-\mathfrak{u}},
\end{equation}
which remains finite at the horizon, allowing these modes to cross smoothly. In contrast, outgoing rays have
\begin{equation}
    k_{\rm out}=\frac{\omega}{\mathfrak{c}+\mathfrak{u}},
\end{equation}
which diverges near the horizon. Linearizing the denominator about $x=x_H$,
\[
\mathfrak{c}+\mathfrak{u}\approx
\eval{\pdv{(\mathfrak{c}+\mathfrak{u})}{x}}_{x_H}(x-x_H)
\triangleq \kappa(x-x_H),
\]
we find
\begin{equation}
    k_{\rm H}\approx\frac{\omega}{\kappa(x-x_H)}.
\end{equation}
As a result, the accumulated phase grows logarithmically,
\[
\int^x\dd{x^\prime}k_{\rm H}\sim\frac{\omega}{\kappa}\ln(x-x_H),
\]
signaling the characteristic phase pile-up responsible for the Hawking effect, with $\kappa$ playing the role of the surface gravity. Quantization of these near-horizon modes then leads to a thermal spectrum with Hawking temperature
\begin{equation}
    T_H=\frac{\kappa}{2\pi}.
\end{equation}

Several remarks are in order. First, the Hawking effect does not depend sensitively on the high-energy completion of the theory and is robust against modifications of short-distance dispersion, although these effects can introduce greybody factors or additional particle-creation channels~\cite{unruh1995DumbHole, CorleyJacobson1996, Unruh2005:Universality}. Second, the energy carried by Hawking quanta is drawn from whatever sustains the horizon---be it the mass of a black hole, the kinetic energy of a flowing medium, or an optical pump. In this sense, classical horizons are generically unstable once quantum effects are included: \textit{Nature abhors horizons}.

\subsection*{Optical analogue white-black holes}

\begin{figure}
    \centering
    \includegraphics[width=\linewidth]{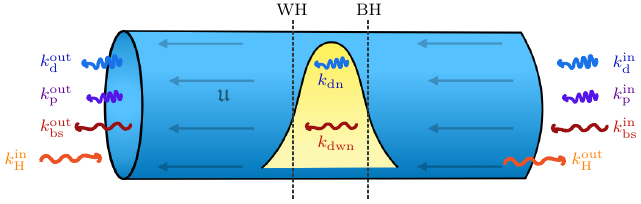}
    \includegraphics[width=.6\linewidth]{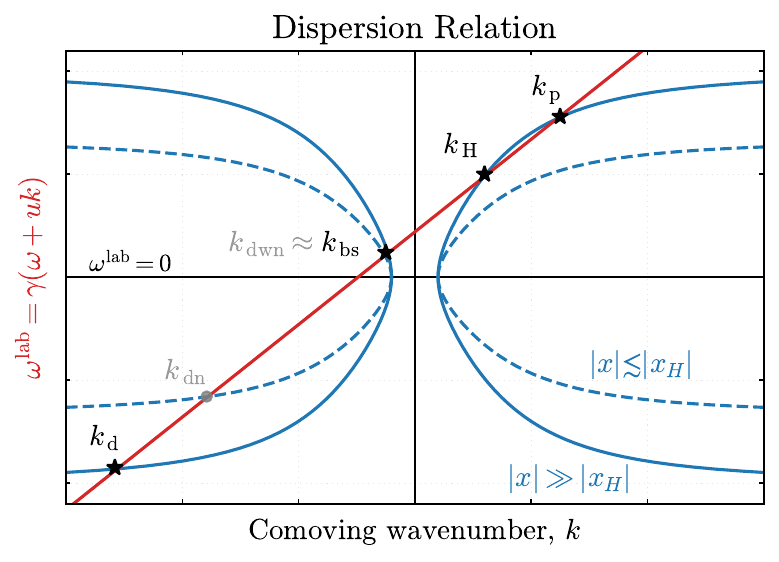}
    \caption{Modes of an analogue optical white-black hole. (Top) Physical illustration of a white-black hole pair generated by a strong pulse in a dielectric medium, as seen in the frame comoving with the pulse at speed $\mathfrak{u}$. (Bottom) Dispersion relation in the comoving frame far away from ($\abs{x}\gg \abs{x_H}$) and in the interior of ($\abs{x}\lesssim \abs{x_H}$) the effective horizons, at fixed comoving frequency $\omega$. Asymptotic modes are $k_{\rm d},k_{\rm p},k_{\rm bs}$, and $k_{\rm H}$. Interior modes $k_{\rm dn},k_{\rm dwn}$ also shown.}
    \label{fig:wbh_pulse}
\end{figure}

We now focus on optical white- and black-hole analogue horizons in dielectric media. The essential ingredient is a strong optical pulse propagating at speed $\mathfrak{u}$ that locally perturbs the refractive index through nonlinear optical effects. This perturbation modifies the propagation speed of weak probe modes. If the pulse is sufficiently strong, a white-hole horizon forms on its trailing edge and a black-hole horizon on its leading edge; see Fig.~\ref{fig:wbh_pulse}. 

For concreteness, we consider the theoretical model introduced by Linder, Sch\"utzhold, and Unruh~\cite{lsu2016hopfield}, together with its extension to pulse-like profiles that support white--black hole pairs~\cite{agullo2022prl, brady2022SympCircs, agullo2023RobustAnalogue}. Our aim here is to highlight the main physical ingredients rather than provide a complete derivation. Experimental realizations of optical analogue horizons are discussed in Refs.~\cite{philbin08, drori19, rosenberg2020optical}.

\paragraph*{Theoretical model and wave equation.}
Following Ref.~\cite{lsu2016hopfield}, we model the dielectric medium as a continuum of polarizable material oscillators, each with resonance frequency $\Omega$, coupled to the electromagnetic field. For simplicity, we work in $1+1$ dimensions. In the continuum limit, the Lagrangian density in the laboratory frame is
\begin{equation}\label{eq:lab_lagrangian}
    \mathcal{L}^{\rm lab}
    =\frac{1}{2}\!\left((\partial_tA)^2-(\partial_xA)^2\right)
    +\frac{1}{2}\!\left((\partial_t\Psi)^2-\Omega^2\Psi^2\right)
    +g\,\Psi\,(\partial_t A),
\end{equation}
where $A$ is the vector potential of weak electromagnetic probe fields (a.k.a., low-energy photonic excitations) and $\Psi$ represents the collective polarization (oscillator) field of the medium. The coupling constant $g$ encodes the linear optical response of the material.\footnote{This may be compared to the Lagrangian of a single polarizable molecule, where the dipolar displacement couples linearly to the electric field.}  

Nonlinear effects induced by the strong pulse are incorporated phenomenologically by allowing the resonance frequency to depend on the comoving coordinate via
\[
\Omega(x-\mathfrak{u}t)=\Omega_0+\delta\Omega(x-\mathfrak{u}t),
\]
where $\delta\Omega$ is localized around the pulse and travels with it at speed $\mathfrak{u}$.

Away from the pulse, the Lagrangian~\eqref{eq:lab_lagrangian} yields a subluminal dispersion relation for electromagnetic waves. In particular, for plane waves of lab-frame frequency $\omega^{\rm lab}$, the refractive index is approximately
\begin{equation}
    n^{\rm lab}(\omega^{\rm lab})
    =\sqrt{1+\frac{g^2}{\Omega^2-(\omega^{\rm lab})^2}}.
\end{equation}
Lower-frequency modes propagate faster, with the maximal phase velocity attained in the $\omega^{\rm lab} \to 0$ limit, $n^{\rm lab}(0)=\sqrt{1+g^2/\Omega^2}$. Local modifications of $\Omega$ induced by the pulse therefore deform the dispersion relation and give rise to optical horizons.

\begin{example}{Wave equations in the lab frame}
From the Lagrangian~\eqref{eq:lab_lagrangian}, the equations of motion for the electromagnetic and polarization fields are
\begin{align}
    \partial_t^2 A - \partial_x^2 A &= -g\,\partial_t \Psi, \\
    \partial_t^2 \Psi + \Omega^2 \Psi &= g\,\partial_t A .
\end{align}
Assuming a slowly varying envelope for $\Psi$,
$\Psi\approx\bar{\Psi}e^{-i\phi}$ with
$\phi=\int^t\dd{t'}\,\omega^{\rm lab}(t')$,
and that $\omega^{\rm lab}$ is the fastest timescale, one finds
\[
\Psi\approx \frac{g\,\partial_t A}{\Omega^2-(\omega^{\rm lab})^2},
\qquad
\partial_t\Psi\approx
\frac{g\,\partial_t^2 A}{\Omega^2-(\omega^{\rm lab})^2}.
\]
Substituting back into the wave equation for $A$ gives
\begin{equation}
    \partial_t^2 A - \partial_x^2 A
    \approx
    -\frac{g^2}{\Omega^2-(\omega^{\rm lab})^2}\,\partial_t^2 A,
\end{equation}
or equivalently
\[
(n^{\rm lab})^2\,\partial_t^2 A-\partial_x^2 A\approx0,
\]
which is the wave equation of a dielectric medium with refractive index $n^{\rm lab}$.
\end{example}

To analyze horizon formation, it is convenient to move to the frame comoving with the pulse. Introducing co-moving coordinates
\begin{align}
    \tau &= \gamma (t-\mathfrak{u}x),\\
    \chi &= \gamma (x-\mathfrak{u}t),
\end{align}
with Lorentz factor $\gamma=(1-\mathfrak{u}^2)^{-1/2}$, the Lagrangian density becomes
\begin{equation}
\mathcal{L}^{\rm co}
=\frac{1}{2}\!\left((\partial_{\tau}A)^2-(\partial_{\chi}A)^2\right)
+\frac{1}{2}\!\left(\gamma^2(\partial_{\tau}\Psi-\mathfrak{u}\partial_{\chi}\Psi)^2-\Omega^2\Psi^2\right)
+\gamma g\,\Psi(\partial_{\tau}A-\mathfrak{u}\partial_{\chi}A).
\end{equation}
In this frame the pulse-induced perturbation $\delta\Omega(\chi)$ is static but spatially varying. The comoving frequency $\omega$ is conserved, while spatial inhomogeneity near the pulse induces scattering between modes with different comoving wavenumbers $k_i(\omega)$.

Eliminating $\Psi$ from the linearized equations yields the comoving-frame dispersion relation
\begin{equation}\label{eq:comoving_dispersion}
    \underbrace{\gamma(\omega+\mathfrak{u}k)}_{=\omega^{\rm lab}} = \Omega\sqrt{1+\frac{g^2}{\omega^2-k^2-g^2}},
\end{equation}
which is transcendental in $k$ for fixed $\omega$. The solutions define the asymptotic modes $(k_{\rm d},k_{\rm p},k_{\rm bs},k_{\rm H})$ shown in Fig.~\ref{fig:wbh_pulse}, which form the in/out basis for the scattering problem.

\begin{example}{Wave equations in the comoving frame}
From $\mathcal{L}^{\rm co}$, the equations of motion read
\begin{align}
    \partial_\tau^2 A - \partial_\chi^2 A
    &= -g\gamma(\partial_\tau-\mathfrak{u}\partial_\chi)\Psi,\label{eq:comoving_waveEq_A} \\
    \gamma^2(\partial_\tau-\mathfrak{u}\partial_\chi)^2\Psi
    +\Omega^2(\chi)\Psi
    &= g\gamma(\partial_\tau-\mathfrak{u}\partial_\chi)A. \label{eq:comoving_waveEq_Psi}
\end{align}
A slowly varying envelope analysis, analogous to that in the lab frame, leads directly to the dispersion relation~\eqref{eq:comoving_dispersion}.
\end{example}

\paragraph*{Effective metric and horizon.}
The asymptotic in/out mode basis at fixed comoving frequency $\omega$ is determined by the real roots $k_i(\omega)$ of the comoving dispersion relation~\eqref{eq:comoving_dispersion}, together with their group velocities
\[
v_g(k_i)=\dv{\omega}{k}\Big|_{k_i}.
\]
With our phase convention $A(\chi,\tau)\propto e^{-i\omega\tau}e^{+ik\chi}$, the sign of $v_g$ fixes the propagation direction: $v_g>0$ corresponds to motion toward increasing $\chi$, while $v_g<0$ corresponds to motion toward decreasing $\chi$. Far away from the pulse, where $\Omega(\chi)$ is constant, one finds four real solutions $k_{\rm d},k_{\rm p},k_{\rm bs},k_{\rm H}$ (Fig.~\ref{fig:wbh_pulse}), and they separate into branches that propagate toward/away from the pulse on each side. Operationally: observers placed at fixed $\chi\to-\infty$ can prepare incoming wavepackets on the branch(es) that propagate toward the pulse and analyze the outgoing radiation on the counterpropagating branches; likewise for observers at $\chi\to+\infty$. We will not need a detailed taxonomy beyond the fact that one branch, $k_H$, plays the role of the outgoing Hawking channel on the black-hole side.

As the pulse is approached, the local resonance $\Omega(\chi)$ is reduced, deforming the dispersion and, in particular, modifying which branches remain propagating inside the pulse region. In the Hopfield-LSU moel, the horizon is most cleanly characterized at low energies: the macroscopic (long-wavelength) mode experiences an effective wave speed $c_{\rm low}(\chi)$ that depends on $\Omega(\chi)$, and a horizon occurs where the pulse velocity matches this speed. In other words, one identifies a black-hole (white-hole) horizon where $c_{\rm low}(\chi)$ crosses $\mathfrak{u}$ with the appropriate sign, so that low-energy signals cannot escape from (enter into) the corresponding side. Dispersion then determines how the low-energy horizon is embedded into the full multibranch scattering problem, leading to greybody effects (and additional mode conversion channels in some regimes). 

Finally, note that in dispersive optical analogues the asymptotic mode norm is determined by the sign of
the laboratory-frame frequency $\omega^{\rm lab}=\gamma(\omega+\mathfrak{u}k)$: asymptotically, modes with
$\omega^{\rm lab}<0$ have negative norm, so any mixing between positive- and negative-norm branches
generically leads to spontaneous particle creation. In the regime of interest, the dominant low-frequency
pair creation can be interpreted as Hawking emission into the outgoing $k_H$ branch, with a partner carried
by the negative-norm channel.

\begin{example}{Effective metric and surface gravity}
To make the horizon concept explicit, it is convenient to pass to a long-wavelength (low-energy) effective description. If $\Omega(\chi)$ varies slowly and we neglect higher-order time derivatives, the polarization field can be adiabatically eliminated from the Hopfield equations, giving the approximate constitutive relation
\[
\Psi \approx \frac{g}{\Omega(\chi)^2}\,\partial_t A
\]
Substituting back into the action yields the effective Lagrangian density for the effective field,
\begin{equation}
\label{eq:Leff_low}
\mathcal{L}_{\rm eff}
=\frac12\Big(1+\frac{g^2}{\Omega(\chi)^2}\Big)\,(\partial_t A)^2
-\frac12(\partial_x A)^2,
\end{equation}
which identifies the low-energy refractive index and wave speed via
\[
\frac{1}{c_{\rm low}(\chi)^2}=1+\frac{g^2}{\Omega(\chi)^2},
\qquad
n_{\rm low}(\chi)=\sqrt{1+\frac{g^2}{\Omega(\chi)^2}}.
\]
In the laboratory frame, Eq.~\eqref{eq:Leff_low} is equivalent (up to the usual subtleties about conformal factors in $1{+}1$ dimensions) to a scalar field propagating in an effective metric. After boosting to the pulse frame $(\tau,\chi)$ with velocity $\mathfrak{u}$, a convenient form of the effective line element is
\begin{equation}
\label{eq:metric_eff_pulse}
ds_{\rm eff}^2
= \gamma^2\!\left(1-\mathfrak{u}^2\Big[1+\frac{g^2}{\Omega(\chi)^2}\Big]\right)d\tau^2
-2\mathfrak{u}\gamma^2\frac{g^2}{\Omega(\chi)^2}\,d\tau\,d\chi 
-\gamma^2\!\left(1+\frac{g^2}{\Omega(\chi)^2}-\mathfrak{u}^2\right)d\chi^2.
\end{equation}
From Eq.~\eqref{eq:metric_eff_pulse}, the horizon condition is simply $g^{\rm eff}_{00}(\chi_H)=0$, i.e.
\[
1-\mathfrak{u}^2\Big(1+\frac{g^2}{\Omega(\chi_H)^2}\Big)=0
\qquad \Longleftrightarrow\qquad
c_{\rm low}(\chi_H)=|\mathfrak{u}|.
\]
Define the gradient parameter
\[
\xi \triangleq \left.\frac{\partial_\chi \Omega}{\Omega}\right|_{\chi=\chi_H}.
\]
We then find, in the co-moving frame, the standard Killing-horizon expression
\[
\kappa=\frac12\left|\frac{\partial_\chi g^{\rm eff}_{00}}{g^{\rm eff}_{01}}\right|_{\chi=\chi_H}
=\mathfrak{u}\,\xi,
\qquad\Rightarrow\qquad
T_H=\frac{\kappa}{2\pi}=\frac{\mathfrak{u}\,\xi}{2\pi}.
\]
Thus the Hawking temperature is set by the relative steepness of the pulse-induced profile, encoded in the relative gradient $\partial_\chi\ln\Omega$ and evaluated at the horizon. 
\end{example}

\paragraph*{Input-output scattering matrix.}
For a realistic pulse profile that generates both white- and black-hole horizons, a fully analytic treatment of mode scattering is generally intractable. In recent work~\cite{agullo2022prl,brady2022SympCircs,agullo2023RobustAnalogue}, my colleagues and I therefore relied on numerical integration of the coupled wave equations to analyze scattering from both sides of the pulse. Guided by the near-horizon kinematics and by the structure of the asymptotic mode basis, it is possible to construct a simple analytic \emph{Ansatz} for the scattering matrix (and corresponding symplectic diagram) that captures the dominant physics of the white-black hole system.

\begin{figure}
    \centering
    \includegraphics[width=.8\linewidth]{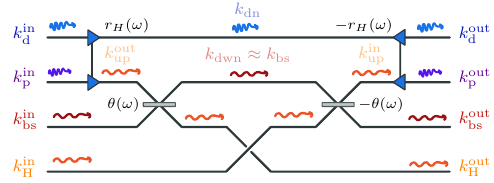}
    \caption{Symplectic diagram for the optical white-black hole pair, representative of the scattering matrix $\bm S_{\rm WB}$. The Hawking squeezing parameter $r_H$ and mixing angle $\theta$ encode spontaneous pair creation and greybody scattering via $\tanh^2 r_H(\omega)=e^{-\omega/T_H}$ and $\cos^2\theta(\omega)=\Gamma_\omega$. The white-hole sector is described by the time-reversed parameters $-r_H$ and $-\theta$. Ingoing and outgoing $k_{\rm H}$ modes do not mix, highlighting the absence of elastic reflection in the Hawking channel. Near-horizon ``up'' modes and interior ``dn'' and ``dwn'' modes are not asymptotically well-defined, but are shown schematically to emphasize correspondence with standard white/black-hole mode structures.}
    \label{fig:wbh_circuit}
\end{figure}

The resulting symplectic diagram, shown in Fig.~\ref{fig:wbh_circuit}, is built by composing elementary transformations that separately describe scattering in black-hole and white-hole geometries. Working in the mode ordering $(k_{\rm d},k_{\rm bs},k_{\rm H},k_{\rm p})$, and using the shorthand ${\rm s}(x)=\sin x$, ${\rm c}(x)=\cos x$, ${\rm sh}(x)=\sinh x$, ${\rm ch}(x)=\cosh x$, the full symplectic transformation mapping incoming to outgoing
quadratures takes the form
\begin{equation}
\label{eq:SWB}
\bm S_{\rm WB}= 
\begin{pmatrix}
 (1+{\rm c}^2\!\theta\,{\rm sh}^2 r_H)\bm{I}_2
 & {\rm c}\theta\,{\rm s}\theta\,{\rm sh} r_H\,\bm Z
 & -{\rm c}\theta\,{\rm sh} r_H\,\bm Z
 & {\rm c}^2\!\theta\,{\rm ch} r_H\,{\rm sh} r_H\,\bm Z\\
 -{\rm c}\theta\,{\rm s}\theta\,{\rm sh} r_H\,\bm Z
 & {\rm c}^2\!\theta\,\bm{I}_2
 & {\rm s}\theta\,\bm{I}_2
 & -{\rm c}\theta\,{\rm s}\theta\,{\rm ch} r_H\,\bm{I}_2\\
 {\rm c}\theta\,{\rm sh} r_H\,\bm Z
 & {\rm s}\theta\,\bm{I}_2
 & 0
 & {\rm c}\theta\,{\rm ch} r_H\,\bm{I}_2\\
 -{\rm c}^2\!\theta\,{\rm ch} r_H\,{\rm sh} r_H\,\bm Z
 & -{\rm c}\theta\,{\rm s}\theta\,{\rm ch} r_H\,\bm{I}_2
 & {\rm c}\theta\,{\rm ch} r_H\,\bm{I}_2
 & (\sin^2\!\theta-{\rm c}^2\!\theta\,{\rm sh}^2 r_H)\bm{I}_2
\end{pmatrix},
\end{equation}
where $\bm Z=\mathrm{diag}(1,-1)$. The parameters $r_H$ and $\theta$ encode Hawking pair creation and greybody
mixing, respectively, via
\[
\tanh^2 r_H(\omega)=e^{-\omega/T_H},
\qquad
\cos^2\theta(\omega)=\Gamma_\omega,
\]
with $T_H$ the Hawking temperature and $\Gamma_\omega$ the greybody factor for the relevant channel.

\begin{figure}
    \centering
    \includegraphics[width=.6\linewidth]{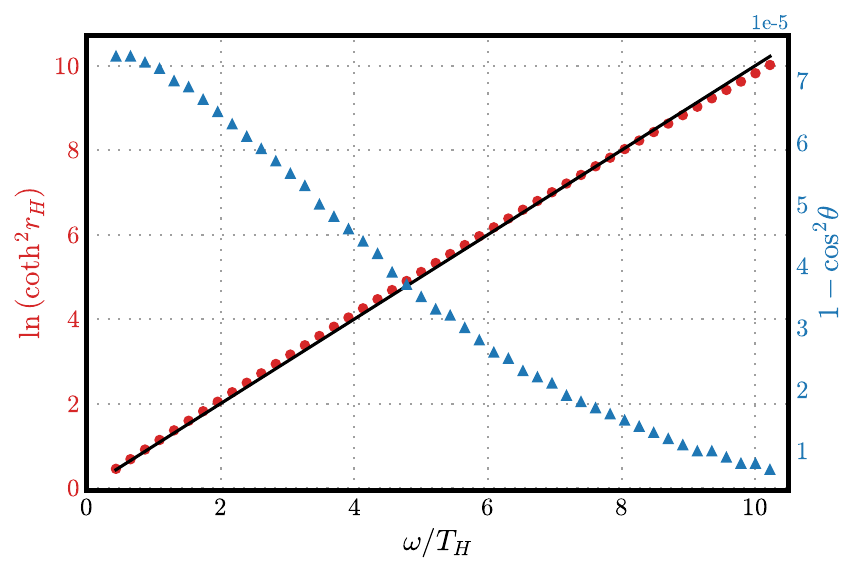}
    \caption{Hawking temperature $T_H$ and greybody factor $\Gamma_\omega=\cos^2\theta$ extracted from numerical scattering data as functions of the comoving frequency $\omega$. The outgoing radiation in the Hawking channel is well described by a thermal spectrum at temperature $T_H$, with deviations captured by the frequency-dependent greybody factor $\Gamma_\omega$. Adapted from Ref.~\cite{agullo2022prl}.}
    \label{fig:TH_data}
\end{figure}

We extract $T_H$ and $\Gamma_\omega$ from the numerical scattering data by fitting the outgoing spectrum at each comoving frequency $\omega$ (see Fig.~\ref{fig:TH_data}). Substituting these numerically obtained parameters into the analytic matrix~\eqref{eq:SWB}, we find excellent agreement between the \textit{Ansatz} and direct numerical results for observables that are invariant under local phase rotations, such as particle numbers and entanglement measures~\cite{agullo2022prl,brady2022SympCircs}. Local phase rotations merely redefine the quadrature basis, e.g.\ $q_i\to q_i\cos\phi+p_i\sin\phi$, and do not affect these quantities.

Beyond its quantitative accuracy, the \textit{Ansatz} offers immediate qualitative insight. For instance, it makes explicit how thermal Hawking radiation emitted by the black-hole horizon seeds subsequent pair creation at the white-hole horizon through the interior partner mode. In this sense, the white-hole emission is not independent, but is driven by the black-hole process.

The fact that this compact symplectic circuit captures the essential scattering physics is highly nontrivial. A general Gaussian transformation acting on four modes corresponds to an element of the real symplectic group ${\rm Sp}(8,\mathbb{R})$, which has dimension $N(2N+1)=36$ for $N=4$ modes. Here, however, the dominant scattering behavior over a broad frequency range is effectively governed by just two parameters, $T_H(\omega)$ and $\Gamma_\omega$. This dramatic reduction reflects the strong constraints imposed by near-horizon kinematics and the universality of the Hawking process.

The analysis thus far is idealized. We have neglected absorption arising from material resonances, which becomes important when $\omega^{\rm lab}\approx\Omega$ and which attenuates optical signals. We have also assumed a regime in which the pulse profile is sufficiently steep that the analogy to a semi-classical spacetime is reliable. If the latter is violated, additional effects, such as partial tunneling of the $k_H$ mode across the white-hole horizon, may occur. While this tunneling is forbidden in genuine spacetime geometries by causal structure, it appears to be a prominent feature in current optical analogue experiments~\cite{philbin08,drori19}.

\paragraph*{Optical horizons in thermal baths.}
In realistic laboratory settings, the Hawking process necessarily occurs in the presence of ambient thermal fluctuations. When the background temperature is comparable to the Hawking temperature, one might expect the entanglement between Hawking quanta and their partner modes to degrade or vanish. This expectation is borne out in simple toy models, where thermal seeding of the ingoing modes rapidly suppresses entanglement generation (cf.\ Part~\ref{lecture:3}, Example~7).

Remarkably though, optical analogue systems evade this conclusion. As shown in Ref.~\cite{agullo2023RobustAnalogue}, the entanglement generated at optical horizons is extraordinarily robust against thermal noise that is stationary in the laboratory frame. The key reason is kinematic: optical horizons move through the medium at velocities close to the speed of light, and dispersion ensures that different asymptotic modes correspond to vastly different laboratory-frame frequencies. As a result, thermal fluctuations populate the relevant ingoing channels unevenly, strongly suppressing noise in the modes that seed Hawking pair creation.

Concretely, although all asymptotic modes share the same comoving frequency $\omega$, their laboratory-frame frequencies $\omega_i^{\rm lab}$ differ by many orders of magnitude. Since thermal occupation depends only on the ratio $|\omega_i^{\rm lab}|/T_{\rm env}^{(\rm lab)}$, this leads to a strong hierarchy in the initial populations: the high-frequency progenitor modes responsible for Hawking emission are essentially unoccupied unless the laboratory temperature is extremely large. In this sense, optical analogues closely mirror the astrophysical situation, where Hawking progenitor modes originate at
exponentially high frequencies in the distant past.

\begin{figure}
    \centering
    \includegraphics[width=.55\linewidth]{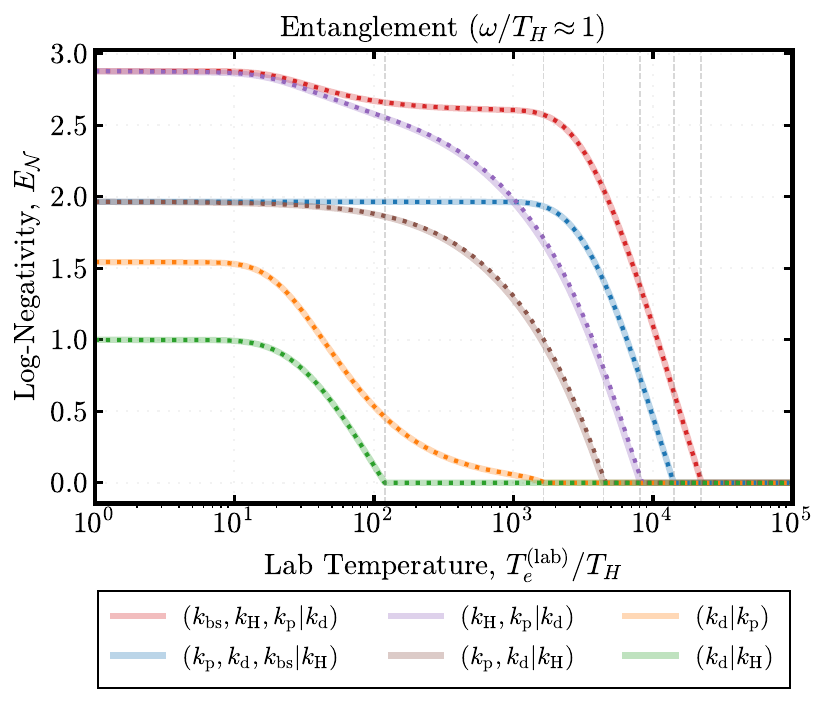}
    \caption{Entanglement, quantified by the log-negativity, as a function of the background laboratory temperature $T_{\rm env}^{(\rm lab)}$ for several mode bipartitions at comoving frequency $\omega\approx T_H$ ($T_H\approx3.51\,\mathrm{K}$). Solid curves are obtained from the \textit{Ansatz} scattering matrix $\bm S_{\rm WB}$, while dotted curves are extracted from numerical scattering data. Adapted from Ref.~\cite{agullo2023RobustAnalogue}.}
    \label{fig:labframe_ln}
\end{figure}

The effect of thermal noise can be quantified by propagating an initial thermal state through the white-black hole scattering matrix $\bm S_{\rm WB}$. While the technical analysis is most conveniently performed in the comoving frame, the results are frame-independent, since Lorentz boosts do not mix the asymptotic modes. Figure~\ref{fig:labframe_ln} shows the resulting entanglement, quantified by the log-negativity, as a function of the laboratory temperature. Strikingly, for a Hawking temperature $T_H\approx3.5\,$K, entanglement persists even when the ambient lab temperature exceeds $T_H$ by several orders of magnitude. In particular, bipartitions involving the negative-norm progenitor mode retain entanglement up to $T_{\rm env}^{(\rm lab)}\sim10^4 T_H$.

This endurance of entanglement has a close analogue in the astrophysical case. For black holes formed by gravitational collapse, the Hawking progenitor modes are exponentially blueshifted relative to outgoing modes at future null infinity. Consequently, thermal fluctuations in the asymptotic past affect Hawking radiation only if their temperature is itself exponentially large. In optical systems, the role of this gravitational blueshift is played by a combination of Lorentz boosting between the lab and comoving frames and material dispersion. Although the resulting frequency separation is not trans-Planckian, it is sufficiently large to protect Hawking-generated entanglement against realistic laboratory noise.

Taken together, these results suggest that optical analogue platforms are not only capable of generating Hawking radiation, but may also allow direct experimental access to horizon-induced entanglement---even in environments far from cryogenic conditions.

\subsection*{Polariton superfluid as a QFCS simulator}

Driven \href{https://en.wikipedia.org/wiki/Polariton_superfluid}{exciton-polariton fluids} provide a versatile platform for emulating QFCS~\cite{Jacquet2020:PolaritonAnalogues}. Experienced experimentalists control the system primarily through the spatial profile---intensity and phase---of the coherent optical pump, which establishes the steady-state density and flow of the polariton fluid. Low-energy collective excitations propagate atop this background and, in suitable regimes, behave as an ``acoustic'' field on an emergent spacetime geometry. 

The platform also offers pristine quantum control and detection capabilities: the photonic component of the polariton grants direct radiative access to the excitation sector, enabling experimentalists to engineer quantum states of the probe light in low-energy modes and to perform high-efficiency photon counting and homodyne measurements on the outgoing radiation fields. At the same time, losses remain relatively small and well characterized, with decoherence arising primarily from the dissipative nature of the semiconductor housing rather than from uncontrolled environmental coupling of the radiation field.

Recent experiments have demonstrated tailored subsonic and supersonic flows and have directly observed negative-norm branches of the excitation spectrum in the supersonic region, providing a sharp diagnostic of horizon/ergoregion physics in a quantum fluid of light~\cite{Falque2025:PolaritonSimulator}. Complementary theoretical analyses of rotational flows predict superradiant mode-mixing accompanied by quantum correlations between outgoing channels, highlighting that radiative decay into observable optical channels plays a stabilizing role rather than acting as a nuisance~\cite{Delhom2024:EntSuperrad}.

\begin{figure}
    \centering
    \includegraphics[width=0.45\linewidth]{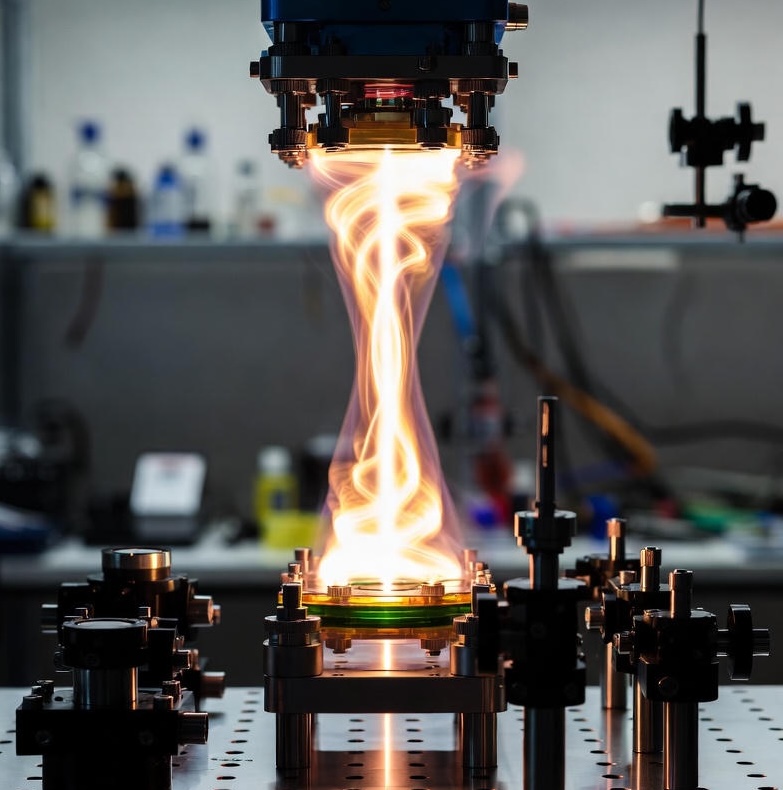}
    \caption{Artistic depiction of an analogue setup in a driven exciton--polariton superfluid, representing a vortex-like flow induced by a structured optical pump. Generated with Grok by xAI.}
    \label{fig:polariton-art}
\end{figure}

From this perspective, the relevant degrees of freedom in this problem are hybrid Bogoliubov–de Gennes (BdG) excitations, which carry the quantum correlations of interest. These excitations decay radiatively into optical output channels that can be directly measured; hence, radiative relaxation provides the mechanism by which we access the BdG sector. At the same time, environmental decoherence, such as lattice vibrations of the semi-conductor housing and thermal fluctuations, acts directly on the BdG excitations and therefore degrade, or even eliminate, the otherwise observable entanglement.

In this section, we do \emph{not} pursue a full microscopic derivation of horizon or ergoregion physics from the underlying exciton-polariton dynamics (see detailed notes of Ref.~\cite{Giacobino2025:BenasqueNotes}). Instead, we adopt a complementary perspective motivated by an outstanding challenge in analogue gravity: Under what conditions can we observe entanglement between superradiant (or Hawking-like) quanta and their partner modes? Accordingly, we take the existence of horizon- or ergoregion-induced pair creation as a working assumption and use an open-system toy model to assess the interplay between radiative decay, phonon-induced decoherence, and spurious thermal excitations in an idealized experimental setup. This allows us to analyze practically relevant scenarios using the tools developed here in a transparent and streamlined manner.

\paragraph*{Wave equation for low-energy excitations.}
We may describe the driven polariton condensate using a driven-dissipative Gross-Pitaevskii equation for the strong, coherent mean-field $\psi(\bm r,t)$. To access low-energy Bogoliubov excitations, we separate the mean field $\psi_0=\sqrt{\rho_0}\,e^{i\phi_0}$ from small fluctuations and work in the hydrodynamic regime. In this limit, the dynamics reduces to a wave equation for a single fluctuation field $\delta \psi$ (a linearized BdG mode) propagating on the background density $\rho_0(\bm r)$ and flow velocity $\bm v_0(\bm r)=\frac{\hbar}{m}\nabla\phi_0$,
\begin{equation}
-\partial_t\!\left[\frac{\rho_0}{c_s^2}\big(\partial_t\delta \psi+\bm v_0\!\cdot\nabla\delta \psi\big)\right]
+\nabla\!\cdot\!\left[\rho_0\nabla\delta \psi-\frac{\rho_0\bm v_0}{c_s^2}\big(\partial_t\delta \psi+\bm v_0\!\cdot\nabla\delta \psi\big)\right]=0,
\label{eq:acoustic-wave}
\end{equation}
where $c_s(\bm r)\triangleq\sqrt{\hbar g\rho_0(\bm r)/m}$ denotes the local speed of sound. We equivalently write this equation as a massless Klein-Gordon equation on an effective curved spacetime,
\begin{equation}
\frac{1}{\sqrt{-g}}\partial_\mu\!\left(\sqrt{-\mathfrak{g}}\,\mathfrak{g}^{\mu\nu}\partial_\nu\delta \psi\right)=0,
\label{eq:KG-acoustic}
\end{equation}
with the acoustic metric $\mathfrak{g}^{\mu\nu}$ fulled determined by $\rho_0$, $c_s$, and $\bm v_0$. The explicit form of the wave equation appears intimidating at first glance, but it is nonetheless linear in $\delta \psi$. Consequently, quantization of the perturbation $\delta \psi$ places this BdG excitation sector squarely within the realm of Gaussian physics.

\paragraph*{Stationary vortex flow.}
We specialize to an axisymmetric steady state with azimuthal flow, $\bm v_0(r)=v_\theta(r)\,\bm e_\theta$, and a negligible radial component. Outside the vortex core, we adopt the standard ansatz $v_\theta(r)=\hbar\ell/(mr)$, up to platform-dependent corrections.\footnote{See, e.g., Ref.~\cite{Carusotto2013:rvw} for discussions of quantized vortices and collective excitations in polariton superfluids.} Assuming stationarity, we separate variables, viz., $\delta\psi(t,r,\theta)=e^{-i\omega t}e^{im\theta}\chi_m(r)$, so that the advection term $\bm v_0\!\cdot\nabla\delta\psi$ produces the familiar Doppler shift $\omega\to\omega-m v_\theta(r)/r$. An \emph{ergoregion} emerges wherever the flow speed exceeds the local sound speed, $|\bm v_0(r)|>c_s(r)$. For fixed $(\omega,m)$, distinct ingoing and outgoing solutions appear on either side of the ergoregion, with positive- and negative-norm branches that mix upon scattering near the ergoregion barrier. This mode mixing underlies analogue quantum superradiance in rotating polariton fluids~\cite{Delhom2024:EntSuperrad}. We do not attempt to solve Eq.~\eqref{eq:acoustic-wave} for any specific profile; instead, we use it only to justify the existence of positive/negative-norm mode mixing and associated pair creation that we treat phenomenologically in the toy open-system model below.

\begin{figure}
    \centering
    \includegraphics[width=0.5\linewidth]{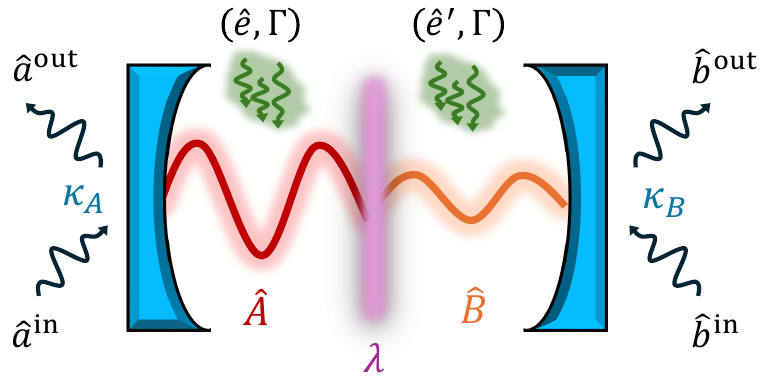}
    \caption{Schematic of the input-output toy model. Trapped (effective BdG) modes $\hat A$ and $\hat B$ represent the dominant quantum superradiant scattering channels associated with an ergoregion at fixed laboratory frequency. Superradiance is modeled phenomenologically as parametric mode mixing (two-mode squeezing) at rate $\lambda$. Radiative decay couples the BdG modes to photonic channels $\hat a$ and $\hat b$ at rates $\kappa_A$ and $\kappa_B$, which dissipatively stabilizes the trapped ergoregion and permits access to quantum correlations of the BdG modes. Phonon-induced diffusion, originating from deformations of the semiconductor housing, are modeled by iid Gaussian baths $\hat e$ and $\hat e'$.}

    \label{fig:toy-model}
\end{figure}

\paragraph*{Radiative readout and decoherence.}
Polariton BdG modes form an intrinsically open system. Radiative decay supplies a direct measurement channel: intra-cavity excitations map to extra-cavity photons, allowing us to access steady-state correlations of the BdG sector in the emitted light~\cite{Delhom2024:EntSuperrad, Falque2025:PolaritonSimulator}. Although, at the same time, BdG excitations couple to lattice vibrations and other material degrees of freedom~\cite{Frerot2023:ThermalPolaritons}; this introduces diffusion-like noise that decoheres the excitation sector and degrades observable entanglement signatures.

\begin{example}{Phonon-induced diffusion (sketch).}
A microscopic source of decoherence in semiconductor polariton platforms is exciton-phonon scattering via deformation-potential coupling~\cite{Frerot2023:ThermalPolaritons},
\begin{equation}
    \hat{\mathcal V}_{xp}
    = i\sum_{\bm q,k_z} g_{xp}(\bm q,k_z)
    \big(\hat c_{\bm q,k_z}-\hat c_{-\bm q,k_z}^\dagger\big)
    \sum_{\bm q'} \hat b_{\bm q'+\bm q}^\dagger \hat b_{\bm q'} ,
\end{equation}
where \(\hat b_{\bm q}\) annihilates an exciton and \(\hat c_{\bm q,k_z}\) annihilates a (longitudinal) phonon. Linearizing about a macroscopically occupied exciton component, \(\hat b_{\bm q}=\alpha_{\bm q}+\delta\hat b_{\bm q}\), and retaining terms first order in \(\delta\hat b\) yields
\begin{equation}
\label{eq:Vxp_bogo_succinct}
    \hat{\mathcal V}_{xp}
    \approx i\sqrt{n_x}\sum_{\bm q,k_z} g_{xp}(\bm q,k_z)
    \big(\hat c_{\bm q,k_z}-\hat c_{-\bm q,k_z}^\dagger\big)
    \Big(\delta\hat b_{\bm q_p-\bm q}
    +\delta\hat b_{\bm q_p+\bm q}^\dagger\Big),
\end{equation}
with \(n_x\) the exciton density and \(\bm q_p\) the pump momentum. Equation~\eqref{eq:Vxp_bogo_succinct} has the generic structure ``phonon quadrature \(\times\) exciton quadrature,'' so tracing over a broadband phonon reservoir produces an effectively Markovian noisy drive on the BdG quadratures, a.k.a., diffusion. 

This motivates the phenomenological coupling that we employ in the toy model below,
\begin{equation}
    \hat H_E(t)
    = i\sqrt{\Gamma}\,
    \big(\hat e_{\rm in}^\dagger(t)-\hat e_{\rm in}(t)\big)
    \big(\hat A(t)+\hat A^\dagger(t)\big),
\end{equation}
(and similarly for \(\hat B\)), where \(\Gamma\) is an effective phonon-induced diffusion rate and \(\hat e_{\rm in}(t)\) is a Gaussian bath operator. To make this correspondence more explicit, we emphasize that excitons do not couple to a single phonon mode but to a continuum (bath) of lattice vibrations. The physically relevant quantity is therefore the associated scattering rate. A Golden-Rule estimate gives
\begin{equation}
\label{eq:scatt_rate}
    \gamma_{xp}
    = \pi g_{xp}^2(\omega)\,\varrho(\omega),
\end{equation}
where \(\varrho(\omega)=\sum_{k_z}\delta(\omega-\omega_p)\) is the phonon density of states. Approximating the longitudinal dispersion as \(\omega_p\approx v_s|k_z|\) yields \(\varrho(\omega)\approx V/(2\pi v_s)\), with \(V=AL\) the effective quantization volume of the quantum well. Linearizing about a condensate of density \(n_x\), this scattering rate translates into an effective diffusion rate for BdG excitations. At the level of the toy model, we identify
\begin{equation}
\label{eq:Gamma_estimate}
    \Gamma \sim n_x\,\gamma_{xp}
    \;\approx\;
    \frac{n_x g_{xp}^2\,V}{2v_s},
\end{equation}
up to prefactors of order unity. This provides a direct physical interpretation of \(\Gamma\) as the rate at which lattice vibrations induce diffusive noise in the BdG sector.
\end{example}

\subsubsection*{Open-system toy model}
To focus on the impact of decoherence in an idealized polaritonic QFCS emulator, we retain two effective (radially ingoing and outgoing) BdG modes with annihilation operators $\hat A$ and $\hat B$, physically representing the dominant superradiant scattering channels at a fixed laboratory frequency. We model quantum superradiance phenomenologically as a parametric mixing process (two-mode squeezing) at rate $\lambda$ between these modes. In analogy with Wald’s basis for rotating black holes (Section~\ref{section:bhs}), $\hat A$ plays the role of an ingoing mode $\varphi_J^{\rm in}$, while $\hat B$ corresponds to the near-horizon ``up'' mode $\varphi_J^{\rm up}$; crucially, in the present setting both modes represent standing waves inside a driven, dissipative microcavity.

Radiative decay couples the trapped BdG modes to photonic output channels $\hat a$ and $\hat b$ at rates $\kappa_A$ and $\kappa_B$, respectively. Rather than acting as a nuisance, radiative relaxation dissipatively stabilizes the parametric dynamics\footnote{Otherwise an instability arises, akin to a black-hole bomb~\cite{press1972floating}!} and provides an optical readout channel for the stationary correlations of BdG sector.

We incorporate incoherent and experimentally inaccessible scattering processes---most notably phonon-induced decoherence---by coupling each BdG mode to an independent Gaussian bath. The bath modes have an effective thermal occupancy $\bar n_e$ and induce diffusion of BdG quadratures at rate $\Gamma$; microscopically, this noise originates from lattice vibrations in the semiconductor host, which couple to the excitonic component of the polariton modes through elastic deformation~\cite{Frerot2023:ThermalPolaritons}. We model this by introducing two independent bath input fields, $\hat e_{\rm in}(t)$ and $\hat e'_{\rm in}(t)$, that couple locally to $\hat A$ and $\hat B$, as illustrated in Fig.~\ref{fig:toy-model}.

In what follows we write the Heisenberg-Langevin equations, state the stability threshold (highlighting dissipation as a stabilization mechanism), and compute a stationary input-output scattering matrix for radiative channels. This lets us estimate when output entanglement survives in the presence of phonon-induced diffusion, including the non-trivial role of \emph{phononic vacuum fluctuations} in the output population and entanglement.

\paragraph*{Heisenberg-Langevin equations.\footnote{See Example 3 in Part~\ref{lecture:1} for a brief background of Heisenberg-Langevin equations.}}
We take the interaction Hamiltonian
\begin{equation}
    \hat H_{\rm int}(t)=\hat H_A(t)+\hat H_B(t)+\hat H_{AB}(t)+\hat H_E(t),
\end{equation}
with
\begin{align}
    \hat H_A(t)
    &= i\sqrt{\kappa_A}\Big(\hat a_{\rm in}^\dagger(t)\hat A(t)-\hat a_{\rm in}(t)\hat A^\dagger(t)\Big),\\
    \hat H_B(t)
    &= i\sqrt{\kappa_B}\Big(\hat b_{\rm in}^\dagger(t)\hat B(t)-\hat b_{\rm in}(t)\hat B^\dagger(t)\Big),\\
    \hat H_{AB}(t)
    &= i\lambda\Big(\hat A^\dagger(t)\hat B^\dagger(t)-\hat A(t)\hat B(t)\Big),\\
    \hat H_E(t)
    &= i\sqrt{\Gamma}\Big[
      \big(\hat e_{\rm in}(t)-\hat e_{\rm in}^\dagger(t)\big)\big(\hat A(t)+\hat A^\dagger(t)\big)
      +
      \big(\hat e_{\rm in}'(t)-\hat e_{\rm in}^{\prime\,\dagger}(t)\big)\big(\hat B(t)+\hat B^\dagger(t)\big)
    \Big].
\end{align}
Here $\kappa_{A,B}$ are radiative decay rates into accessible photonic channels, $\lambda$ is the pair-production rate between the effective BdG modes, and $\Gamma$ is an effective phonon-induced diffusion rate. We treat the bath operators as Markov input fields, with commutators
\begin{equation}
[\hat a_{\rm in}(t),\hat a_{\rm in}^\dagger(t')]=\delta(t-t'),\quad
[\hat b_{\rm in}(t),\hat b_{\rm in}^\dagger(t')]=\delta(t-t'),\quad
[\hat e_{\rm in}(t),\hat e_{\rm in}^\dagger(t')]=\delta(t-t'),
\end{equation}
(and similarly for $\hat e'_{\rm in}$), and all cross-commutators zero.

With these conventions, the input-output boundary conditions take the standard form
\begin{equation}
\label{eq:ab_inout_flip}
    \hat a_{\rm out}(t)=\hat a_{\rm in}(t)+\sqrt{\kappa_A}\,\hat A(t),
    \qquad
    \hat b_{\rm out}(t)=\hat b_{\rm in}(t)+\sqrt{\kappa_B}\,\hat B(t).
\end{equation}
We do not introduce $\hat e_{\rm out}$ operators since the phonon channels are inaccessible, acting purely as incoherent (noise) drives.

Deriving the Heisenberg equations of motion and invoking the input-output conditions~\ref{eq:ab_inout_flip} (see Example~3 in Part~\ref{lecture:1} for a similar calculation), we find the Heisenberg-Langevin equations for the effective BdG modes:
\begin{align}
\label{eq:HL-eqn-AB}
    \partial_t\hat A
    &= -\frac{\kappa_A}{2}\hat A + \lambda\,\hat B^\dagger
    - \sqrt{\kappa_A}\,\hat a_{\rm in}
    + \sqrt{\Gamma}\,\big(\hat e_{\rm in}-\hat e_{\rm in}^\dagger\big),\\
    \partial_t\hat B
    &= -\frac{\kappa_B}{2}\hat B + \lambda\,\hat A^\dagger
    - \sqrt{\kappa_B}\,\hat b_{\rm in}
    + \sqrt{\Gamma}\,\big(\hat e_{\rm in}'-\hat e_{\rm in}^{\prime\,\dagger}\big).
\end{align}
The linear drift matrix of \eqref{eq:HL-eqn-AB} has eigenvalues with negative real part iff the parametric gain does not overwhelm radiative relaxation:\footnote{To see this quickly, adiabatically eliminate $\hat B$ and reduce the dynamics to a single-mode decay equation for $\hat A$ with effective dissipation rate $\tilde{\kappa}=\kappa_A-4\lambda^2/\kappa_B$, which is only positive under the stability condition.}
\begin{equation}
\label{eq:stability_condition}
    4\lambda^2 \le \kappa_A\kappa_B.
\end{equation}
Radiative decay therefore plays a dual role. It dissipatively stabilizes the parametric squeezing process and also provides an optical readout channel.

\paragraph*{Stationary (zero-frequency) scattering matrix.}
We now derive the stationary input-output map relating the accessible outputs $(\hat a_{\rm out},\hat b_{\rm out})$ to the inputs $(\hat a_{\rm in},\hat b_{\rm in},\hat e_{\rm in},\hat e_{\rm in}')$. For transparency, we set $\kappa_A=\kappa_B\triangleq\kappa$ in what follows; the asymmetric case is analogous but notationally heavier.

In the zero-frequency (stationary) limit we enforce $\partial_t\hat A=\partial_t\hat B=0$
in \eqref{eq:HL-eqn-AB} (a convenient trick) and solve the coupled linear equations using the input-output relations~\eqref{eq:ab_inout_flip}. Defining
\begin{equation}
    \mathcal{C}\triangleq \frac{4\lambda^2}{\kappa^2},
    \qquad
    \mathcal{D}\triangleq \frac{\Gamma}{\kappa},
\end{equation}
with $\mathcal{C}<1$ by stability, we obtain
\begin{align}
\label{eq:aout_expanded_flip}
    \hat a_{\rm out}
    &= \frac{1+\mathcal{C}}{1-\mathcal{C}}\,\hat a_{\rm in}
     -\frac{2\sqrt{\mathcal{C}}}{1-\mathcal{C}}\,\hat b_{\rm in}^\dagger
     -\frac{2\sqrt{\mathcal{D}}}{1-\mathcal{C}}\big(\hat e_{\rm in}-\hat e_{\rm in}^\dagger\big)
     +\frac{2\sqrt{\mathcal{D}\mathcal{C}}}{1-\mathcal{C}}\big(\hat e_{\rm in}'-\hat e_{\rm in}^{\prime\,\dagger}\big),\\
\label{eq:bout_expanded_flip}
    \hat b_{\rm out}^\dagger
    &= \frac{1+\mathcal{C}}{1-\mathcal{C}}\,\hat b_{\rm in}^\dagger
     -\frac{2\sqrt{\mathcal{C}}}{1-\mathcal{C}}\,\hat a_{\rm in}
     -\frac{2\sqrt{\mathcal{D}}}{1-\mathcal{C}}\big(\hat e_{\rm in}'-\hat e_{\rm in}^{\prime\,\dagger}\big)
     +\frac{2\sqrt{\mathcal{D}\mathcal{C}}}{1-\mathcal{C}}\big(\hat e_{\rm in}-\hat e_{\rm in}^\dagger\big).
\end{align}
The first two terms establish the familiar two-mode-squeezing transformation between the photonic channels; the remaining terms quantify phonon-induced diffusion. Crucially, these noise terms include $\hat e^\dagger$ operators and
therefore contribute even at $\bar n_e=0$. \emph{In other words, phononic vacuum fluctuations emerge as measurable output radiation and, furthermore, degrade entanglement in the BdG sector without any thermal seeding.}

Packaging \eqref{eq:aout_expanded_flip}-\eqref{eq:bout_expanded_flip} into a scattering matrix provides the compact form
\begin{equation}
\label{eq:scattering_ab_flip}
    \begin{pmatrix}
        \hat a_{\rm out} \\[.3em] \hat b_{\rm out}^\dagger
    \end{pmatrix}
    =
    \underbrace{
    \begin{pmatrix}
        \frac{1+\mathcal{C}}{1-\mathcal{C}} & -\frac{2\sqrt{\mathcal{C}}}{1-\mathcal{C}}\\[.6em]
        -\frac{2\sqrt{\mathcal{C}}}{1-\mathcal{C}} & \frac{1+\mathcal{C}}{1-\mathcal{C}}
    \end{pmatrix}}_{\bm S}
    \begin{pmatrix}
        \hat a_{\rm in} \\[.3em] \hat b_{\rm in}^\dagger
    \end{pmatrix}
    + \hat{\bm f}_{\rm out},
\end{equation}
where
\begin{equation}\label{eq:fout_noise}
        \hat{\bm f}_{\rm out}\triangleq
        \frac{2\sqrt{\mathcal{D}}}{1-\mathcal{C}}
        \begin{pmatrix}
            1 & -\sqrt{\mathcal{C}}\\[.4em]
          \sqrt{\mathcal{C}} & -1
        \end{pmatrix}
        \begin{pmatrix}
            \hat{e}_{\rm in}-\hat{e}_{\rm in}^\dagger \\[.4em]
            \hat{e}_{\rm in}'-\hat{e}_{\rm in}^{\prime\,\dagger}
        \end{pmatrix}
\end{equation}
denotes the phononic noise operator, which entirely describes the open-system character of the setup. The gain of the two-mode squeezing process is $G=(1+\mathcal{C})^2/(1-\mathcal{C})^2$.

\paragraph*{Outgoing flux, entanglement, and the role of phonon noise.}
Take the photonic inputs to be vacuum and the phonon inputs to be thermal and uncorrelated,
\begin{equation}
    \langle \hat e_{\rm in}(t)\hat e_{\rm in}^\dagger(t')\rangle
    =
    \langle \hat e_{\rm in}'(t)\hat e_{\rm in}^{\prime\,\dagger}(t')\rangle
    =
    (1+\bar n_e)\,\delta(t-t').
\end{equation}
Then $\langle \hat a_{\rm out}^\dagger \hat a_{\rm out}\rangle
=
\langle \hat b_{\rm out}^\dagger \hat b_{\rm out}\rangle$, and we compute the radiative flux (photons per second per bandwidth) 
\begin{equation}
\label{eq:flux_2resonator_rewrite}
    \langle \hat a_{\rm out}^\dagger \hat a_{\rm out}\rangle
    =
    \underbrace{\frac{4\mathcal{C}}{(1-\mathcal{C})^2}}_{\bar n_s}
    +
    \underbrace{\frac{4\mathcal{D}(1+\mathcal{C})}{(1-\mathcal{C})^2}\,(1+2\bar n_e)}_{\bar n_E}.
\end{equation}
The superradiant contribution is $\bar n_s$, while $\bar n_E$ is the dressed phonon contribution. Importantly, $\bar n_E$ remains nonzero as $\bar n_e\to 0$ due to phononic vacuum fluctuations. [Physically these effective fluctuations are not intrinsic to the phonon bath alone, but arise from backaction: lattice vibrations couple to a macroscopically occupied excitonic component, effectively converting excitonic coherence into diffusive noise in the BdG sector. In the limit of vanishing exciton fraction, this contribution disappears~\cite{Frerot2023:ThermalPolaritons}.]

From the scattering matrix and applying techniques from Part~\ref{lecture:3} (see example box below), we quantify entanglement and, furthermore, derive a necessary and sufficient \emph{separability condition}.  In the radiative-decay dominated regime $\lambda,\Gamma\ll\kappa$, we find that entanglement between the observable photonic channels vanishes iff
\begin{equation}
\label{eq:rate_condition_rewrite}
    \qq{(separability condition)} \qquad 2\lambda \;\lesssim\; \Gamma\,(1+2\bar n_e).
\end{equation}
Equivalently, in terms of mean occupation numbers, a convenient proxy is
\begin{equation}
\label{eq:occupation_condition_rewrite}
    \qq{(separability condition)} \qquad 2\sqrt{\bar n_s}\;\lesssim\;\bar n_E.
\end{equation}
This is drastically less restrictive than a naive ``balanced emission'' condition $\bar n_s\sim \bar n_E$ that one might expect to be necessary for quantum correlations to be present and observable.

To be more quantitative, take $\Gamma\approx 0.6~{\rm GHz}$ ($2.5~\mu{\rm eV}$)~\cite{Frerot2023:ThermalPolaritons} and assume $\bar n_e\ll 1$. The rate criterion \eqref{eq:rate_condition_rewrite} suggests a threshold $\lambda \gtrsim \lambda_{\star}\triangleq \Gamma/2 \approx 0.3~{\rm GHz}$. Given a cavity decay rate $\kappa\approx 10~{\rm GHz}$ ($41.7~\mu{\rm eV}$), this corresponds to a critical cooperativity
\[
\mathcal{C}_{\star}=\frac{4\lambda_\star^2}{\kappa^2}
=\frac{\Gamma^2}{\kappa^2}\approx 3.6\times 10^{-3}.
\]
Since $G=(1+\mathcal{C})^2/(1-\mathcal{C})^2$, this implies $G_\star-1\approx 4\mathcal{C}_\star\approx 1.4\%$ as an amplification threshold to observe entanglement in the radiative output modes. For comparison, a naive condition based on balanced emission $\bar n_s\approx \bar n_E$ occurs at $\mathcal{C}_{\rm bal}\sim \Gamma/\kappa\approx 0.06$, which implies
\[
G_{\rm bal}-1=\frac{4\mathcal{C}_{\rm bal}}{(1-\mathcal{C}_{\rm bal})^2}
\approx 27\%,
\]
substantially more demanding than the rigorous estimate from separability criteria.

\begin{example}{Separability Criterion in Quadrature Variables}
We recast the stationary input--output map
\eqref{eq:aout_expanded_flip}--\eqref{eq:bout_expanded_flip} in canonical
quadratures. We may then use techniques developed in Parts~(\ref{lecture:1})-(\ref{lecture:3}) to analyze entanglement in the system. For any mode $\hat c$, let $\hat q^{(c)}\triangleq(\hat c+\hat c^\dagger)/\sqrt2$ and $\hat p^{(c)}\triangleq i(\hat c^\dagger-\hat c)/\sqrt2$. Introduce the
quadrature vectors
\[
\hat{\bm r}_{\rm in}\triangleq(\hat q_{\rm in}^{(a)},\hat p_{\rm in}^{(a)},\hat q_{\rm in}^{(b)},\hat p_{\rm in}^{(b)})^\top,\quad
\hat{\bm r}_{\rm out}\triangleq(\hat q_{\rm out}^{(a)},\hat p_{\rm out}^{(a)},\hat q_{\rm out}^{(b)},\hat p_{\rm out}^{(b)})^\top,
\]
and analogously $\hat{\bm r}_{e}$ for the two phonon baths $(e,e')$. Given $\mathcal{C}=4\lambda^2/\kappa^2$ and $\mathcal{D}=\Gamma/\kappa$,
the input-output quadrature relations read
\begin{equation}
\label{eq:r_out_linear_example}
\hat{\bm r}_{\rm out}
=
\bm{S}_{\rm tms}\,\hat{\bm r}_{\rm in}
+
\bm{B}\,\hat{\bm r}_{e},
\end{equation}
where $\bm{S}_{\rm tms}$ is the standard two-mode-squeezing matrix [Eq.~\eqref{eq:tms_transform}] and $\bm{B}$ encodes the diffusion drives. In this toy model, phonons drive only the $p$ quadratures, so $\bm{B}$ has support on rows $2$ and $4$. By Eq.~\eqref{eq:moments-inout}, the covariance matrix obeys the input-output rule
\begin{equation}
\label{eq:sigma_out_example}
\bm\sigma_{\rm out}
=
\bm{S}_{\rm tms}\,\bm\sigma_{\rm in}\,\bm{S}_{\rm tms}^{\top}
+
\bm V,
\qquad
\bm V\triangleq \bm{B}\,\bm{\sigma}_e\,\bm{B}^{\top},
\end{equation}
where $\bm{\sigma}$ denotes the bare covariance matrix of the phononic bath modes.  Taking vacuum photonic inputs ($\bm\sigma_{\rm in}=\bm I_4$) and independent thermal phonons with occupancy $\bar n_e$ ($\bm{\sigma}_e=(\bar n_e+\tfrac12)\bm I_4$), we obtain an additive noise matrix of the form
\begin{equation}
\label{eq:V_form_example}
\bm V=
\begin{pmatrix}
0&0&0&0\\
0&\sigma^2&0&V_\times\\
0&0&0&0\\
0&V_\times&0&\sigma^2
\end{pmatrix},
\qquad
\sigma^2=\left(\frac{8\mathcal{D}(1+\mathcal{C})}{(1-\mathcal{C})^2}\right)(1+2\bar n_e),
\quad
V_\times=-\left(\frac{2\sqrt{\mathcal{C}}}{1+\mathcal{C}}\right)\sigma^2.
\end{equation}
The key point is that $\bm V\neq 0$ even at $\bar n_e=0$: phononic vacuum fluctuations inject noise in the outgoing photonic fields. 

\vspace{1em}

We evaluate entanglement according to the PPT criterion introduced in Part~\ref{lecture:3}. Let $\tilde\nu_-$ be the smallest symplectic PPT eigenvalue. The two-mode output state is separable iff $\tilde\nu_-\ge 1$ (and entangled iff $\tilde\nu_-<1$). For the present diffusion model, we find:
\begin{equation}
\label{eq:nu_tilde_example}
\tilde{\nu}_-^2=
\frac{(1-\sqrt{\mathcal{C}})^6\big(1+\mathcal{C}-2\sqrt{\mathcal{C}}+8\mathcal{D}(1+2\bar n_e)\big)}{(1-\mathcal{C})^4}.
\end{equation}
For fast radiative decay $(\mathcal{C},\mathcal{D}\bar n_e)\ll 1$, this yields the separability condition
\begin{equation}
\label{eq:sep_approx_example}
\sqrt{\mathcal{C}}\;\lesssim\;\mathcal{D}(1+2\bar n_e)
\qquad\Longleftrightarrow\qquad
2\lambda\;\lesssim\;\Gamma(1+2\bar n_e),
\end{equation}
which makes explicit that phonon vacuum fluctuations ($\bar n_e=0$) can extinguish entanglement when $\Gamma$ and $\lambda$ are commensurate.
\end{example}

\paragraph*{Closing remarks.}
The simplified toy model deliberately strips the polariton QFCS emulator down to its essential ingredients: superradiant (two-mode-squeezing) pair production, radiative readout, and spurious decoherence effects. Despite its minimalism, the model captures a key and nontrivial lesson for analogue gravity in quantum fluids of light: Open-system effects (e.g., phonon-induced diffusion) can inject sufficient noise to suppress entanglement in the outgoing superradiant- or Hawking-pair modes, even when thermal fluctuations are presumably negligible. Radiative relaxation plays a dual role by stabilizing the dynamics and enabling direct optical access to quantum correlations in the BdG sector.

Suffice to say that this analysis is not a substitute for a full microscopic treatment. A complete open-system BdG description, including realistic mode structure, spatial inhomogeneity, and detailed decoherence effects, remains an important task. More pressing, however, is the actual challenge of \emph{verifying} entanglement in any practical analogue gravity experiment: identifying robust observables, measurement strategies, and operating regimes in which genuine quantum correlations persist.

\clearpage

\begin{nutshell}{Gaussian physics at horizons: from black holes to polaritons}

\textbf{Semi-classical black holes.}
\begin{itemize}
    \item In the semi-classical regime---quantum fields propagating on a fixed classical spacetime---the dynamics of free bosonic fields during gravitational collapse is fully \emph{Gaussian}. All physical content resides in first and second moments.

    \item Once the geometry settles to a stationary black hole, each mode sector
    $J=\{\omega,\ell,m\}$ evolves independently through a single-mode Gaussian channel,
    \[
      \mathcal{N}_{J}=
      \begin{cases}
          \mathcal{L}_{1-\Gamma_J,\bar n_H}, & \varpi>0,\\[3pt]
          \mathcal{A}_{G_J,\bar n_H}, & \varpi<0,
      \end{cases}
    \]
    where $\varpi=\omega-m\Omega_H$ determines whether the channel is superradiant, $\Gamma_J$ is the greybody factor,
    $G_J$ the amplification gain, and
    $\bar n_H=(e^{|\varpi|/T_H}-1)^{-1}$ the Hawking occupation number.

    \item Because the evolution is Gaussian, particle fluxes, amplification, entropy production etc. follow from simple transformations of the ingoing covariance matrix.

    \item Tensions emerge when the semi-classical picture is pushed to its limits: An isolated black hole steadily loses mass via Hawking emission, producing mixed radiation entangled with interior modes that disappear behind the horizon—suggesting, within semi-classical theory, an apparent loss of quantum information. Furthermore, the exterior entropy of radiation grows, while the Bekenstein-Hawking entropy shrinks. Reconciling this apparent conundrum lies at the heart of the information problem. 
\end{itemize}

\textbf{Analogue gravity.}
\begin{itemize}
    \item The Hawking effect relies on two ingredients: quantum fields and an effective horizon. Near a horizon, logarithmic phase accumulation leads to spontaneous pair creation with a characteristic temperature (the Hawking temperature) set by the flow gradient.

    \item This mechanism is universal and appears in diverse systems: classical fluids, Bose-Einstein condensates, optical media, and driven exciton-polariton superfluids. In all cases, linearized (BdG) excitations obey Gaussian dynamics.

    \item In optical analogues, intense pulses engineer black- and white-hole horizons via
    traveling refractive-index perturbations. A large relative boost between the comoving frame of the pulse and the laboratory frame renders Hawking correlations remarkably robust against background thermal fluctuations.

    \item Exciton-polariton platforms realize a complementary analogue-gravity regime in which effective horizons or ergoregions form in driven quantum fluids of light. The relevant Bogoliubov excitations are linear (Gaussian) and intrinsically open, with engineered pumping and controlled radiative decay providing both dynamical stabilization and direct optical access to the excitation sector.

    \item A distinctive feature of polariton systems is phonon-induced diffusion arising from the solid-state environment. Remarkably, \emph{phononic vacuum fluctuations} suppress effective spacetime-induced entanglement even at zero temperature.

\end{itemize}

\textbf{Perspective.}
Across astrophysical black holes, optical horizons, and polariton quantum fluids, the same
Gaussian structure underlies spacetime-induced particle creation and entanglement. Understanding and verifying these quantum correlations in realistic, open analogue quantum systems remains a central challenge for QFCS emulators.
\end{nutshell}


\acknowledgements
I am grateful to Ivan Agullo for invigorating conversations and insight over the years, and to the organizers of \textit{Analogue Gravity 2023} for the invitation to lecture in the beautiful town of Benasque, Spain. I also thank Adri\`a Delhom for fruitful discussions and supplying the data used to produce Figs.~\ref{fig:dMdL_spectra} and~\ref{fig:negativity_entropy_Tin}. A.J.B. acknowledges support from the NRC Research Associateship Program at NIST.


\bibliography{main}
\end{document}